\documentclass[useAMS,usenatbib]{mn2e}
\usepackage{graphicx}
\usepackage{amsmath,amssymb}
\usepackage{hyperref}
\usepackage{enumitem}
\usepackage{epstopdf}
\usepackage{color}
\usepackage{xcolor}

\newcommand{\Msol}{M_{\odot}}
\newcommand{\gccm}{\textrm{g cm}^{-3}}
\newcommand{\ms}{\textrm{ms}}
\newcommand{\erg}{\textrm{erg}}
\newcommand{\ergs}{\textrm{erg} \, \textrm{s}^{-1}}

\newcommand{\mbi}[1]{\textbf{\em #1}}

\newcommand{\eps}{\epsilon}
\newcommand{\cl}{c}
\newcommand{\eul}{\textrm{e}}
\newcommand{\hn}{h}
\newcommand{\vecx}{\textbf{\em x}}
\newcommand{\vecp}{\textbf{\em p}}
\newcommand{\vecv}{\textbf{\em v}}

\newcommand{\vecn}{\textbf{\em n}}
\newcommand{\vecF}{\textbf{\em F}}

\newcommand{\vect}[1]{\textbf{\em #1}}
\newcommand{\half}{\frac{1}{2}}

\newcommand{\n}{\mathsf{n}}
\newcommand{\hi}{\mathsf{i}}
\newcommand{\hj}{\mathsf{j}}
\newcommand{\hk}{\mathsf{k}}

\newcommand{\aap}{A\&A}
\newcommand{\apss}{Ap\&SS} 

\newcommand{\apjs}{ApJS}
\newcommand{\apj}{ApJ}

\newcommand{\prd}{Phys. Rev. D}

\newcommand{\mnras}{MNRAS}

\newcommand{\physrep}{Phys. Rep.}

\newcommand{\jqsrt}{J.~Quant.~Spec.~Radiat.~Transf.}

\def\ga{\,\,\raise0.14em\hbox{$>$}\kern-0.76em\lower0.28em\hbox
{$\sim$}\,\,}
\def\la{\,\,\raise0.14em\hbox{$<$}\kern-0.76em\lower0.28em\hbox
{$\sim$}\,\,}

\title[A new multidimensional neutrino-hydrodynamics code]
{A new multidimensional, energy-dependent two-moment transport code for neutrino-hydrodynamics}

\author[Just, Obergaulinger, \& Janka]
{O. Just,$^{1,2}$ M. Obergaulinger,$^3$ and H.-T.~Janka$^1$ \\
  $^1$Max-Planck-Institut f\"ur Astrophysik, Postfach 1317, 85741 Garching, Germany \\
  $^2$Max-Planck/Princeton Center for Plasma Physics (MPPC) \\
  $^3$Departament d{\'{}}Astronomia i Astrof{\'i}sica, Universitat de Val{\`e}ncia, \\
  Edifici d{\'{}}Investigaci{\'o} Jeroni Mu{\~n}oz, C/ Dr.~Moliner, 50, E-46100 Burjassot (Val{\`e}ncia), Spain
}

\date{Released 2015 Xxxxx XX}

\pagerange{\pageref{firstpage}--\pageref{lastpage}} \pubyear{2015}

\begin{document}

\label{firstpage}

\maketitle

\begin{abstract}
  We present the new code ALCAR developed to model multidimensional, multi energy-group neutrino
  transport in the context of supernovae and neutron-star mergers. The algorithm solves the
  evolution equations of the 0th- and 1st-order angular moments of the specific intensity,
  supplemented by an algebraic relation for the 2nd-moment tensor to close the system. The scheme
  takes into account frame-dependent effects of order $\mathcal{O}(v/c)$ as well as the most
  important types of neutrino interactions. The transport scheme is significantly more efficient
  than a multidimensional solver of the Boltzmann equation, while it is more accurate and consistent
  than the flux-limited diffusion method. The finite-volume discretization of the essentially
  hyperbolic system of moment equations employs methods well-known from hydrodynamics. For the time
  integration of the potentially stiff moment equations we employ a scheme in which only the local
  source terms are treated implicitly, while the advection terms are kept explicit, thereby allowing
  for an efficient computational parallelization of the algorithm. We investigate various problem
  setups in one and two dimensions to verify the implementation and to test the quality of the
  algebraic closure scheme. In our most detailed test, we compare a fully dynamic, one-dimensional
  core-collapse simulation with two published calculations performed with well-known Boltzmann-type
  neutrino-hydrodynamics codes and we find very satisfactory agreement.
\end{abstract}



\section{Introduction}\label{sec:intro}

In various astrophysical scenarios involving matter in a hot and dense phase, neutrino interactions
take place in a way that the full transport problem -- which consistently follows the emission,
propagation and absorption of neutrinos -- needs to be taken into account to correctly describe
these systems. A prominent example is a core-collapse supernova (CCSN), in which according to the
present standard model the explosion is essentially only made possible by the energy deposition due
to the re-absorption of neutrinos just produced in the proto-neutron star (proto-NS; see
\citealp{Janka2012, Burrows2013} for recent reviews). Genuine neutrino-transport effects can also be
crucial for determining the properties of potentially nucleosynthesis-relevant outflows and may even
give rise to these outflows to begin with. Such outflows are believed to occur during a CCSN in the
form of a neutrino-driven wind expelled from the surface of the proto-NS
\citep[e.g.][]{Qian1996}. Another example is a massive NS formed during the merger of two NSs
\citep{Dessart2009}, or similarly a black-hole (BH) torus configuration also produced by such a
merger or by the merger of a NS and a BH \citep[e.g][]{Wanajo2012, Fernandez2013}. Another
astrophysical scenario in which neutrino transport may be crucial is the launching of a gamma-ray
burst jet, which could be powered to some degree by neutrino pairs annihilating in the polar regions
of a BH-torus system \citep[e.g.][]{Popham1999}.

Unfortunately, most multidimensional results for the aforementioned scenarios stem from more or less
idealized investigations, mainly owing to the enormous computational requirements of a
time-dependent, multidimensional treatment of neutrino transport. The level of simplification is
typically chosen to provide the optimal balance between accuracy and computational expense, given
the constraints of the available computational resources and the considered physical effects to be
captured to a sufficient degree. The most accurate neutrino-transport schemes follow the full
spatial, energetic and directional dependence of the neutrino distribution, described by the
Boltzmann equation. However, multidimensional applications of these ``Boltzmann solvers'' are
severely constrained by their complexity and computational expense and force modern simulations to
employ miscellaneous restrictions, such as the ``ray-by-ray'' approach \citep[e.g.][]{Buras2006,
  Hanke2012}, the omission of (a subset of) velocity-dependent terms and the decoupling of neutrinos
with different energies (\citealp[e.g.][]{Livne2004, Ott2008}), or the investigation of only
individual static background configurations of matter \citep{Sumiyoshi2015}. A computationally
cheaper alternative to a Boltzmann solver is the ``flux-limited diffusion'' (FLD) method \citep[see,
e.g.,][]{Levermore1981, Mihalas1984}, in which the radiation energy density (as represented by the
0th angular moment of the specific intensity, see Sec.~\ref{sec:equations} for explicit definitions)
is evolved assuming that the radiation flux density (as represented by the 1st angular moment) can
be written as a function of the radiation energy density. This method is the simplest realization of
a ``truncated moment scheme''. However, FLD implies several physical as well as technical drawbacks.

In this paper we present a genuinely multidimensional, fully energy-dependent
radiation-hydrodynamics (RHD) scheme where both the energy density and the flux density of neutrinos
are evolved quantities. The system of equations is closed by an algebraic expression for the
Eddington tensor (defined as the normalized 2nd angular moment of the specific intensity) as
function of the lower-order moments. This makes the scheme a specific realization of the class of
``variable Eddington factor'' methods and it will therefore be denoted here as ``algebraic Eddington
factor'' (AEF) method.

In both methods, AEF and FLD, the closure (i.e. the corresponding angular moment which closes the
set of moment equations) is assumed to be given solely as a function of the local values of the
evolved radiation moments (or additionally of the local gas properties in the case of FLD). Since in
general the closure is a non-local function of the surrounding radiation sources, it is clear that a
universal closure relation between the local angular moments, corresponding to an equation of state
as in hydrodynamics, does not exist for arbitrarily shaped radiation fields. It is nevertheless
conceivable that for not too complex geometries (e.g. single-source configurations like a PNS or a
NS-merger remnant) the radiation field tends to be arranged such that typical relations between the
lowest angular moments are fulfilled up to a sufficient degree. Keeping also in mind that not the
full angular information but rather only the angular moments of the radiation field enter the
hydrodynamics part of the full RHD system, the fluid may only be secondarily affected by the error
introduced by the approximate closure. Certainly, the actual performance and applicability of either
method, FLD or AEF, is problem dependent and has to be examined individually for a given problem
setup. However, the main advantage of an AEF scheme (and also of an FLD scheme) compared to a more
accurate Boltzmann solver is its computational efficiency, which allows to perform genuinely
multidimensional RHD simulations and larger sets of model calculations with reasonable computational
effort in the first place.

Compared to a standard FLD solver, the AEF method as presented in this paper features the following
advantages: (1) It is potentially more accurate, simply on account of the fact that an FLD scheme is
essentially an AEF scheme in which a certain collection of terms is dropped
\citep[e.g.][]{Levermore1981, Cernohorsky1990, Dgani1991}. (2) It is more consistent: A particular
consequence of retaining the evolution equations for the 1st moments in AEF is that the conservation
of the total (radiation plus fluid) momentum and therefore also of the total energy can be ensured
in the case of momentum-exchanging interactions between neutrinos and the gas \citep{Baron1989,
  Cernohorsky1990}. Another, related advantage is that the 1st-moment vectors can in principle point
into arbitrary directions in an AEF scheme and are not forced to be parallel to the gradient of the
0th moments as in FLD. This allows, for example, to describe shadows behind illuminated opaque
structures (see the test problem in Sec.~\ref{sec:shad-cast-probl}). (3) The different mathematical
nature of the evolution equations in AEF (hyperbolic as opposed to parabolic in FLD) enables the
application of time-explicit methods based on Riemann-solvers for the advection part of the
system. Such methods are well-known from hydrodynamics. In contrast, in a time-explicit treatment of
FLD the time step would in principle be unbound from below, which in practice forces one to employ a
fully time-implicit scheme. An explicit scheme compared to an implicit one is particularly
advantageous in multidimensional simulations, however, because first, its computational requirements
only scale linearly with the size of the grid, and second, the computational parallelization is
generally more efficient and straightforward.

The strategy of using an AEF scheme for radiation transport is not new\footnote{In the literature,
  these schemes are sometimes simply denoted as ``$M_1$ schemes'', which actually only refers to the
  specific $M_1$ closure being used to express the Eddington tensor in a truncated two-moment scheme
  (see Sec.~\ref{sec:analyt-eddingt-fact} for details).}. Multidimensional applications regarding
photon transport exist in a number of realizations \citep[see, e.g., ][]{Audit2002, Hayes2003,
  Gonz'alez2007, Aubert2008, Vaytet2011, Scadowski2013, Skinner2013, McKinney2014}. Until recently,
however, only a few investigations considered the AEF scheme in the context of neutrino transport
\citep{Schinder1989, Dgani1991, Koerner1992, Cernohorsky1992} , although several studies concerning
aspects of the closure prescription \citep{Cernohorsky1994, Bludman1995, Smit2000} and the solution
strategies \citep{Smit1997, Pons2000} elucidated its capabilities. In recent works by
\citet{Shibata2011a} and \citet{Cardall2013a} the truncated two-moment scheme was extended to a
general relativistic framework.  \citet{Shibata2012} made use of this framework in the ``grey''
approximation (i.e. averaging over the neutrino energies) and presented axisymmetric
neutrino-magnetohydrodynamics simulations of BH accretion tori as models for central engines of
gamma-ray bursts.  \citet{Kuroda2012, Kuroda2014} combined the relativistic AEF scheme with an
isotropic diffusion source approximation (IDSA, as developed in \citealp{Liebendorfer2009}) to
perform three-dimensional CCSN simulations. A general relativistic AEF scheme was implemented in the
GR1D code by \citet{OConnor2013}, who conducted one-dimensional, energy-dependent CCSN simulations
neglecting all velocity-dependent terms and energy-coupling interactions. Recently,
\citet{OConnor2014} released an improved version of this scheme, including velocity-dependent terms
and energy-coupling interactions, but still constrained to spherical symmetry. Another
implementation of the AEF scheme, in special relativity and using a grey approximation, was
presented in \citet{Takahashi2013a}.

In contrast to the schemes used in the aforementioned studies, the neutrino-transport code presented
here, which will be designated the name ``ALCAR'' (\textbf{A}lgebraic \textbf{L}ocal
\textbf{C}losure \textbf{A}pproach for \textbf{R}adiation), combines all of the following
features: It is genuinely multidimensional, fully energy dependent and it takes into account all
velocity-dependent terms up to order $\mathcal{O}(v/c)$ as well as energy-coupling due to Doppler
shift and inelastic neutrino-matter interactions. To validate the numerical implementation of the
algorithm and each of its aforementioned features we examine several test problems. However, the
test problems are not only conducted to check the correct numerical implementation of the algorithm
but also serve to examine the central approximation of the scheme embodied in the algebraic
closure. In particular, two tests focus on the question how the AEF method performs in a CCSN setup
and how sensitive the results are to specific choices of the closure. In one test problem, only the
neutrino field is evolved in a fixed hydrodynamic background, making it possible to test the AEF
scheme against the FLD and a Boltzmann scheme independently of the hydrodynamic treatment. In the
other test problem, the results of a fully dynamic CCSN calculation are compared with results
published in \citet{Liebendorfer2005}, which were obtained with the two well-established
Boltzmann-RHD codes VERTEX-PROMETHEUS \citep{Rampp2002} and AGILE-BOLTZTRAN
\citep{Liebendorfer2004}. Our main findings from these tests are that the AEF scheme is sufficiently
accurate to represent a competitive, though computationally much simpler, alternative to Boltzmann
solvers in simulations of CCSNe. Note that this conclusion is in agreement with \citet{OConnor2014},
who conducted a similar comparison between 1D CCSN simulations performed with his general
relativistic AEF code and with the two aforementioned Boltzmann codes.

This article is structured as follows: In Sec.~\ref{sec:equations}, we outline the steps that lead
from the equation of radiative transfer to the moment equations that are solved in our code, and we
motivate and present the approximations and assumptions contained in the AEF method. Subsequently,
in Sec.~\ref{sec:numerics}, we describe the methods used to discretize the moment equations in
space, energy and time. In Sec.~\ref{sec:tests}, we present a suite of test problems, in which we
investigate the quality of the AEF approximation and study features such as the velocity dependence,
the correct behavior in the static and dynamic diffusion limits, and the ability of the scheme to
describe radiation shadowing. Finally, in Sec.~\ref{sec:conc} we summarize our presentation. 

The conventions regarding our notation are as follows: We use lower-case, italic letters
$i,j,k\ldots$ to denote spatial tensor components, lower-case roman letters $\hi,\hj,\hk\ldots$ for
grid indices and $\n$ for the time index. Moreover, we make use of the Einstein notation to write
sums of products of tensor components. Symbols with hats, as for instance $\hat X$, refer to
discretized quantities, while symbols with bars, as for instance $\bar{X}$, denote versions of the
corresponding quantities integrated over the whole energy spectrum. Vectors in spatial and momentum
space are denoted as $\vecx$ and $\vecp$, and $t$ refers to the time coordinate. The symbols $\cl,h$
and $k_{\mathrm{B}}$ mean the speed of light, the Planck constant and the Boltzmann constant,
respectively.


\section{The $\mathcal{O}(v/c)$ Equations of Radiation Hydrodynamics}\label{sec:equations}

In this section we briefly define the basic quantities and present the RHD equations used in our
code.

\subsection{The equation of radiative transfer in the comoving frame}

Both the equations of hydrodynamics and of radiative transfer have their origin in the Boltzmann
equation for the corresponding particle distribution function $\mathcal{F}$, in terms of which
\begin{equation}
  \mathrm{d} N \, = \, \frac{g}{\hn^3} \mathcal{F}(\vecx,\vecp,t)
  \mathrm{d}^3x\,\mathrm{d}^3p
\label{eq:defif}
\end{equation}
is the number of particles within the phase-space volume $\mathrm{d}^3x\,\mathrm{d}^3p$, where $g$
is the statistical weight of the species. ``Radiation'' in the present context is defined as a
distribution of particles that move with the speed of light $\cl$ and that are not subject to an
external force in an inertial frame ($\dot{\vecp} \equiv 0$). The Boltzmann equation for radiative
transfer in a fixed frame then becomes ($\vecn\equiv\vecp/|\vecp|$):
\begin{equation}\label{eq:BE1}
\frac{1}{\cl}\frac{\partial}{\partial t}\mathcal{F} + \vecn\cdot\nabla_{\vecx}\mathcal{F}
\, = \, \mathcal{B} \, .
\end{equation}
Here, and in several following cases, we suppress the functional dependencies. The collision
integral $\mathcal{B}\equiv \mathcal{B}(\vecx,\vecp,t)$ generally contains explicit integrals in
momentum space, making Eq.~(\ref{eq:BE1}) an integro-partial differential equation. Instead of
working with the distribution function directly, for the macroscopic view one prefers using the
frame-dependent specific (i.e. monochromatic) intensity
\begin{equation}
\mathcal{I}(\vecx,\vecn,\eps,t)=(\eps/\hn\cl)^3\,\cl\,\mathcal{F}(\vecx,\vecp,t) \, ,
\label{eq:defI}
\end{equation}
where\footnote{We will use the terms ``energy'' and ``frequency'' interchangeably when referring to
  the corresponding degree of freedom in phase space.} $\eps=|\vecp|\cl$. Bearing in mind that an
essential part of the collision integral depends on the distributions of fluid particles it is often
preferable to measure $\mathcal{I}$ in the frame comoving with the fluid (``comoving frame'',
``fluid frame''), since in that frame the isotropy of the fluid particle distributions\footnote{We
  implicitly assume the fluid to be in local thermodynamic equilibrium (but not necessarily in
  equilibrium with neutrinos).} induces symmetries in the collision integral that make it
computationally most feasible. Using arbitrary, but fixed, Eulerian spatial coordinates defined in a
frame we denote as the laboratory frame (``lab-frame'') and momentum space coordinates (i.e. $\eps$
and $\vecn$) defined in the fluid frame, the comoving-frame equation of radiative transfer up to
order $\mathcal{O}(v/c)$ ($v\equiv|\vecv|$ is the velocity of the fluid as measured in the lab
frame) becomes \citep[e.g.][]{Buchler1979,Kaneko1984,Munier1986a}:
\begin{eqnarray}
  & & \frac{1}{\cl}\frac{\partial\mathcal{I}}{\partial t} + \frac{\mbi{v}\cdot\vecn}{\cl^2}
  \frac{\partial\mathcal{I}}{\partial t} + n^j\frac{\partial\mathcal{I}}{\partial x^j}
  + \frac{v^j}{\cl}\frac{\partial\mathcal{I}}{\partial x^j} \nonumber\\
  & & + \frac{\partial}{\partial\eps} \left[ \mathcal{I}\eps \left( \frac{\mbi{a}\cdot\mbi{n}}{\cl^2} 
      + \frac{1}{\cl}n^j n^k\nabla_j v_k \right) \right] \nonumber\\
  & & + \frac{\partial}{\partial n^i} \left[ \mathcal{I} \left( \frac{\mbi{a}\cdot\mbi{n}}{\cl^2}n^i
      - \frac{a^i}{\cl^2} + \frac{1}{\cl}n^i n^j n^k\nabla_j v_k - \frac{1}{\cl}n^j\nabla_j v^i
       \right.\right. \nonumber\\
  & & \hspace{1cm}\left.\left. - \Gamma^i_{jk}n^j n^k 
      - \frac{1}{\cl}\Gamma^i_{jk}v^j n^k \right) \right] \nonumber\\
  & & + \,\mathcal{I}\left[ 2\frac{\mbi{a}\cdot\mbi{n}}{\cl^2} + \frac{1}{\cl}\nabla_i v^i
    +\Gamma^i_{ij}n^j + \frac{1}{\cl}n^i n^j\nabla_i v_j \right]
  = \mathcal{C} \, ,
\label{eq:transfer1}
\end{eqnarray}
where $\mbi{a}\equiv \partial_t \mbi{v}$, $\Gamma^i_{jk}$ are the Christoffel symbols associated
with the spatial coordinates and $\mathcal{C} \equiv (\eps/\hn\cl)^3\cl \mathcal{B}$.
Equation~(\ref{eq:transfer1}) can be derived from Eq.~(\ref{eq:BE1}) using Eq.~(\ref{eq:defI}) and
the $\mathcal{O}(v/c)$ versions of the Lorentz transformations for $\mathcal{I}$, $\eps$ and
$\vecn$.

\subsection{Angular moments of the transfer equation}\label{sec:radi-moment-equat}

In order to reduce the dimensionality of the radiative transfer problem and to construct the link to
the hydrodynamics equations, one utilizes the fact that the specific intensity is related to the
specific (frequency-integrated) energy density $E$ ($\bar{E}$), energy flux density $F^i$
($\bar{F}^i$) and pressure tensor $P^{ij}$ ($\bar{P}^{ij}$) of radiation by virtue of its angular
moments of increasing order, defined by
\begin{equation}
  \{ \cl E,F^i,\cl P^{ij}, Q^{ijk} \} = \int\mathrm{d}\Omega \,\mathcal{I} \,
  \{ 1,n^i,n^i n^j,n^i n^j n^k \} \, , \label{eq:momdef1} \\
\end{equation}
and
\begin{equation}
  \{ \bar{E},\bar{F}^i,\bar{P}^{ij}, \bar{Q}^{ijk} \} = 
  \int\mathrm{d}\eps \, \{ E,F^i,P^{ij},Q^{ijk} \} \, , \label{eq:momdefgrey1} \\
\end{equation}
where $Q^{ijk}$ and $\bar{Q}^{ijk}$ are the analog 3rd-moment quantities. 

In the following, we neglect terms including the acceleration $a^i$ and the second term containing
the time derivative in Eq.~(\ref{eq:transfer1}). These terms are effectively of order
$\mathcal{O}(v^2/c^2)$ for temporal changes of the velocity and radiation fields occuring on a fluid
timescale $l/v$, where $l$ is a characteristic length scale of changes in the hydrodynamic
background and $v$ a typical fluid velocity (\citealt{Mihalas1984}; see, however,
\citealt{Rampp2002} and \citealt{Lowrie2001} for comments on the second term of
Eq.~(\ref{eq:transfer1})). Temporal changes of these fields on the radiation timescale $l/\cl$ would
enhance the importance of the aforementioned terms. In that case, however, the preceding validity
assumption of the $\mathcal{O}(v/c)$ equation may become questionable to begin with anyway. It is
worth noting that the results of one test in Sec.~\ref{sec:fully-dynam-evol} suggest that for the
conditions in CCSNe the omission of the aforementioned terms is justified.

\subsubsection{Moment equations of energy transport}

The system for the first two moments of Eq.~(\ref{eq:transfer1}), excluding the aforementioned terms
that are effectively of order $\mathcal{O}(v^2/c^2)$, is obtained by performing the angular
integrations as in Eq.~\eqref{eq:momdef1} and it reads
\begin{subequations}\label{eq:tmt1}
\begin{align}
  &\partial_tE + \nabla_jF^j + \nonumber\\
  &\hspace{0.5cm} \underbrace{\nabla_j(v^j E)}_{\mathrm{(I)}} + \underbrace{(\nabla_j v_k)P^{jk}}_{\mathrm{(II)}} -  
    \underbrace{(\nabla_j v_k)\partial_\eps (\eps P^{jk})}_{\mathrm{(IV)}} = C^{(0)} \, , \label{eq:tmte1}\\
  &\partial_tF^i + \cl^2\nabla_jP^{ij} +  \nonumber\\
  &\hspace{0.5cm} \underbrace{\nabla_j(v^j F^i)}_{\mathrm{(I)}} + 
   \underbrace{F^j\nabla_j v^i}_{\mathrm{(III)}}
   - \underbrace{(\nabla_j v_k)\partial_\eps (\eps Q^{ijk})}_{\mathrm{(IV)}} =C^{(1),i} \, , \label{eq:tmtf1}
\end{align}
\end{subequations}
where $C^{(0)} \equiv \int \mathrm{d}\Omega\, \mathcal{C}$ and
$C^{(1),i} \equiv \int\mathrm{d}\Omega\,n^i\,\mathcal{C}$. The labeling by Roman numerals denotes
the different types of velocity-dependent terms: The (I)-terms account for the change of the
comoving-frame moments owing to advection. The (II)-terms account for the change of radiation energy
due to compressional work against the radiation pressure. The (III)-terms account for changes of the
radiation fluxes due to aberration. The energy-coupling (IV)-terms are responsible for the change of
the spectral shape of the radiation field associated with the Doppler shift. Note that in the grey
formulation of the moment equations (i.e. after integrating Eqs.~\eqref{eq:tmt1} over energy $\eps$)
the (IV)-terms vanish. For the explicit form of Eqs.~\eqref{eq:tmt1} in spherical polar coordinates
(additionally including the aforementioned terms that are omitted in our presentation) the reader
may consult the Appendix of \citet{Buras2006}.

Equations~\eqref{eq:tmt1} are the radiation evolution equations used in our code. Since the source
terms in general depend on the energy and species of neutrinos, we have to solve a set of moment
equations for each energy group (after discretizing the energy space into a finite set of energy
groups/bins, cf. Sec.~\ref{sec:numerics}) and for each species. Hence, given $N_{\eps}$
energy bins, $N_{\mathrm{spe}}$ species and taking into account $N_{\mathrm{dim}}$ components of the
flux density $F^i$, we have to process $(N_{\mathrm{dim}}+1)\times N_{\mathrm{spe}} \times N_{\eps}$
equations in total in our multidimensional, multi-group radiation transport scheme. For the
following presentation, however, we will only indicate individual species or the energy dependence
if it is demanded by the context.

\subsubsection{Moment equations of number transport}

The moments connected to the number transport (number density, number flux density etc.) are given
by
\begin{equation}
  \label{eq:nummomdef}
  \{ \: N,F^i_{N},P^{ij}_{N}, Q^{ijk}_{N}, \ldots \: \} \equiv
   \eps^{-1}\{ \: E,F^i,P^{ij}, Q^{ijk}, \ldots \: \}. 
\end{equation}
Although we do not directly use them in our code, we list the equations describing the neutrino
number evolution for completeness here. They are structurally similar to
Eqs.~\eqref{eq:tmt1} except for terms associated with the energy derivatives:
\begin{subequations}\label{eq:tmtn1}
\begin{align}
  &\partial_tN + \nabla_jF^j_{N} + 
  \nabla_j(v^j N) - \nabla_j v_k\partial_\eps (\eps P^{jk}_{N}) = \eps^{-1}C^{(0)} \, ,\label{eq:tmten1}\\
  &\partial_tF^i_{N} + \cl^2\nabla_jP^{ij}_{N} + \nabla_j(v^j F^i_{N}) + \nonumber\\
  & \hspace{0.5cm} F^j_{N}\nabla_j v^i - \nabla_j v_k\left[Q^{ijk}_{N} + 
    \partial_\eps (\eps Q^{ijk}_{N})\right] = \eps^{-1}C^{(1),i} \, . \label{eq:tmtfn1}
\end{align}
\end{subequations}

\subsubsection{Transformation into lab-frame}

The transformation of energy-integrated moments from the comoving into the lab-frame can be
performed by referring to their intrinsic tensorial structure which dictates the way the Lorentz
transformation has to be applied. The energy-associated moments $\bar{E},\bar{F}^i,\bar{P}^{ij}$ are
components of a 2nd-rank tensor, namely the energy--momentum tensor of radiation, while the
number-associated 0th and 1st moments combine to a 4-vector. This results in the following
transformation rules correct to order $\mathcal{O}(v/c)$ for the energy-related moments:
\begin{subequations}\label{eq:momtrafoe}
\begin{align}
  \label{eq:momtrafo}
  \bar{E}_{\mathrm{lab}} &= \bar{E} + 2\cl^{-2}\,v_j \bar{F}^j \, , \\
  \bar{F}^{i}_{\mathrm{lab}} &= \bar{F}^i + v^i \bar{E} 
  + v_j \bar{P}^{ij}\label{eq:momtrafoflux} \, , \\
  \bar{P}^{ij}_{\mathrm{lab}} &= \bar{P}^{ij} + \frac{1}{\cl^2}
  (v^i \bar{F}^j + v^j \bar{F}^i)  \, , \label{eq:momtrafopres}
\end{align}
\end{subequations}
and for the number-related moments:
\begin{subequations}\label{eq:momtrafon}
\begin{align}
  \label{eq:momtrafonumber}
  \bar{N}_{\mathrm{lab}} &= \bar{N} + \cl^{-2}\,v_i \bar{F}_{N}^i \, , \\
  \label{eq:momtrafonumberflux}
  \bar{F}_{N,\mathrm{lab}}^{i} &= \bar{F}_{N}^i + v^i \bar{N} \, .
\end{align}
\end{subequations}
Note that these transformation rules only apply for the grey quantities but not for the
monochromatic moments\footnote{However, corresponding $\mathcal{O}(v/c)$ expressions for the
  monochromatic moments can be formulated in terms of Taylor expansions of the moments in energy
  space \citep[e.g.][]{Mihalas1984, Hubeny2007}.}. The energy-integrated source terms
$\bar{C}^{(0)},\bar{C}^{(1),i}$ transform into the lab-frame source terms
$\bar{C}^{(0)}_{\mathrm{lab}},\bar{C}^{(1),i}_{\mathrm{lab}}$ in a similar way as $\bar{N}$ and
$\bar{F}_{N}$ (cf. Eq.~\eqref{eq:momtrafon}), i.e. as a 4-vector, since they are defined to form the
right-hand side of a conservation law of a 2nd-rank tensor in its original relativistic formulation.

\subsection{Interaction source terms and coupling to hydrodynamics}\label{ssec:sourcesandhydro}

The interaction source terms are the actual terms that introduce the microphysical properties and
the coupling of matter and radiation into the transport problem. The source terms
$C^{(0)},C^{(1),i}$ for the neutrino moments give rise to corresponding hydrodynamic source terms
$Q_{\mathrm{M}}$, $Q_{\mathrm{E}}$ that account for the change of fluid momentum and gas internal
energy, respectively, due to the interaction with neutrinos. We restrict ourselves here to the basic
Euler equations and neglect additional physics, such as viscosity, magnetic fields or the
co-evolution of a set of nuclear species. The evolution of the baryonic density $\rho$, momentum
density $\rho v^i$, total gas-energy density $e_{\mathrm{t}}\equiv e_{\mathrm{i}}+\rho\vecv^2/2$
(where $e_{\mathrm{i}}$ is the internal energy density) and electron fraction\footnote{As usual,
  this quantity is defined as $Y_e\equiv (n_{e^-}-n_{e^+})/n_{\mathrm{B}}$ with the number densities
  $n_{e^\pm}$ of electrons and positrons and the baryon number density $n_{\mathrm{B}}$.}  $Y_e$ is
then dictated by the system
\begin{subequations}\label{eq:hydevo}
\begin{align}
   \partial_t \rho + \nabla_j(\rho v^j) & = 0  \, , \\
   \partial_t (\rho Y_e) + \nabla_j(\rho Y_e v^j) &= Q_{\mathrm{N}} \, , \label{eq:hydyeevo}\\
   \partial_t (\rho v^i) + \nabla_j(\rho v^i v^j + P_{\mathrm{g}} ) 
   & = Q_{\mathrm{M}}^i  \, , \\
   \partial_t e_{\mathrm{t}} + \nabla_j\left(v^j(e_{\mathrm{t}} + P_{\mathrm{g}})
  \right) \
   & = Q_{\mathrm{E}} 
  + v_j Q_{\mathrm{M}}^j  \, . \label{eq:hydetotevo}
\end{align}
\end{subequations}
The gas pressure $P_{\mathrm{g}}$ is obtained by invoking an equation of state (EOS), which at the
same time provides the quantities required to compute the opacities (such as the temperature,
chemical potentials, or particle densities in case of nuclear statistical equilibrium). By virtue of
the physical meaning of the moments $E$ and $F^i$, the source terms for the hydrodynamic equations
can immediately be identified with
\begin{subequations}\label{eq:hydsource}
\begin{align}
  Q_{\mathrm{E}}   &=  - \sum_{\mathrm{species}} \bar{C}^{(0)} \, ,\label{eq:einsource} \\
  Q_{\mathrm{M}}^i &=  - \frac{1}{\cl^2} \sum_{\mathrm{species}} \bar{C}^{(1),i}  \, , \label{eq:momsource}\\
  Q_{\mathrm{N}}   &= - m_{\mathrm{B}}\int_0^\infty \left[\left(\frac{C^{(0)}}{\eps}\right)_{\nu_e}
  - \left(\frac{C^{(0)}}{\eps}\right)_{\bar\nu_e}\right] \mathrm{d}\eps \, , \label{eq:yesource}
\end{align}
\end{subequations}
where $m_{\mathrm{B}}$ is the atomic mass unit and the sums contain all contributions from
individual neutrino species.

At present, the most important types of (electron-) neutrino interactions are implemented, namely
the capture of (anti-) neutrinos and (anti-) electrons by free nucleons and nuclei, isoenergetic
scattering of (anti-) neutrinos off free nucleons and nuclei, and inelastic scattering of (anti-)
neutrinos off (anti-) electrons. All corresponding source terms are adopted as described in
\citet{Rampp2002}, which is mostly based on the compilation given by \citet{Bruenn1985}.

For introducing further concepts in a way that keeps the presentation as essential as possible, in
the following we will for simplicity assume that only isoenergetic scattering and
absorption/emission reactions take place. In this case the source terms in the moment equations,
Eqs.~\eqref{eq:tmt1}, can be written as \citep[e.g.][]{Bruenn1985}
\begin{subequations}\label{eq:sourceterms}
\begin{align}
  C^{(0)} &= \cl \kappa_{\mathrm{a}} \left( E^{\mathrm{eq}} - E \right) \label{eq:sourceE} \, , \\
  C^{(1),i} &= - \cl ( \kappa_{\mathrm{a}} + \kappa_{\mathrm{s}} ) F^i \label{eq:sourceF} \, ,
\end{align}
\end{subequations}
where $\kappa_{\mathrm{a}}$ and $\kappa_{\mathrm{s}}$ are the combined absorption (corrected for
final-state Fermion-blocking) and scattering opacities, and $E^{\mathrm{eq}}$ is the equilibrium
energy density associated with the Fermi-Dirac distribution $\mathcal{F}_{\mathrm{FD}}$,
\begin{align}
  \label{eq:fermidirac}
  E^{\mathrm{eq}}(\eps,\mu_\nu,T) &\equiv \int \mathrm{d}\Omega 
  \left(\frac{\eps}{\hn c}\right)^3 \mathcal{F}_{\mathrm{FD}} \nonumber\\
  &\equiv\left(\frac{\eps}{\hn c}\right)^3 \frac{1}{\exp\{(\eps-\mu_\nu)/(k_{\mathrm{B}}T)\} + 1} \, ,  
\end{align}
which is a function of the fluid temperature $T$ and the chemical potential $\mu_\nu$ of the
corresponding neutrino species. The transport opacity $\kappa_{\mathrm{tra}}$, which is given by
$\kappa_{\mathrm{tra}}=\kappa_{\mathrm{a}}+\kappa_{\mathrm{s}}$, is related to the mean free path
$\lambda_\nu$ between two momentum-exchanging reactions via
$\kappa_{\mathrm{tra}}=\lambda_\nu^{-1}$.

\subsection{Algebraic closure methods}\label{ssec:closures}

The full information contained in the Boltzmann equation can be captured equally well by an infinite
series of conservation equations for the angular moments, in which the evolution equation for a
moment of rank $m$ contains the moment of rank $m+1$ within the divergence operator. Instead of
solving the infinite series of moment equations, the series can be truncated at the level of the
$(m+1)$-th moment, provided the $(m+1)$-th moment is available to close the set of $m$
equations. However, in order to determine the $(m+1)$-th moment that is consistent with the
Boltzmann equation, it is inevitable to solve the latter in some approximation or the other, using
computationally expensive methods such as discrete ordinate or (long- or short-) characteristics
schemes. A computationally much cheaper, though more approximate option is to assume that an
algebraic closure relation holds between the evolved moments and the $(m+1)$-th moment. This is what
defines the algebraic closure methods, such as FLD and AEF. Essentially, this corresponds to
imposing additional conditions or symmetries on the local radiation field. The consequence is that
two (out of seven in the general case) independent variables describing the angular dependence of
the radiation field disappear from the treatment. Evidently, the tradeoff for this computational
simplification is that an algebraic closure method may strongly vary in quality between different
physical setups. For example, since in an algebraic closure method the consistent evolution of
higher-order angular moments is ignored, it appears likely that the quality of the scheme
appreciably depends on the geometric complexity of the radiation field, or equivalently on the shape
and number of the individual radiation sources. Moreover, connected to this issue is the
circumstance that in the optically thin limit of vanishing source terms an algebraic moment scheme
is generally not able to accurately describe the unperturbed superposition of multiple radiation
fronts, simply on account of the closure being a local, non-linear function of the evolved
quantities. Despite these conceptual deficiencies, which have to be individually tested for in each
case of application, algebraic closure methods can in many cases offer an excellent compromise
between efficiency and accuracy when performing energy-dependent, multidimensional RHD simulations.

Independent of the rank at which the scheme is truncated, any closure prescription should agree with
certain consistency requirements that directly follow from the definition of the moments or from the
Boltzmann equation. Using the normalized moments $\vect{f}\equiv\vecF/(\cl E)$, where
$f\equiv|\vect{f}|$ is the flux-factor, $D^{ij}\equiv P^{ij}/E$, the Eddington tensor, and
$q^{ijk}\equiv Q^{ijk}/(\cl E)$, the normalized 3rd-moment tensor, it follows from the definition of
the angular moments, Eqs.~\eqref{eq:momdef1}, that
\begin{subequations}\label{eq:momrelations}
\begin{align}
  |\vect{f}|   &\leq 1  \, , \label{eq:momrel1} \\
  D^{ij}        &\leq 1     \, , \label{eq:momrel2} \\
  \sum_j D^{jj} &= 1 \, , \label{eq:momrel3}  \\
  |q^{ijk}| & \leq 1 \, , \label{eq:momrel4} \\
  \sum_jq^{ijj} & = \sum_jq^{jij} = \sum_jq^{jji} =f^i \, , \label{eq:momrel5}
\end{align}
\end{subequations}
must hold at any time. In the ``free-streaming limit'', which is approached far away from regions
of radiation--matter interaction, all of the radiation propagates into a single direction away from
its source and it must hold
\begin{equation}
  \label{eq:freelimit}
  f     \, = \, 1  \quad , \quad
  D^{ij} \, = \, n^i_\vecF \, n^j_\vecF \, , 
  \quad , \quad q^{ijk} \, = \, n^i_\vecF \, n^j_\vecF \, n^k_\vecF \, ,
\end{equation}
with $n^i_\textbf{\em F}\equiv F^i/|\textbf{\em F}|$ denoting the direction of the flux density. In
the opposite limit of very frequent interactions, i.e. in the ``diffusion limit'', the specific
intensity is approximately isotropic. Ignoring velocity terms, the radiation moment equations
degenerate in this limit to the diffusion equation
\begin{equation}
\partial_t E + \nabla_i\left(-\frac{\cl}{3\kappa_{\mathrm{tra}}}\nabla^i E\right) = C^{(0)} \label{eq:diffe}
\end{equation}
and the relations
\begin{equation}
  \label{eq:difflimit}
  \vect{f} \, = \, - \frac{1}{3\kappa_{\mathrm{tra}}}\, \frac{\nabla E}{E}\quad , \quad
  D^{ij}    \, = \,\frac{1}{3} \, \delta^{ij} \, .
\end{equation}
In the following sections, we will outline the basic concepts of FLD and AEF schemes and present
several closure prescriptions. In Sec.~\ref{sec:neutr-radi-field}, we will explicitly compare both
methods and the presented closures on the basis of the neutrino radiation field produced by a
proto-NS.

\subsubsection{Flux-limited diffusion method}\label{sec:flux-limit-diff}

The approach of FLD \citep[e.g.][]{Wilson1975a, Levermore1981} is to truncate the set of moment
equations at the level of the 1st-moment equation and to derive an expression for the flux density
based on the diffusion limit described by Eqs.~\eqref{eq:diffe}
and~\eqref{eq:difflimit}. Introducing the ``Knudsen number'' $R=|\vect{R}|$, with
\begin{equation}
  \label{eq:knudsen}
  \vect{R} \equiv \frac{1}{\omega\,\kappa_{\mathrm{tra}}} \frac{\nabla E}{E}  \, ,
\end{equation}
where
\begin{equation}
  \label{eq:albedo}
  \omega\equiv (\kappa_{\mathrm{s}}E + \kappa_{\mathrm{a}}E^{\mathrm{eq}})/(\kappa_{\mathrm{tra}}E)
\end{equation}
is the ``effective albedo'', the flux density $\vecF_{\mathrm{FLD}}$ is prescribed as
\begin{equation}
  \label{eq:fldflux}
  \vecF_{\mathrm{FLD}} = - \,  \Lambda(R) \, \vect{R} \cl E \, ,
\end{equation}
in which $\Lambda(R)$ is called the ``flux-limiter''. The latter is constructed such that the flux
density correctly preserves the constraints of Eqs.~\eqref{eq:freelimit}, \eqref{eq:difflimit}. To
this end the limits
\begin{equation}
  \label{eq:fluxlimthin}
  \lim_{R\rightarrow\infty} \Lambda(R)\,R\, = \, 1  , \\
\end{equation}
and
\begin{equation}
  \label{eq:fluxlimthick}
  \lim_{R\rightarrow 0} \Lambda(R) \, = \,\frac{1}{3} \, ,
\end{equation}
respectively, have to be fulfilled (note that $\omega\rightarrow 1$ in the diffusion limit). Three
prescriptions are widely used in the literature \citep{Wilson1975a, Liebendorfer2004,
  Levermore1981}:
\begin{subequations}\label{eq:fluxlimiters}
\begin{align}
  \label{eq:wilsonlimiter}
  & \Lambda_{\mathrm{Wilson}}(R)\, = \,\frac{1}{3 + R} \, , \\
  \label{eq:bruennlimiter}
  & \Lambda_{\mathrm{Bruenn}}(R)\, = \nonumber\\
  & \quad \begin{cases}
    \min\left(\Lambda_{\mathrm{Wilson}}(R), \frac{1 + \sqrt{1-(r_\nu/r)^2}}{2R} \right) & , r > r_\nu  \\
    \Lambda_{\mathrm{Wilson}}(R)                                                         & \textrm{, else} \, ,
    \end{cases} \\
  \label{eq:leverlimiter}
  & \Lambda_{\mathrm{Levermore}}(R)\, = \,\frac{1}{R}\left( \coth R - \frac{1}{R} \right) \, .
\end{align}
\end{subequations}
The limiter in Eq.~\eqref{eq:bruennlimiter} is only designed for the spherically symmetric case in
which $r$ is the radius coordinate and $r_\nu$ is the radius of a (properly defined)
neutrinosphere. The limitation for $r>r_\nu$ is intended to ensure that the neutrino flux cannot be
higher than if the neutrinos were distributed isotropically into a finite cone subtending the sphere
of radius $r_\nu$.

The main drawbacks of FLD are: First, the prescription of the flux density is in general not
consistent with the 1st-moment equation. As a direct consequence, the full RHD system suffers from
momentum and therefore energy non-conservation \citep{Bruenn1985, Baron1989} whenever momentum
transfer between matter and radiation takes place. In the case of CCSNe, the violation is found
\citep{Cernohorsky1990} to be particularly significant in the semi-transparent region where
$\Lambda(R)$ undergoes the main part of the transition $1/3\rightarrow 0$. Even though interim
solutions of this shortcoming can be formulated, e.g. by introducing an artificial opacity
\citep{Janka1991a, Dgani1992} that contains the missing information of the 1st-moment equation, they
introduce further degrees of freedom, hence rendering the resulting method rather tuned to special
cases.

Second, in more than one dimension a complication arises from the fact that the flux density vector
is always directed opposite to the gradient of the energy density since the pressure is effectively
isotropic (cf. Eq.~\eqref{eq:difflimit}): Radiation in the free-streaming limit will not keep its
original flux direction after closely passing opaque objects. Instead it behaves like a gas and
fills up space in every direction, unable to form persistent shadows.

The third issue is a purely computational aspect: The energy equation evolved in FLD is -- at least
whenever $f\neq 1$ -- of parabolic mathematical nature. As such, it comes with the property that the
operator $\nabla\cdot\vecF$ needs to be treated time-implicitly in most practical cases. This is
because the characteristic timescale $\tau_{\mathrm{FLD}}$ associated with $\nabla\cdot\vecF$ can
become extremely short in the optically thin limit $\kappa_{\mathrm{tra}} \rightarrow 0$. One can
roughly estimate the local timescale $\tau_{\mathrm{FLD}}$ by thinking of the operator
$\nabla\cdot\vecF$ as being locally a linear convex combination of the advection operator
$\alpha\vect{f}\cl\nabla E$ and the diffusion operator
$(1-\alpha)(-\Lambda\cl/\kappa_{\mathrm{tra}})\nabla^2 E$, with some weighting factor
$0<\alpha < 1$. A heuristic dimensional analysis then gives
\begin{subequations}
\begin{align}
  \frac{E}{\tau_{\mathrm{FLD}}} &\sim\: \alpha \frac{\cl}{\Delta x} f E + 
    (1-\alpha ) \frac{\cl}{\kappa_{\mathrm{tra}}\Delta x^2} \Lambda E \, \\
\Rightarrow \tau_{\mathrm{FLD}} &\sim\: \left( \frac{\alpha f}{\tau_{\mathrm{adv}}} 
  + \frac{(1-\alpha)3\Lambda}{\tau_{\mathrm{diff}}} \right)^{-1} \, , \label{eq:FLDtimestep}
\end{align}
\end{subequations}
where $\Delta x$ is the local grid size and
\begin{subequations}\label{eq:tautimescales}
\begin{align}
  \tau_{\mathrm{adv}} &\equiv  \Delta x / \cl \, , \label{eq:tautimescale_adv}\\
  \tau_{\mathrm{diff}} &\equiv 3\kappa_{\mathrm{tra}}\Delta x^2 / \cl \label{eq:tautimescale_diff}
\end{align}
\end{subequations}
are the characteristic timescales of advection and diffusion, respectively. Hence, owing to the fact
that $\tau_{\mathrm{diff}}\rightarrow 0$ in optically thin regions, $\tau_{\mathrm{FLD}}$ can drop
significantly below $\tau_{\mathrm{adv}}$.

\subsubsection{Algebraic Eddington factor method}\label{sec:analyt-eddingt-fact}

In the AEF method the truncation of the moment equations takes place at the level of the 2nd-moment
equation, i.e. the Eddington tensor is expressed as a function of evolved quantities. Besides
resolving by construction the first two drawbacks of the FLD scheme mentioned in
Sec.~\ref{sec:flux-limit-diff}, another important computational difference to the FLD scheme is that
the equations solved in the AEF scheme are intrinsically hyperbolic, which allows to use explicit
time integration methods (at least for all but the source terms, see Sec.~\ref{sec:numerics}) with
time steps not lower than the minimum of $\tau_{\mathrm{adv}}$ over the computational domain.

Historically, algebraic closures have been constructed and analyzed most often in the context of
one-dimensional systems and up to the present day quite a number of one-dimensional closures have
been proposed in the literature. The algebraic Eddington factor $\chi\equiv P/E$ is typically
expressed as $\chi=\chi(e,f)$, where
$e = (4\pi)^{-1} \int\mathrm{d}\Omega \mathcal{F} = (\hn\cl)^3/(4\pi\eps)^3E$. Note that in
constrast to the FLD scheme, the closure only depends on radiation moments and not on the
opacities. In the course of this paper we consider a variety of closures, of which the Eddington
factors are given by
\begin{subequations}\label{eq:closures}
\begin{align}
  \chi_{\mathrm{Minerbo}}(f)  &= 
  \frac{1}{3} + \frac{1}{15}\left( 6f^2 - 2f^3 + 6f^4 \right)           \, , \label{eq:minclos}\\
  \chi_{M_1}(f)               &= \frac{3 + 4f^2}{5 + 2\sqrt{4 - 3f^2}}                       \, ,\label{eq:m1closure} \\
  \chi_{\mathrm{Janka}}(f) &= \frac{1}{3} \left( 1 + \frac{1}{2} f^{1.31} + \frac{3}{2} f^{3.56} \right)  \, , \\
  \chi_{\mathrm{Max. Ent.}}(e,f) &= \frac{2}{3} (1-e)(1-2e) \sigma\left(\frac{f}{1-e}\right) + \frac{1}{3} \, , 
  \label{eq:maxentclosure}
\end{align}
\end{subequations}
where $\sigma(x)\equiv x^2(3-x+3x^2)/5$ in Eq.~\eqref{eq:maxentclosure}. The statistical closure
$\chi_{\mathrm{Minerbo}}(f)$ by \citet{Minerbo1978} assumes a fermionic particle distribution with
low density ($e\rightarrow 0$) to be in the state of maximum entropy. The generalization of this
closure is $\chi_{\mathrm{Max. Ent.}}(e,f)$, derived by \citet{Cernohorsky1994}, which additionally
to $\chi_{\mathrm{Minerbo}}$ takes into account the effects of fermion blocking for $e>0$. In
Eqs.~\eqref{eq:minclos}, \eqref{eq:maxentclosure} we employed the polynomial approximation derived
by \citet{Cernohorsky1994} to circumvent the numerical inversion of the Langevin function occurring
in the original formulations of both closures. The $M_1$ closure can either be derived from the
assumption of the radiation field being isotropic in some unspecified frame of reference
\citep{Levermore1984} or from maximizing a photon entropy functional \citep{Dubroca1999}. Note that
in both aforementioned derivations the $M_1$ closure actually relates the energy-integrated
moments. In our present treatment, in contrast, we apply all closure relations in
Eqs.~\eqref{eq:closures} individually for each neutrino energy $\eps$. Finally, the closure by
\citet{Janka1991a} was obtained by fitting the neutrino radiation profile around a proto-NS
calculated with Monte-Carlo methods. For further detailed discussions about specific one-dimensional
closures and their properties, see, e.g., \citet{Smit2000} and \citet{Pons2000}.

To extend the one-dimensional Eddington factor $\chi$ to the multidimensional Eddington tensor
$D^{ij}$ and the 3rd-moment tensor $q^{ijk}$, we make use of the underlying assumption that these
tensors are only a function of the scalar $E$ and the vector $\vecF$\footnote{Mathematically
  speaking, we assume $D^{ij}$ and $q^{ijk}$ to be \emph{isotropic tensor functions}
  \citep[e.g.][]{Pennisi1987} of $E$ and $\vecF$, which in particular implies that these functions
  may not depend on derivatives of $E$ or $\vecF$.}. From symmetry arguments it follows that
$D^{ij}$ must be symmetric with respect to rotation around the flux direction,
$\vecn_\vecF\equiv \vecF/|\vecF|$, which is only fulfilled if $D^{ij}$ is a linear combination of
the two tensors $\delta^{ij}$ and $n_\vecF^in_\vecF^j$ \citep[e.g.][]{Pennisi1987}. After using the
trace condition of Eq.~\eqref{eq:momrel3} the two remaining coefficients can be expressed as
functions of a single parameter, $\chi$, which is the multidimensional generalization of the
Eddington factor and is defined as
\begin{equation}
  \label{eq:defchi}
  \chi \equiv \frac{ \int \mathrm{d}\Omega\, \left(\vecn\cdot\vecn_\vecF\right)^2 \mathcal{F} }
                { \int \mathrm{d}\Omega \, \mathcal{F}  } \, .
\end{equation}
The Eddington tensor then reads \citep[e.g.][]{Levermore1984}:
\begin{equation}\label{eq:eddingtonmultidim}
  D^{ij} = \frac{1-\chi}{2}\,\delta^{ij} + \frac{3\chi - 1}{2}\, n^i_\textbf{\em F}\,n^j_\textbf{\em F} \, .
\end{equation}

The energy-dependent comoving-frame moment equations, Eqs.~\eqref{eq:tmt1}, also contain the 3rd
moments, $q^{ijk}$. In analogy to the derivation of $D^{ij}$ above, the condition that this
  tensor only depends on $E$ and $\vecF$ must result in $q^{ijk}$ being invariant under rotation
  around $\vecn_\vecF$, which is only fulfilled if $q^{ijk}$ is a linear combination of
  $n^i_\vecF\delta^{jk}, n^j_\vecF\delta^{ik}$ and $n^k_\vecF\delta^{ij}$ as well as
  $n^i_\vecF n^j_\vecF n^k_\vecF$ \citep[e.g.][]{Pennisi1992}. The corresponding coefficients can be
  eliminated using the trace conditions of Eq.~\eqref{eq:momrel5} in favor of a single parameter,
  $q$, defined as
\begin{equation}
  q \equiv \frac{ \int \mathrm{d}\Omega \,\left(\vecn\cdot\vecn_\vecF\right)^3 \mathcal{F} }
                { \int \mathrm{d}\Omega \, \mathcal{F}  } \, ,
\label{eq:defqfac}
\end{equation}
yielding for the 3rd-moment tensor:
\begin{align}
  q^{ijk}  &=   \frac{f-q}{2} (n^i_\vecF\,\delta^{jk}+
    n^j_\vecF\,\delta^{ik}+n^k_\vecF\,\delta^{ij} ) + \frac{5q-3f}{2}\, n^i_\vecF \,n^j_\vecF\, n^k_\vecF \, .
\label{eq:defQ}
\end{align}

The 3rd-moment factor $q$ explicitly depends on the distribution function. Therefore, only
closures that dictate an explicit functional form of the distribution function are suited for the
computation of the 3rd moment, unless additional assumptions are made in the construction of the
closure. For the Minerbo closure, the factor $q$ can be calculated in a straightforward manner in
analogy to the derivation of $\chi$ \citep{Minerbo1978} and reads (again using the polynomial
approximation of the Langevin function as given in \citealp{Cernohorsky1994})
\begin{equation}
  q_{\mathrm{Minerbo}}(f) = \frac{f}{75} (45 + 10f -12f^2 - 12f^3 + 38f^4 - 12f^5 + 18f^6 )  \, .
\label{eq:defqsmall}
\end{equation}
We outline the derivation of Eq.~\eqref{eq:defqsmall} in Appendix \ref{sec:deriv-3rd-moment}.
Recently, \citet{Vaytet2011} calculated the 3rd-moment factor $q$ also for the $M_1$ closure.


\section{Numerical Method}\label{sec:numerics}

\subsection{Motivation of the integration scheme}\label{sec:gener-cons}

Before presenting the details of the numerical implementation of the AEF scheme outlined in the
previous section, we briefly summarize and motivate the general framework of the discretization
scheme. Owing to the fact that the advection-type operators on the left-hand side of the two-moment
system, Eqs.~\eqref{eq:tmt1}, are of hyperbolic mathematical nature, we can employ Godunov-type
finite-volume methods commonly used in numerical hydrodynamics to discretize these
operators. However, in regions of strong coupling with matter the source terms become stiff and the
moment equations approach the parabolic diffusion limit. Hence, the time integration is performed in
a mixed explicit--implicit manner, in which all terms on the left-hand side of Eqs.~\eqref{eq:tmt1}
are treated explicitly in time while the source terms on the right-hand side of Eqs.~\eqref{eq:tmt1}
are handled implicitly (at least whenever being in the stiff regime). In that way the overall
time step used for integration of the whole scheme is constrained by the Courant condition to be on
the order of the advection timescale $\tau_{\mathrm{adv}}\equiv \Delta x / \cl$, i.e. the
light-crossing time of grid cells with width $\Delta x$. The alternative would be to integrate the
full two-moment system implicitly in time. In that case the computational cost per time step would
be significantly higher (particularly in the multidimensional case) but on the other hand this would
allow to employ a larger time step which is closer to the fluid-dynamical time step $\Delta x /
v$. We opted for the former version of integration, mainly for the following reasons:

(1) Since the velocities in CCSNe and NS mergers are typically rather high,
$v\sim \mathcal{O}(0.1c)$, the characteristic hydrodynamics time step and therefore the implicit
radiation time step turn out to be only a factor of a few greater than the explicit radiation time
step. (2) Since all operators containing spatial derivatives are treated explicitly, the common
parallelization methods can be applied with very high efficiency. This is particularly advantageous
in the multidimensional case in which a fully implicit scheme would become increasingly expensive
(as its computational cost typically increases faster than linear with the number of grid
zones). (3) Light fronts and discontinuities in the radiation field can be sharply resolved, which
tend to be smeared out in an implicit method, unless a time step comparable to the explicit one is
used. (4) The overall numerical implementation is less involved than for an implicit scheme because
inversions of large matrices that couple spatial grid points are avoided. (5) While high-order
spatial reconstruction methods can be implemented in a straightforward manner in explicit schemes,
they are usually too prohibitive to be used in implicit schemes.

For other implementations of a similar explicit--implicit integration method
see,e.g. \citet{Scadowski2013}, \citet{Skinner2013}, \citet{OConnor2014}. For implementations of
fully implicit schemes see, e.g., \citet{Audit2002}, \citet{Hayes2003}.

\subsection{Basic discretization scheme}\label{ssec:basic-discr-scheme}

The spatial and energy-space discretization scheme for all quantities is based on the finite-volume
approach. The set of spatial coordinates can be varied between Cartesian $(x,y,z)$, cylindrical and
spherical polar $(r,\theta,\phi)$ coordinates. However, for outlining the discretization scheme in
this section we will restrict ourselves to Cartesian coordinates. The extension to curvilinear
coordinates is realized by adding the appropriate geometric source terms to the presented
discretized derivatives. The geometric source terms are purely algebraic functions of the evolved
quantities and the grid coordinates and are discretized simply by replacing the arguments of these
functions with the corresponding discretized quantities.

The spatial grid is composed of volume cells $(\hi,\hj,\hk)$ that are obtained after discretizing a
given domain in each coordinate direction $\{x,y,z\}$ into $\{N_x,N_y,N_z\}$ zones, of which each is
defined by the cell boundaries $x_{\hi\pm\half},y_{\hj\pm\half},z_{\hk\pm\half}$, with
$\{\hi,\hj,\hk\} = \{1\ldots N_x,1\ldots N_y,1\ldots N_z\}$. The cell-center coordinates are
computed as $x_\hi\equiv 1/2(x_{\hi-\half}+x_{\hi+\half})$ and analog for the other directions. The
volume of cell $(\hi,\hj,\hk)$ is denoted by $\Delta V_{\hi,\hj,\hk}$, and the area of the surface
$(\hi+\half,\hj,\hk)$, located between cells $(\hi,\hj,\hk)$ and $(\hi+1,\hj,\hk)$, is denoted by
$\Delta A_{\hi+\half,\hj,\hk}$.

For the grid in energy space, given by $N_\eps$ energy bins, we use $\eps_{\xi\pm\half}$ to denote
the boundaries of the $\xi$'th bin, with $\xi=1\ldots N_\eps$. Furthermore, $\eps_{\xi} \equiv 1/2
(\eps_{\xi+\half} + \eps_{\xi-\half})$ and $\Delta\eps_{\xi} \equiv \eps_{\xi+\half} -
\eps_{\xi-\half}$ define the center and width of the $\xi$'th bin, respectively.

We define the radiation fields as well as the fluid quantities on the same spatial grid. The
discrete representations
$\hat{U} \in \{\hat{\rho}, \hat{(\rho Y_e)}, \hat{(\rho v^i)}, \hat{e}_{\mathrm{t}} \}$ of the
hydrodynamic quantities $U\in \{\rho,\rho Y_e, \rho v^i, e_{\mathrm{t}}\}$ are interpreted as
cell-volume averages in space, i.e. as
\begin{equation}\label{eq:hyddiscrete}
  \hat{U}_{\hi,\hj,\hk} \equiv \frac{1}{\Delta V_{\hi,\hj,\hk}}\int_{\Delta V_{\hi,\hj,\hk}} \mathrm{d}V \, U \, ,
\end{equation}
while the discrete representations $\hat{X}\in\{\hat{E},\hat{F}^i,\hat{P}^{ij},\hat{Q}^{ijk}\}$ of
the radiation moments $X\in\{E,F^i, P^{ij},Q^{ijk}\}$ are interpreted as averages in space and
integrals in energy space in the following way
\begin{equation}\label{eq:momdiscrete2}
  \hat{X}_{\hi,\hj,\hk,\xi} \equiv \frac{1}{\Delta V_{\hi,\hj,\hk}}\int_{\Delta V_{\hi,\hj,\hk}} \mathrm{d}V
  \int_{\Delta\eps_{\xi}} \mathrm{d}\eps \, X \, .
\end{equation}
The spatial and energy discretization of the moment equations, Eqs~\eqref{eq:tmt1}, is realized by
applying the spatial averaging and the energy integration as in Eq.~\eqref{eq:momdiscrete2} to the
moment equations.

The discretization of the spatial derivatives demands it to reconstruct quantities from cell-volume
averages to call-face averages, which for example on the cell-face $(\hi+\half,\hj,\hk)$ are given
by
\begin{equation}
  \hat{X}_{\hi+\half,\hj,\hk,\xi} \equiv \frac{1}{\Delta A_{\hi+\half,\hj,\hk}}\int_{\Delta A_{\hi+\half,\hj,\hk}} \mathrm{d}A
  \int_{\Delta\eps_{\xi}} \mathrm{d}\eps X \, .
\end{equation}
The reconstruction algorithms that we use are adopted from the hydrodynamics part of the code and
can be switched between piecewise-constant, piecewise-linear and high-order ``monotonicity
preserving'' (MP) schemes \citep{Suresh1997} or ``weighted essentially non-oscillatory'' (WENO)
schemes \citep{Liu1994}.  In what follows, we symbolically use $\hat{X}^{\mathrm{L}}$ and
$\hat{X}^{\mathrm{R}}$ to denote the reconstructed values of a quantity $\hat{X}$ on the left- and
right-hand side of an interface, respectively. 

\subsection{Summary of the integration algorithm}\label{ssec:update}

In order to formulate the algorithm to integrate the evolution equations for the radiation
  moments, $\mathbf{X}\equiv\left(E,\vecF\right)$ (cf. Eqs.~\eqref{eq:tmt1}), and for the fluid
  quantities, $\mathbf{U} \equiv \left(\rho,\rho Y_e, \rho\vecv, e_{\mathrm{t}}\right)$
  (cf. Eqs.~\eqref{eq:hydevo}), we decompose these equations in the following way:
\begin{subequations}\label{eq:evosummary}
\begin{align}
  \partial_t\mathbf{X} + (\delta_t \mathbf{X})_{\mathrm{hyp}} + (\delta_t \mathbf{X})_{\mathrm{vel}} 
  &= (\delta_t \mathbf{X})_{\mathrm{src}} \, , \label{eq:tmtsplit}\\
  \partial_t\mathbf{U} + (\delta_t \mathbf{U})_{\mathrm{hyd}}  &= (\delta_t \mathbf{U})_{\mathrm{src}} \, . \label{eq:hydsplit}
\end{align}
\end{subequations}
In Eq.~\eqref{eq:tmtsplit},
$(\delta_t \mathbf{X})_{\mathrm{hyp}} \equiv \left(\nabla_j F^j,c^2\nabla_jP^{ij}\right)$ represents
the velocity-independent hyperbolic advection terms, while all velocity-dependent terms are subsumed
in $(\delta_t \mathbf{X})_{\mathrm{vel}}$, and the interaction source terms are represented by
$(\delta_t \mathbf{X})_{\mathrm{src}}$. In Eq.~\eqref{eq:hydsplit},
$(\delta_t \mathbf{U})_{\mathrm{src}}$ are the radiative source terms given by
Eqs.~\eqref{eq:hydsource}, while all remaining terms representing non-radiative physics are
contained in $(\delta_t \mathbf{U})_{\mathrm{hyd}}$. We first summarize the overall integration
scheme of Eqs.~\eqref{eq:evosummary}. Subsequently, in Secs.~\ref{ssec:hyperbolic}--\ref{ssec:hydro}
we will explicitly describe how the individual terms are computed.

The following steps are performed to evolve the RHD equations, Eqs.~\eqref{eq:evosummary}, from time
$t^\n$ to $t^{\n+1}= t^{\n}+\Delta t$. Note that we use the shorthand notation
$\hat{A}^\n\equiv \hat{A}(t^\n)$ to refer to a quantity $\hat{A}$ at some time step $t^\n$:
\begin{enumerate}[label=\emph{\arabic*)},leftmargin=*,noitemsep]
\item Compute the global integration time step $\Delta t$ used for both hydrodynamics and radiation
  transport as ($i\in\{x,y,z\}$):
  \begin{subequations}\label{eq:timestep}
    \begin{align}
      \Delta t_{\mathrm{rad}} &= \min_{\hi,\hj,\hk,i}
      \left\{ \frac{(\Delta x)_{\hi,\hj,\hk}^i}{|v^i_{\hi,\hj,\hk}|+\max|(\lambda_{\pm})^i_{\hi,\hj,\hk}|}\right\} \, , \\
      \Delta t_{\mathrm{hyd}} &= \min_{\hi,\hj,\hk,i}
      \left\{ \frac{(\Delta x)_{\hi,\hj,\hk}^i}{\max|(\lambda_{\mathrm{fluid}})^i_{\hi,\hj,\hk}|}\right\} \, , \\
      \Delta t &= \mathrm{CFL}\cdot\min\left\{\Delta t_{\mathrm{rad}},\Delta t_{\mathrm{hyd}}  \right\} \, , 
      \label{eq:timestep3}
    \end{align}
  \end{subequations}
  where $\lambda_{\pm}$ and $\lambda_{\mathrm{fluid}}$ are the characteristic velocities of the
  radiation system (cf. Eqs.~(\ref{eq:sigspeedsalpha}, \ref{eq:sigspeeds})) and of the fluid,
  respectively, CFL is the chosen Courant-Friedrichs-Lewy number and $(\Delta x)_{\hi,\hj,\hk}^i$ is
  the length of cell $(\hi,\hj,\hk)$ in coordinate direction $i$.
\item Construct the advection operator
  $(\delta_t \hat{\mathbf{X}})_{\mathrm{adv}}^\n \equiv (\delta_t
  \hat{\mathbf{X}})_{\mathrm{hyp}}^\n+(\delta_t\hat{\mathbf{X}})_{\mathrm{vel}}^\n$
  as a function of the radiation moments $\hat{\mathbf{X}}^\n$ and fluid quantities
  $\hat{\mathbf{U}}^\n$ using the discretization rules described in Secs.~\ref{ssec:hyperbolic} and
  \ref{ssec:transport}.
\item Compute the hydrodynamics evolution operator $(\delta_t \hat{\mathbf{U}})_{\mathrm{hyd}}^\n$
  as a function of $\hat{\mathbf{U}}^\n$. Eventually, add other properly discretized terms to
  $(\delta_t \hat{\mathbf{U}})_{\mathrm{hyd}}^\n$ corresponding to additional, non-radiative physics
  (such as gravitation or magnetic fields).
\item Perform an intermediate update to the time $t^{\n+\half}\equiv t^\n + \Delta t/2$ by solving
  \begin{subequations}\label{eq:updthalf}
    \begin{align}
      \hat{\mathbf{X}}^{\n+\half} &= \hat{\mathbf{X}}^{\n} + \frac{\Delta t}{2} \,
      \left[-(\delta_t \hat{\mathbf{X}})_{\mathrm{adv}}^\n + 
        (\delta_t \hat{\mathbf{X}})_{\mathrm{src}}^{\n,\n+\half}\right] \, , \\
      \hat{\mathbf{U}}^{\n+\half} &= \hat{\mathbf{U}}^{\n} + \frac{\Delta t}{2} \,
      \left[-(\delta_t \hat{\mathbf{U}})_{\mathrm{hyd}}^\n + 
        (\delta_t \hat{\mathbf{U}})_{\mathrm{src}}^{\n,\n+\half}\right]
    \end{align}
  \end{subequations}
  for $\hat{\mathbf{X}}^{\n+\half}$, $\hat{\mathbf{U}}^{\n+\half}$, where the two comma-separated
  superscripts indicate that the source terms can generally depend on the hydro- and radiation
  variables at both the old and the new time step, accounting for the fact that the integration can
  be performed explicitly and/or implicitly as the circumstances require
  (cf. Sec.~\ref{ssec:physsource}).
\item In analogy to steps~\emph{2)} and~\emph{3)} use $\hat{\mathbf{X}}^{\n+\half}$ and
  $\hat{\mathbf{U}}^{\n+\half}$ to compute $(\delta_t \hat{\mathbf{X}})_{\mathrm{adv}}^{\n+\half}$
  and $(\delta_t \hat{\mathbf{U}})_{\mathrm{hyd}}^{\n+\half}$, respectively.
\item The quantities at $t^{\n+1}$ are finally obtained by solving
  \begin{subequations}\label{eq:updtfull}
    \begin{align}
      \hat{\mathbf{X}}^{\n+1} &= \hat{\mathbf{X}}^{\n} + \Delta t \,
      \left[-(\delta_t \hat{\mathbf{X}})_{\mathrm{adv}}^{\n+\half} + 
        (\delta_t \hat{\mathbf{X}})_{\mathrm{src}}^{\n+\half,\n+1}\right] \, , \\
      \hat{\mathbf{U}}^{\n+1} &= \hat{\mathbf{U}}^{\n} + \Delta t \,
      \left[-(\delta_t \hat{\mathbf{U}})_{\mathrm{hyd}}^{\n+\half} + 
        (\delta_t \hat{\mathbf{U}})_{\mathrm{src}}^{\n+\half,\n+1}\right]
    \end{align}
  \end{subequations}
  for $\hat{\mathbf{X}}^{\n+1}$, $\hat{\mathbf{U}}^{\n+1}$ in an analog manner as in step~\emph{4)}.
\end{enumerate}
The above update scheme is formally unsplit\footnote{By \emph{unsplit}, we refer to the property
  that the advection and source terms are integrated within a single step, in contrast to which in a
  \emph{split} scheme the quantities would be updated first for one set of terms (including the
  recomputation of primitive variables such as temperatures, opacities and Eddington factors) before
  calculating the other set of terms.} and is 2nd-order accurate in time with respect to all
explicit (advection or source) terms and 1st-order accurate with respect to implicit source
terms. Although realizations of higher-order implicit--explicit (IMEX) schemes exist
\citep[e.g.][]{Ascher1997,Pareschi2005,McKinney2014} we found that the method described above is
sufficiently robust and accurate in all applications we considered so far.

\subsection{Velocity-independent hyperbolic part}\label{ssec:hyperbolic}

Our basic treatment of the velocity-independent, hyperbolic part of the two-moment system,
Eqs.~\eqref{eq:tmt1}, follows along the lines of \citet{Pons2000} and \citet{Audit2002}. The notion
is to exploit a Godunov method \citep{Godunov1959} as the basis for a high-resolution shock
capturing scheme that solves the local Riemann problems between discontinuous states at the
interfaces of grid cells. We start the presentation of its working method by considering the
one-dimensional system
\begin{equation}\label{eq:redtmt1}
  \partial_t \left(\begin{array}{c} E \\ F \end{array}\right)
  + \partial_x\left(\begin{array}{c} F \\ \cl^2\chi E \end{array}\right) = 0 \, ,
\end{equation}
where the algebraic closure $\chi=\chi(e,f)$ is a function of $e$ and $f$. This system is hyperbolic
if the Jacobian matrix $\mathcal{J}$ of the vector of fluxes $(F,\cl^2\chi E)^{\mathrm{T}}$,
\begin{equation}
  \mathcal{J} = \left(\begin{array}{cc} 0 & 1 \\ \cl^2(\chi + e\frac{\partial\chi}{\partial e} 
      - f \frac{\partial\chi}{\partial f}) & \cl \frac{\partial\chi}{\partial f} \end{array}\right) \, ,
\end{equation}
has real eigenvalues $\lambda_{\pm}^{\mathrm{1D}}$, given by
\begin{equation}\label{eq:propspeeds}
  \lambda_\pm^{\mathrm{1D}} = \frac{\cl}{2}\frac{\partial\chi}{\partial f} \pm \frac{\cl}{2}\sqrt{\frac{\partial\chi}{\partial f}^2 
      + 4(\chi + e\frac{\partial\chi}{\partial e} -f\frac{\partial\chi}{\partial f} )} \, .
\end{equation}
All of the closures listed in Eqs.~\eqref{eq:closures} fulfill the condition of hyperbolicity and
lead to the following properties: In the free-streaming limit, $f\rightarrow 1$, we have
\begin{equation}
  \chi = 1 \; , \; \lambda_+^{\mathrm{1D}} = +\cl \; , \;  \lambda_-^{\mathrm{1D}} = (\frac{\partial\chi}{\partial f}-1)\cl  \, ,
\end{equation}
while in the diffusion limit, $f\rightarrow 0$, one obtains
\begin{equation}\label{eq:lambdadiff}
  \chi = \frac{1}{3} \quad , \quad \lambda_\pm^{\mathrm{1D}} = \pm \frac{1}{\sqrt{3}}\cl \, .
\end{equation}
That is, the limiting cases for the Eddington factor and the wave speeds are consistent with
what is dictated by the Boltzmann equation.

In the multidimensional generalization of Eq.~\eqref{eq:redtmt1} the matrix eigenvalues contain an
additional dependence on the direction cosine $\mu\equiv\cos\alpha_\vecF$, where $\alpha_\vecF$ is
the angle between the direction of the radiation flux vector $\vecF$ and the coordinate direction
with respect to which the derivative is taken. The wave speeds are now obtained as roots of a cubic
polynomial leading, at least in terms of a general closure, to rather large
expressions\footnote{See, however, \citet{Skinner2013} who found for the particular case of the
  $M_1$ closure comparably compact expressions for the wave speeds as functions of $\mu$ and
  $f$.}. For practical purposes we do not take into account the exact angular dependence of the
eigenvalues\footnote{Note that also a third eigenvalue $\lambda_0$ appears in the multidimensional
  case which fulfills $\lambda_-<\lambda_0 <\lambda_+$.  However, this eigenvalue is not relevant
  for our present purpose.} $\lambda_{\pm}^{\mathrm{exact}}(\mu)$ but we instead approximate the
latter using the following 1st-order expansion in $\mu$:
\begin{equation}\label{eq:sigspeedsalpha}
  \lambda_{\pm}(\mu) = \lambda_{\pm}^{\mathrm{exact}}(0) + |\mu| \left(
    \lambda_{\pm}^{\mathrm{exact}}(1) - \lambda_{\pm}^{\mathrm{exact}}(0) \right)  \, ,
\end{equation}
where
\begin{subequations}\label{eq:sigspeeds}
  \begin{align}
    \lambda_{\pm}^{\mathrm{exact}}(1) &= \lambda_{\pm}^{\mathrm{1D}} \, , \label{eq:lambda1}\\
    \label{eq:lambdaorthog}
    \lambda_{\pm}^{\mathrm{exact}}(0) &= \pm \frac{c}{2} \sqrt{2(1 - \chi - e\frac{\partial\chi}{\partial e}) +
      \frac{1}{f}\frac{\partial\chi}{\partial f} (1+2f^2-3\chi) } \, .
  \end{align}
\end{subequations}
In Appendix~\ref{sec:mult-char-wave} we provide the components of the Jacobian in terms of a general
closure $\chi(e,f)$ and show plots of the exact and linearized wave speeds for some specific
closures. These plots indicate that the linearized wave speeds reproduce the exact wave speeds
sufficiently well for the former to be used instead of the latter as estimates for signal speeds of
an approximate Riemann solver (see below). It is worth to note that the qualitative behavior of the
angular dependence of the wave speeds, $\lambda_{\pm}$, is physically consistent with the underlying
Boltzmann equation: In the diffusion regime, $f\ll 1$, the wave speeds become nearly isotropic with
$|\lambda_{\pm}|\rightarrow \cl/\sqrt{3}$, while in the free-streaming regime, $f\rightarrow 1$, the
wave speeds become forward-peaked with $\lambda_{\pm}\rightarrow c$ in the direction of $\vecF$ and
$\lambda_{\pm}\rightarrow 0$ orthogonally to $\vecF$.

In a fashion that is commonly employed in numerical hydrodynamics, we use the above velocities as
signal speeds for an approximate Riemann solver in order to compute the numerical fluxes through
each cell interface. We use the two-wave solver by Harten, Lax and van Leer \citep[HLL,
][]{Harten1983}, which approximates the final numerical interface fluxes as functions of the
left-/right-hand side fluxes $\mathsf{F}^{\mathrm{L/R}}$ (with $\mathsf{F}\in\{F^i, \cl^2 P^{ij}\}$)
and states $\mathsf{U}^{\mathrm{L/R}}$ (with $\mathsf{U} \in \{E, F^i\}$) as
\begin{equation}\label{eq:HLLhyp}
  \mathsf{F}^{\mathrm{HLL}} \equiv \frac{\lambda^{\mathrm{HLL}}_+\mathsf{F}^{\mathrm{L}} -
        \lambda^{\mathrm{HLL}}_- \mathsf{F}^{\mathrm{R}}}{\lambda^{\mathrm{HLL}}_+ - \lambda^{\mathrm{HLL}}_-}
      + \frac{\lambda^{\mathrm{HLL}}_+\lambda^{\mathrm{HLL}}_- 
        \left( \mathsf{U}^{\mathrm{R}} - \mathsf{U}^{\mathrm{L}} \right)}
            {\lambda^{\mathrm{HLL}}_+ - \lambda^{\mathrm{HLL}}_-} \, ,
\end{equation}
with the signal speeds
$\lambda^{\mathrm{HLL}}_+=\max(0,\lambda_+^{\mathrm{L}},\lambda_+^{\mathrm{R}})$ and
$\lambda^{\mathrm{HLL}}_-=\min(0,\lambda_-^{\mathrm{L}},\lambda_-^{\mathrm{R}})$. All quantities
labeled by L/R in this flux formula are computed using the cell-interface reconstructed moments
$\hat{E}^{\mathrm{L/R}}, \hat{F}^{i,\mathrm{L/R}}$. Applying this solver, the final
spatially-discretized version of the operator $(\delta_t X)_{\mathrm{hyp}}$ of
Eq.~\eqref{eq:tmtsplit} reads (using $\hat{X}\in \{\hat{E},\hat{F}\}$)
\begin{align}\label{eq:operhyp}
  \left(\delta_t\hat{X}_{\hi,\hj,\hk,\xi}\right)_{\mathrm{hyp}} &=
  \frac{\Delta A_{\hi+\half,\hj,\hk}\mathsf{F}^{\mathrm{HLL}}_{\hi+\half,\hj,\hk,\xi} -
  \Delta A_{\hi-\half,\hj,\hk}\mathsf{F}^{\mathrm{HLL}}_{\hi-\half,\hj,\hk,\xi}}
{\Delta V_{\hi,\hj,\hk}} \nonumber \\
  & \hspace{0.5cm}+ ``y" + ``z" \, ,
\end{align}
where we symbolically indicated the contributions from the $y$- and $z$-directions, which are
obtained in an analog manner.

Yet, there is a caveat we have to face when approaching the parabolic diffusion limit
(cf. Eq.~\eqref{eq:difflimit}) with the scheme described above, since the latter is originally
designed only for hyperbolic systems. In contrast to the hyperbolic system, the parabolic diffusion
equation is not associated with characteristic waves propagating information between cells with
finite speeds. Hence, the ansatz of using a Riemann solver that tracks characteristics via upwinding
and captures shocks by adding diffusivity is no longer justified in the parabolic diffusion
regime. Instead, the fluxes in the diffusion regime should be of central type (i.e. symmetric with
respect to the cell interface) and they should lead to as little as possible numerical diffusivity
in order not to spoil the effects of the physical diffusivity. To handle this issue, we employ a
simple switch between the two types of fluxes according to:
\begin{equation}
  \mathsf{F}^{\mathrm{HLL},\ast}_{\hi+\half}  =
  \begin{cases}
    \mathsf{F}^{\mathrm{HLL}}_{\hi+\half} & \textrm{if } \mathcal{P}_{\hi+\half} < 1 \, , \\
    \half \left( \mathsf{F}^{\mathrm{L}}_{\hi+\half} + \mathsf{F}^{\mathrm{R}}_{\hi+\half} \right)
    & \textrm{if } \mathcal{P}_{\hi+\half} > 1 \, ,
  \end{cases}
\label{eq:pecflux}
\end{equation}
where the index ``$\hi$'' denotes a representative grid index for any coordinate direction and
the ``stiffness parameter''
\begin{equation}\label{eq:pecdef}
  \mathcal{P} \equiv \kappa_{\mathrm{tra}}\Delta x = \frac{\Delta x}{\lambda_\nu} \, ,
\end{equation}
is a measure of the degree of neutrino--matter coupling relative to numerically resolved scales of
length and time: For $\mathcal{P}\ga 1$ neutrino interactions proceed on spatial and temporal scales
smaller than the grid scale $\Delta x$ and shorter than the numerical time step $\Delta x/c$,
respectively. Hence, for $\mathcal{P}$ exceeding unity the source terms become stiff and thereby
cause the two-moment system to undergo the transition from a hyperbolic to a parabolic system. Our
experience from several tests (cf. Sec.~\ref{sec:tests}, particularly
Sec.~\ref{sec:stat-dynam-diff}) has shown that the seemingly discontinuous jump between both flux
formulations in Eq.~\eqref{eq:pecflux} has no significant influence on the solution. This is because
at the point of transition, $\mathcal{P}=1$, one often has the situation that (1) the flux factors
$f$ are sufficiently small to lead to nearly equal contributions of $\mathsf{F}^{\mathrm{L}}$ and
$\mathsf{F}^{\mathrm{R}}$ in $\mathsf{F}^{\mathrm{HLL}}$, and (2) the relative importance of the
diffusive part of $\mathsf{F}^{\mathrm{HLL}}$ (i.e. the second term in Eq.~\eqref{eq:HLLhyp}) is
still negligible, particularly when using high-order spatial reconstruction.

\subsection{Velocity-dependent terms}\label{ssec:transport}

In the following we present the recipes used to discretize the velocity-dependent terms
$(\delta_t X)_{\mathrm{vel}}$. In order to discretize the terms containing velocity derivatives, we
reconstruct the velocities to obtain $\hat{v}^{i,\mathrm{L/R}}$ located at each cell interface by
using the same reconstruction algorithm as for the radiation moments.

\subsubsection{Fluid-advection terms}\label{ssec:quasiadv}

As fluid-advection terms we denote the (I)-terms in Eqs.~\eqref{eq:tmt1}. For their discretization
we also employ an HLL-type Riemann solver for each coordinate direction analog to
Eqs.~\eqref{eq:HLLhyp} and \eqref{eq:operhyp}. Specifically, we first compute fluxes
$\mathsf{F}^{\mathrm{HLL,adv}}$ defined by the right-hand side of Eq.~\eqref{eq:HLLhyp}, but with
the numerical interface fluxes
$\mathsf{F}^{\mathrm{L/R}} = \hat{v}^{\mathrm{L/R}}\hat{X}^{\mathrm{L/R}}$ and the signal velocities
$\lambda^{\mathrm{HLL}}_+=\max(0,\hat{v}^{\mathrm{L}},\hat{v}^{\mathrm{R}} )$ and
$\lambda^{\mathrm{HLL}}_-=\min(0,\hat{v}^{\mathrm{L}},\hat{v}^{\mathrm{R}} )$, where the
$\hat{v}^{\mathrm{L/R}}$ are the reconstructed velocity components normal to the interface at which
the corresponding numerical flux is computed, and
$\hat{X}^{\mathrm{L/R}}\in\{\hat{E}^{\mathrm{L/R}},\hat{F}^{i,\mathrm{L/R}}\}$ are the reconstructed
moments defined at the same interface. The final fluid-advection terms are then discretized
exemplarily in $x$-direction as
\begin{equation}
  \left(\delta_t\hat{X}_{\hi,\hj,\hk,\xi}\right)_{\mathrm{adv}} =
  \frac{\Delta A_{\hi+\half,\hj,\hk}\mathsf{F}^{\mathrm{HLL,adv}}_{\hi+\half,\hj,\hk,\xi} -
  \Delta A_{\hi-\half,\hj,\hk}\mathsf{F}^{\mathrm{HLL,adv}}_{\hi-\half,\hj,\hk,\xi}}{\Delta V_{\hi,\hj,\hk}}
\end{equation}
and analogously in the other coordinate directions.

\subsubsection{Velocity derivatives}

The remaining velocity derivatives are discretized exemplarily in $x$-direction as
\begin{equation}
  \partial_x v^i \longrightarrow
  \frac{\Delta A_{\hi+\half,\hj,\hk}\hat{v}^i_{\hi+\half,\hj,\hk} - \Delta A_{\hi-\half,\hj,\hk}\hat{v}^i_{\hi-\half,\hj,\hk}}
  {\Delta V_{\hi,\hj,\hk}} \, ,
\end{equation}
while to obtain unique interface velocities $\hat{v}^i_{\hi+\half,\hj,\hk}$ we arithmetically
average the reconstructed velocities:
\begin{equation}
 \hat{v}^i_{\hi+\half,\hj,\hk} =
 \half \left( \hat{v}^{i,\mathrm{L}}_{\hi+\half,\hj,\hk} + \hat{v}^{i,\mathrm{R}}_{\hi+\half,\hj,\hk} \right)\, .
\end{equation}
The discretization of the remaining components of $\partial_j v^i$ is given by analog expressions.

\subsubsection{Doppler shift terms}\label{ssec:ebincoupling}

In our multi-group treatment of comoving-frame radiation transport, we allow radiation energy to be
redistributed between energy groups via the (IV)-terms in Eqs.~\eqref{eq:tmt1} describing Doppler
shift. From the computational point of view an important property of the Doppler shift terms is that
they have a different functional structure in the energy-based moment equations,
Eqs.~\eqref{eq:tmt1}, than in the number-based moment equations, Eqs.~\eqref{eq:tmtn1}, albeit being
physically equivalent. As a consequence, a naive discretization of the Doppler terms in the energy
equation will generally lead to non-conservation of neutrino number and therefore of lepton
number. Although the non-conservation could be avoided by solving the number-based moment equations
in addition to the energy-based versions \citep[e.g.][]{Rampp2002}, this would at least double the
computational expense. We therefore implemented the number-conservative method developed by
\citet{Muller2010}. For a detailed description of this scheme we refer the reader to the original
paper; in the following we only briefly summarize the key features.

Suppressing spatial grid and tensor indices, we write the combined Doppler shift terms of the
0th-moment equation for the $\xi$'th energy bin as
\begin{align}
  \left(\delta_t\hat{E}_\xi\right)_{\mathrm{Doppler}} &= w \int_{\Delta\eps_\xi}
  \left( P - \frac{\partial\eps P}{\partial\eps}\right)\mathrm{d}\eps \nonumber\\
  &= w \, ( \hat{P}_\xi + \mathbb{F}_{\xi-\half} - \mathbb{F}_{\xi+\half}  ) \, ,
\end{align}
where $w$ subsumes the discretized velocity derivatives, $\hat{P}_\xi$ denotes a discretized
component of the 2nd-moment tensor obtained by applying the closure relation to the discretized
moments $\hat{E}_\xi$ and $\hat{F}_\xi$, and $\mathbb{F}_{\xi\pm\half}$ are the discretized versions
of the effective fluxes $\eps P$ located at the energy-bin interfaces $\eps_{\xi\pm\half}$. With the
constraint that the energy-integrated number density shall be conserved, i.e. that
\begin{equation}
  \sum_\xi \eps_\xi^{-1}\left(\delta_t\hat{E}_\xi\right)_{\mathrm{Doppler}} \stackrel{!}{=} 0\, ,
\end{equation}
the interface fluxes $\mathbb{F}_{\xi\pm\half}$ can be written as\footnote{Note that these interface
  fluxes are not uniquely determined but chosen as a specific set fulfilling the imposed constraint
  of total number conservation.}
\begin{equation}\label{eq:doppl_fluxsplit}
  \mathbb{F}_{\xi+\half} = \mathbb{F}^{\mathrm{L}}_\xi + \mathbb{F}^{\mathrm{R}}_{\xi+1} \, ,
\end{equation}
with
\begin{equation}\label{eq:doppl_flux1}
  \mathbb{F}^{\mathrm{L}}_{\xi} = \frac{1}{1-\eps_\xi\eps_{\xi+1}^{-1}}\hat{P}_\xi\gamma_\xi
  \quad , \quad
  \mathbb{F}^{\mathrm{R}}_{\xi} = \frac{1}{\eps_\xi\eps_{\xi-1}^{-1}-1}\hat{P}_\xi(1-\gamma_\xi) \, .
\end{equation}
The weighting factor $\gamma_\xi\in [0,1]$ can be chosen manually and we fix it similarly as in
\citet{Muller2010}. At the lower boundary in energy space we usually have the minimum energy
$\eps_{1-\half}=0$ and therefore $\mathbb{F}_{1-\half}=0$. For the upper boundary at
$\eps_{N_\xi+\half}$ we either use an exponentially extrapolated energy distribution for the
high-energy tail or the condition that the numerical flux vanishes. For the energy derivative of the
3rd moments occurring in the 1st-moment equation, Eq.~\eqref{eq:tmtf1}, we use an analog
prescription as the one given above for the 2nd moments. Specifically, the Doppler shift terms
  for one component $\hat{F}_\xi$ of the 1st-moment vector (again suppressing tensor indices) is
  discretized as 
\begin{align}
  \left(\delta_t\hat{F}_\xi\right)_{\mathrm{Doppler}} &= w \int_{\Delta\eps_\xi}
  \left( - \frac{\partial\eps Q}{\partial\eps}\right)\mathrm{d}\eps \nonumber\\
  &= w \, ( \mathbb{F}_{\xi-\half} - \mathbb{F}_{\xi+\half}  ) \, ,
\end{align}
where $w$ again represents velocity derivatives and $\mathbb{F}_{\xi-\half}$ are calculated via
Eqs.~\eqref{eq:doppl_fluxsplit} and~\eqref{eq:doppl_flux1} exactly as explained above but with
$\hat{P}_\xi$ replaced by $\hat{Q}_\xi$.

\subsection{Interaction source terms}\label{ssec:physsource}

As already mentioned before we intend to use a time step $\Delta t$ that is close to the
radiative advection timescale $\tau_{\mathrm{adv}}\equiv \Delta x / \cl$ to integrate the full
system of moment equations. However, the numerical integration of the interaction source terms
$(\delta_t X)_{\mathrm{src}}$, Eq.~\eqref{eq:tmtsplit}, deserves special care because the
characteristic neutrino-interaction timescale,
\begin{equation}\label{eq:tautimeste}
\tau_{\mathrm{int}} \equiv ( \cl \kappa_{\mathrm{tra}} )^{-1} = \lambda_\nu/\cl \, ,
\end{equation}
can become shorter than $\tau_{\mathrm{adv}}$ by up to many orders of magnitude. In this case, i.e.
for stiffness parameters $\mathcal{P}>1$ (cf. Eq.~\eqref{eq:pecdef}), the moment equations are stiff
and a fully explicit time integration would lead to numerical instability. Hence, the source terms
make an implicit treatment indispensable. However, it is important to note that the characteristic
timescales of the advection-type terms $(\delta_t X)_{\mathrm{hyp}}$ and
$(\delta_t X)_{\mathrm{vel}}$ (cf. Eq.~\eqref{eq:tmtsplit}) are longer than $\tau_{\mathrm{adv}}$
also for $\mathcal{P}>1$: In the diffusion limit the characteristic timescale of
$(\delta_t X)_{\mathrm{hyp}}$ is the diffusion timescale
$\tau_{\mathrm{diff}}\equiv 3\kappa_{\mathrm{tra}}\Delta x^2 / \cl = 3\, \mathcal{P} \cdot
\tau_{\mathrm{adv}}>\tau_{\mathrm{adv}}$,
while the characteristic timescale of $(\delta_t X)_{\mathrm{vel}}$ is independent of the stiffness
parameter and roughly approximated by\footnote{This estimate disregards fluxes in energy space
  mediated by the Doppler shift terms (see Sec.~\ref{ssec:ebincoupling}), which in principle can
  cause these terms to change on timescales shorter than $\Delta x / v$. In practice, however, these
  timescales are usually longer than $\tau_{\mathrm{adv}}$ such that an explicit integration of the
  Doppler terms with time step $\sim \tau_{\mathrm{adv}}$ turns out to be unproblematic.}
$\sim \Delta x / v$. Hence, we can safely apply a scheme in which the operators
$(\delta_t X)_{\mathrm{hyp}}$ and $(\delta_t X)_{\mathrm{vel}}$ are treated explicitly, while the
local contributions from the interaction source terms are integrated implicitly in time. Compared to
an implicit treatment of the full system, this mixed-type integration greatly reduces the size of
the matrix that needs to be inverted since all spatial derivatives that couple neighbouring cells on
the spatial grid are handled by the explicit part. 

Depending on the included types of interactions, the source terms
$(\delta_t \hat{\mathbf{X}})_{\mathrm{src}}$ and $(\delta_t \hat{\mathbf{U}})_{\mathrm{src}}$ can in
general each depend on all evolved radiation and hydrodynamic quantities. However, (standard)
neutrino interactions do not change the baryonic mass density $\rho$, and they only have a marginal
impact on the fluid momenta $\rho\vecv$, at least in typical situations where neutrino transport is
relevant; hence these quantities may be treated explicitly in time. Still, if we were to integrate
all but the aforementioned quantities implicitly -- which means expressing the source terms
$(\delta_t \hat{\mathbf{X}})_{\mathrm{src}}, (\delta_t \hat{\mathbf{U}})_{\mathrm{src}}$ as
functions of these variables defined at the new time step in Eqs.~\eqref{eq:updthalf}
and~\eqref{eq:updtfull} -- we would generally need to solve a non-linear system of equations of rank
$(N_{\mathrm{dim}}+1)\times N_\eps\times N_{\mathrm{spe}}+2$ (recalling that $N_{\mathrm{dim}}$ and
$N_{\mathrm{spe}}$ are the number of evolved 1st-moment components and neutrino species,
respecively, and $N_\eps$ is the number of energy bins), in which all radiation moments $E, \vecF$
as well as the gas-energy density $e_{\mathrm{i}}$ and the electron-number density
$n_e\equiv\rho Y_e/m_{\mathrm{B}}$ are handled implicitly. However, an implicit treatment of all of
these quantities is not always necessary. That is, under certain conditions a significant reduction
of computational expense can be achieved by treating a subset of variables explicitly, i.e. by using
the quantities defined at the old time step in the update formulae, Eqs.~\eqref{eq:updthalf}
and~\eqref{eq:updtfull}. Below we list the different modes of the source-term treatment, which we
implemented in order to avoid a fully implicit integration whenever this appears justified:
\begin{enumerate}[label=\emph{\alph*)},leftmargin=*,noitemsep]
\item \label{enu:srca} All radiation moments $\hat{E},\hat{\vecF}$ plus the hydrodynamic variables
  $\hat{e}_{\mathrm{i}}, \hat{n}_e$ appearing in the source terms
  $(\delta_t \hat{\mathbf{X}})_{\mathrm{src}}$ and $(\delta_t \hat{\mathbf{U}})_{\mathrm{src}}$ are
  defined at $t^{\n+1}$, i.e. the systems of Eqs.~\eqref{eq:updthalf} and~\eqref{eq:updtfull} are
  solved fully implicitly. Consistently, also most of the primitive variables (temperature,
  opacities etc.) are handled implicitly. Only the normalized 2nd- and 3rd-radiation moments $\chi$
  and $q$ (cf. Sec.~\ref{sec:analyt-eddingt-fact}), respectively, and the Legendre-coefficient
  matrices for reactions coupling multiple energy bins \citep[see, e.g.,][]{Rampp2002} are taken
  from the old time step.
\item \label{enu:srcb} Like~\emph{a)}, but the gas-energy density $\hat{e}_{\mathrm{i}}$ and
  electron-number density $\hat{n}_e$ are taken from the old time step. This reduces not only the
  dimensionality of the coupled non-linear system by 2, but also alleviates the computational
  expense by the demands of re-computing the temperatures and opacities within each iteration step
  of the root finding procedure.
\item \label{enu:srcc} Like~\emph{b)}, but all energy-coupling interactions (e.g. neutrino--electron
  scattering) are treated explicitly in time. The remaining source terms corresponding to
  emission/absorption and isoenergetic scattering can then be written as in
  Eqs.~\eqref{eq:sourceterms}, which results in the source terms to completely decouple from each
  other, allowing for a straightforward implicit integration without any matrix inversion.
\end{enumerate}
Since the criteria for selecting a certain integration mode are highly problem dependent, we do not
specify them here; see, e.g., Sec.~\ref{sec:fully-dynam-evol} for a particular choice in the CCSN
context. In any case where a \emph{coupled} non-linear system of equations has to be solved, we make
use of the routine \textsf{nag\_nlin\_sys\_sol} from the NAG
library\footnote{\textsf{www.nag.co.uk}}, which employs the ``Broyden method'' for root finding.

Based on our experience (see, e.g., last paragraph in Sec.~\ref{sec:fully-dynam-evol}) and in
agreement with \citet{OConnor2014}, the use of integration mode~\emph{a)} (i.e. an implicit
treatment of $\hat{e}_{\mathrm{i}}$ and $\hat{n}_e$) is barely necessary in practice, even in
regions of very high stiffness parameters, $\mathcal{P}\gg 1$. This is because in most situations
where $\mathcal{P}\gg 1$ neutrinos are trapped in the gas and very close to weak equilibrium, which
means that the net (absorption minus emission) source terms for $\hat{e}_{\mathrm{i}}$ and
$\hat{n}_e$ are effectively small. Consequently, under these conditions $\hat{e}_{\mathrm{i}}$ and
$\hat{n}_e$ essentially change only on grounds of fluid-dynamical effects on fluid timescales, the
latter being longer than the radiative timescale $\tau_{\mathrm{adv}}\sim\Delta x / c $ used for
time integration. Conversely, an explicit treatment of $\hat{e}_{\mathrm{i}}$ and $\hat{n}_e$ would
not be feasible if we would integrate the radiation moment equations fully implicitly using a time
step longer than $\sim \Delta x / v$ (as it is done, for instance, in most existing
neutrino-hydrodynamics codes using FLD or a Boltzmann-solver). Additionally, it is worth to note
that although an explicit treatment of $\hat{e}_{\mathrm{i}}$ and $\hat{n}_e$ may slightly reduce
the accuracy it does not significantly harm the numerical stability of the overall scheme, because
the effective source terms for $\hat{e}_{\mathrm{i}}$ and $\hat{n}_e$ are computed via
Eqs.~\eqref{eq:hydsource} using the time-discretized source terms for $\hat{E}_{\nu,\xi}$, which
themselves result from an implicit (and hence stable) integration in $\hat{E}_{\nu,\xi}$.

Once after appropriately time-discretized expressions for the source terms of the 1st radiation
moments are found, the changes of the fluid-momentum and kinetic-energy densities due to momentum
transfer with neutrinos obtained using Eq.~\eqref{eq:momsource} as:
\begin{subequations}\label{eq:srcmom}
\begin{align}
  \left(\delta_t \hat{\rho} \hat{v}^i \right)_{\mathrm{src}} &=    
  -\frac{1}{\cl^2}\sum_{\nu, \xi}\left(\delta_t \hat{F}^i_{\nu,\xi} \right)_{\mathrm{src}} \, , \\
  \left(\delta_t \hat{e}_{\mathrm{k}} \right)_{\mathrm{src}} &=    
  -\frac{\hat{v}_j}{\cl^2}\sum_{\nu, \xi}\left(\delta_t \hat{F}^j_{\nu,\xi} \right)_{\mathrm{src}} \, . \label{eq:srcmomekin}
\end{align}
\end{subequations}

Finally, since this is a non-trivial and important aspect, we now demonstrate that the
time-integration algorithm presented in Sec.~\ref{ssec:update} in combination with the
stiffness-parameter dependent flux formulation, Eq.~\eqref{eq:pecflux}, allows the radiation flux
$\vecF$ in diffusive regions (in which $\mathcal{P}\gg 1$) to relax to the corresponding diffusive
flux $\vecF_{\mathrm{D}} \equiv -\cl \nabla E / (3\kappa_{\mathrm{tra}})$ in a numerically stable
and non-oscillatory manner. To this end, we may neglect velocity-dependent terms -- which are
reduced by a factor of $\mathcal{O}[(v/c)(1/\mathcal{P})]$ compared to the dominant terms in our
comoving-frame formulation and therefore subdominant in the diffusion regime -- and we only consider
the (usually dominant) emission-/absorption and isoenergetic scattering reactions, for which the
1st-moment source terms can be written as
$(\delta_t \vecF)_{\mathrm{src}}=-c\kappa_{\mathrm{tra}}\vecF$. Since in the diffusion regime we
expect the pressure tensor to be almost isotropic, $P^{ij}\simeq \delta^{ij} E/3$, the hyperbolic
operator for the flux density then reads exemplarily in $x$-direction
(cf. Eqs.~\eqref{eq:operhyp},~\eqref{eq:pecflux}):
\begin{align}\label{eq:hypdiffflux}
  (\delta_t\hat{F}_{\hi}^x)^{\n}_{\mathrm{hyp}} &= 
   \textstyle{\frac{c^2}{3\Delta V_{\hi}}\bigg(\Delta A_{\hi+\half} \frac{\hat{E}_{\hi+\half}^{\mathrm{L}}+\hat{E}_{\hi+\half}^{\mathrm{R}}}{2}
     - \Delta A_{\hi-\half} \frac{\hat{E}_{\hi-\half}^{\mathrm{L}}+\hat{E}_{\hi-\half}^{\mathrm{R}}}{2} \bigg)} \nonumber\\
   &\simeq -\cl\hat{\kappa}_{\mathrm{tra}}(\hat{F}_{\mathrm{D}}^x)^{\n}_\hi \, ,
\end{align}
where $\hat{F}_{\mathrm{D}}^x$ is a proper numerical representation of the diffusive flux
$F_{\mathrm{D}}^x$. The first partial update (step~\emph{4)} in Sec.~\ref{ssec:update}) for
$\hat{\vecF}$ then results to:
\begin{align}
  \hat{\vecF}^{\n+\half} &= \hat{\vecF}^{\n} + \Delta t^\ast \left[
    -(\delta_t\hat{\vecF})^{\n}_{\mathrm{hyp}} + (\delta_t\hat{\vecF})^{\n+\half}_{\mathrm{src}}\right] \nonumber \\
  &\simeq  \hat{\vecF}^{\n} + c\hat{\kappa}_{\mathrm{tra}}\Delta t^\ast \hat{\vecF}^\n_{\mathrm{D}}
  - c\hat{\kappa}_{\mathrm{tra}}\Delta t^\ast \hat{\vecF}^{\n+\half} \nonumber\\
  &\simeq  \frac{\hat{\vecF}^{\n}} { 1 + \cl\hat{\kappa}_{\mathrm{tra}}\Delta t^\ast } +
  \frac{\cl\hat{\kappa}_{\mathrm{tra}}\Delta t^\ast \, \hat{\vecF}_{\mathrm{D}}^{\n}} 
  { 1 + \cl\hat{\kappa}_{\mathrm{tra}}\Delta t^\ast } \, ,
\end{align}
where $\Delta t^\ast \equiv \Delta t / 2$. Hence, for $\cl\kappa_{\mathrm{tra}}\Delta t^\ast \gg 1$
the flux density $\hat{\vecF}$ consistently relaxes to $\hat{\vecF}_{\mathrm{D}}$ within one
(partial) time step without any numerical overshooting.

\subsection{Hydrodynamics}\label{ssec:hydro}

The equations of hydrodynamics are integrated using the finite-volume high-resolution
shock-capturing scheme developed in \citet{Obergaulinger2008}. The scheme evolves the conserved
hydrodynamic variables $(\rho,\rho\vecv,e_{\mathrm{t}})$, to which end a variety of procedures for
spatial reconstruction (cf. Sec.~\ref{ssec:basic-discr-scheme}) and Riemann solvers (Lax-Friedrich,
HLL, HLLC, HLLD, see, e.g., \citealp{Toro1997}) can be selected. Moreover, a magnetic-field solver
and various models of viscosity and magnetic diffusivity are implemented. In the case of the model
setup requiring the co-evolution of a set of different fluid species a simplified version of the
``Consistent Multifluid Advection'' scheme \citep{Plewa1999} is utilized. For more details about the
non-radiative part of the code, we refer the reader to \citet{Obergaulinger2008}.


\begin{figure*}
  \includegraphics[width=0.99\textwidth]{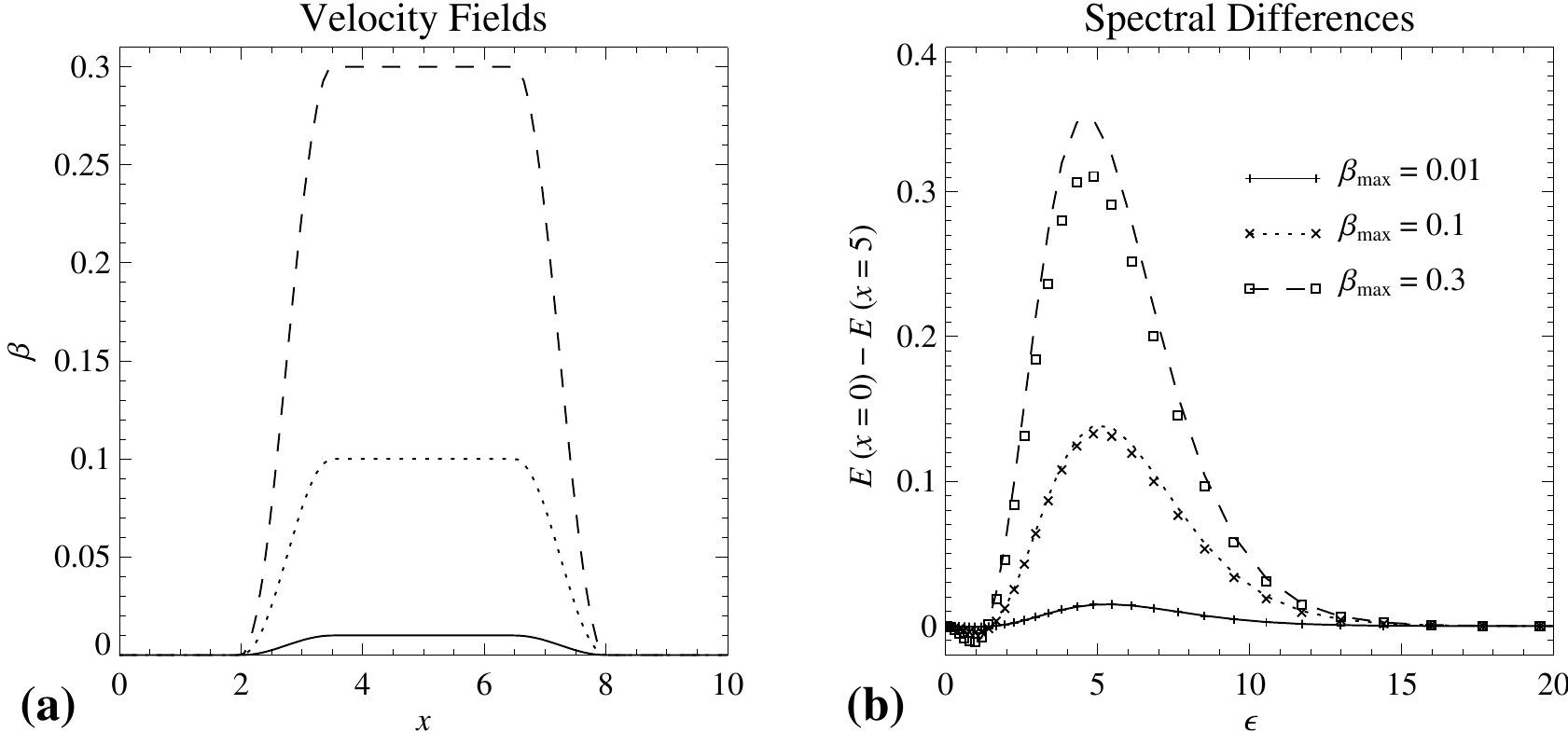}
  \caption{Doppler shift of free-streaming radiation. Panel~(a) shows the three velocity fields as
    functions of the location $x$ and panel~(b) depicts the differences between the energy
    distributions in the two frames with $\beta=0,\beta_{\mathrm{max}}$. The lines denote the fully
    relativistic, analytic solutions, while the symbols show the AEF results.\label{fig:doppler}}
\end{figure*}

\section{Test problems}\label{sec:tests}

In this section, we present a variety of problems to test the methods described in the previous
sections. Several idealized, non-microphysical tests in 1D and 2D are performed which, although they
are not directly related to typical scenarios where neutrino transport plays a role, serve to assess
the quality of the two-moment closure approximation and the coupling of the radiation moments to the
velocity field and to the hydrodynamics part of the code. Subsequently, we present two
one-dimensional test problems related to CCSNe in which we test the AEF scheme both for different
closure prescriptions and against corresponding results from FLD and Boltzmann schemes.

In order to avoid excessive repetitions we list some recurring properties and parameters that
various following tests are equipped with:
\begin{itemize}[noitemsep,leftmargin=*]
\item Regarding the numerical treatment, we employ a 5th-order MP reconstruction method and an HLL
  Riemann solver for both the radiation transport and the hydrodynamics part of the system. For the
  tests in 1D we take a global CFL factor of CFL$=0.7$ while for the 2D tests we set CFL$=0.5$.
\item We apply boundary conditions (BCs) by fixing the values in the boundary (ghost) zones
  surrounding the computational domain according to a given prescription. For a reflective boundary,
  e.g. at $x=x_0$, we copy scalar quantities, e.g.
  $\left. E\right|_{x_0+\delta x}=\left. E\right|_{x_0-\delta x}$, and apply the reflection operator
  to vectorial quantities, e.g.
  $\left.(F^x,F^y,F^z)\right|_{x_0+\delta x} = \left.(-F^x,F^y,F^z)\right|_{x_0-\delta x}$. For a
  non-reflective outflow boundary, e.g. at $x=x_0$ we employ the usual 0th-order extrapolation for
  all quantities, e.g. we set $E(x_0+\delta x)=E(x_0)$.
\item In all of the subsequent tests in which dimensionless equations and quantities are employed
  the speed of light is set to $\cl=1$ and for the velocity the symbol $\beta$ is used.
\item Except for the tests in Secs.~\ref{ssec:supercr-radi-shock},~\ref{sec:neutr-radi-field} and
  \ref{sec:fully-dynam-evol}, we will exclusively use the Minerbo closure, which is expressed by the
  Eddington factor as in Eq.~\eqref{eq:minclos} and the 3rd-moment factor as in
  Eq.~\eqref{eq:defqsmall}.
\end{itemize}

\begin{figure*}
  \includegraphics[width=0.99\textwidth]{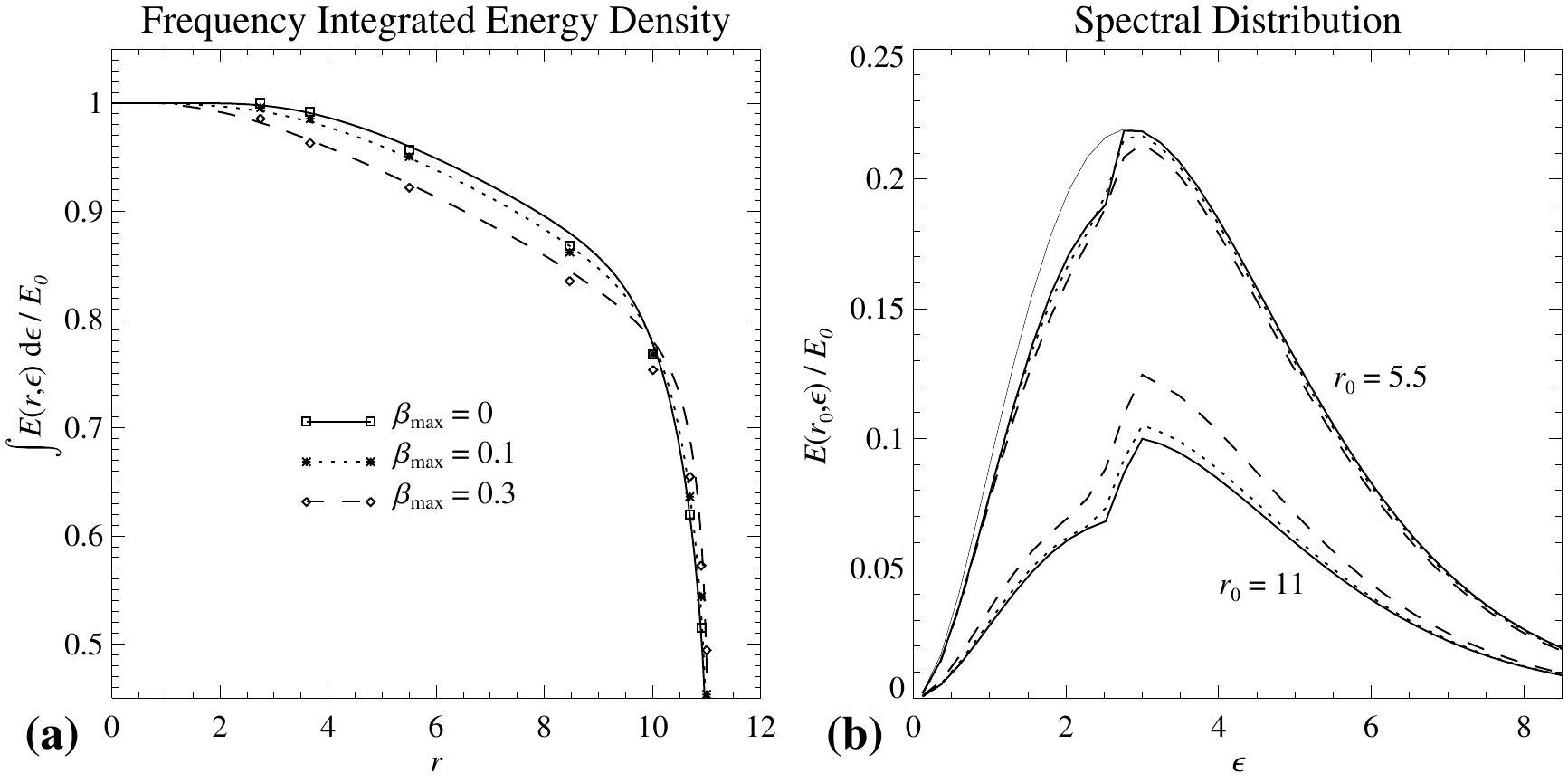}
  \caption{Differentially expanding isothermal atmosphere. In panel~(a) the frequency-integrated
    energy densities, normalized by $E_0\equiv\bar{E}(r=0)$, as function of radius are shown for
    different maximum velocities $\beta_{\mathrm{max}}$. The symbols denote the reference solution
    computed by \citet{Mihalas1980}, where the points are read off their Fig.~5. In panel~(b) we show
    for the same models the spectral distributions of the normalized energy densities at the two radii
    $r=5.5$ and $11$, at which the optical depth for photons with energy $\eps<\eps_0$ is $\approx 1$
    and $\approx 0$, respectively. The thin solid line displays the equilibrium distribution
    $E^{\mathrm{eq}}(\eps)$.} \label{fig:expatmo}
\end{figure*}

\subsection{One-dimensional idealized test problems}\label{sec:one-dimens-ideal}

\subsubsection{Doppler shift of free-streaming radiation}\label{ssec:doppl-redsh-free}

The energy-coupling scheme describing the Doppler shift (cf. Sec.~\ref{ssec:ebincoupling}) can
immediately be tested by comparing the spectra of a free-streaming radiation field in two different
frames with non-vanishing relative velocity. We adopt the basic setup from \citet{Vaytet2011}, but
we use dimensionless units and take higher values for the maximum velocities
$\beta_{\mathrm{max}}\in\{0.01,0.1,0.3\}$. The Cartesian spatial domain covers $x\in[0,10]$ and is
resolved by 100 equidistant grid points, while the energy space is discretized between
$\eps \in [0,50]$ using a logarithmic grid with 40 bins and a bin-enlargement factor of
$\Delta\eps_{\xi+1}/\Delta \eps_{\xi}=1.1$. At $x=0$ we inject a beam of radiation by fixing the
radiation quantities in the boundary zones according to $E(x=0)=\eps^3/(\eul^\eps-1)=F(x=0)$. The
boundary at $x=10$ is set to outflow. The radiation field traverses velocity fields with the shape
of smoothed step-functions as shown in panel~(a) of Fig.~\ref{fig:doppler}. Within regions where
$\beta>0$ the redshifted (i.e. Lorentz-boosted) spectrum of the comoving-frame specific energy
density is analytically given by
\begin{equation}
  E_\beta = \frac{1}{s}\frac{(s\eps)^3}{\eul^{s\eps}-1} \, , \quad \mathrm{where}\;
  s \equiv \sqrt{\frac{1+\beta}{1-\beta}} \, ,
\end{equation}
which is valid for $0\leq\beta\leq1$.

The differences between the spectra in the frames $\beta=0,\beta_{\mathrm{max}}$ for both our
numerical and the analytic solution are shown in panel~(b) of Fig.~\ref{fig:doppler}. The Doppler
shift is captured well by our scheme: The agreement with the analytic solution converges for
decreasing values of $\beta_{\mathrm{max}}$. For high velocities, $\beta_{\mathrm{max}}=0.3$, the
$\mathcal{O}(v/c)$ approximation leads to errors of about 10\,\% with respect to the relativistic
solution.

\subsubsection{Differentially expanding isothermal atmosphere}\label{sec:expanding-atmosphere}

To test the algebraically closed two-moment transport in combination with frame-dependent effects
and (idealized) radiation--matter interactions we examine a scenario that was also investigated by
\citet{Mihalas1980} (in full relativity) and by \citet{Rampp2002} (in $\mathcal{O}(v/c)$), both
using accurate Boltzmann techniques. The scenario includes an expanding, isothermal atmosphere
that expands with velocity $beta$ as function of radius $r$ as
\begin{equation}
  \beta(r) = \beta_{\mathrm{max}}\frac{r-r_{\mathrm{min}}}{r_{\mathrm{max}}-r_{\mathrm{min}}}
\end{equation}
and which exhibits an absorption opacity $\kappa_{\mathrm{a}}$ that varies in $r$ and (photon)
energy $\eps$ as
\begin{equation}
  \kappa_{\mathrm{a}}(r,\eps) =
  \begin{cases}
    \frac{10\alpha}{r^2}\eul^{-(\eps-\eps_0)^2/\Delta^2} +
    \frac{\alpha}{r^2}\left(1 - \eul^{-(\eps-\eps_0)^2/\Delta^2}\right) & , \, \eps\leq\eps_0 \\
    \frac{10\alpha}{r^2} & , \, \eps>\eps_0 \, .
  \end{cases}
\end{equation}
That is, for fixed radius $r$, the opacity is a smoothed step-function in energy space with the
transition at energy $\eps_0$ from a low opacity to a 10 times higher opacity with transition width
$\Delta$. The model parameters are
$\{r_{\mathrm{min}},r_{\mathrm{max}},\eps_0,\Delta,\alpha\} = \{ 1, 11, 3, 0.2,10.9989\}$ and the
maximum velocity $\beta_{\mathrm{max}}$ is varied between $\beta_{\mathrm{max}}\in\{0,0.1,0.3\}$. We
use dimensionless units in which the temperature $T=1$ such that the photon equilibrium energy
density is given by $E^{\mathrm{eq}}(\eps)=\eps^3/(\eul^\eps - 1)$. We set up a uniform radial grid
of 400 zones to cover a region of $r\in[0.1,15]$. The additional (with respect to the reference
calculations) region $[11,15]$, wherein the opacities are set to zero, merely serves as a transition
zone for the radiation field to reach near free-streaming conditions in order to avoid unphysical
feedback from the outer radial boundary, at which outflow BCs are employed. At $r=0.1$, a reflective
BC is applied. The energy grid is composed of 50 equidistant bins within $\eps\in[0,12]$.

For comparison, we show similar plots as in \citet{Rampp2002}, see Fig.~\ref{fig:expatmo}. The total
radiation energy density as function of radius is shown in panel~(a), while the radiation spectrum
at two representative radii is shown in panel~(b). A remarkable fact is that the case with no
expansion is already reproduced well with an accuracy of $\lesssim1\%$ by the approximate AEF
scheme. By switching to $\beta>0$ we introduce the following effects: Due to the expansion the
comoving-frame energy- and flux densities of photons created deeper within the atmosphere decreases
on their way to the surface, as can be seen by monotonic decline of the energy densities with
increasing $\beta_{\mathrm{max}}$ up to $r\approx 10$. However, this trend is competed by the effect
that the overall fraction of photons originally created in the high-opacity band is lifted with
increasing velocity (and radius), since a fraction of photons are redshifted from $\eps>\eps_0$ to
$\eps<\eps_0$ on their way to the surface. Hence, the opacity jump is effectively redshifted by the
expansion, leading to higher integral values of the energy density at $r\gtrsim 10$. Both effects
are captured well by the AEF method. We notice an increasing difference between the
$\mathcal{O}(v/c)$ results and the fully relativistic results for increasing
$\beta_{\mathrm{max}}$. However, for $\beta_{\mathrm{max}}=0.3$ the maximum error in the integral
energy density, cf. panel~(a) of Fig.~\ref{fig:expatmo}, is still less than $\sim 5\,\%$.

\begin{figure}
  \includegraphics[width=84mm]{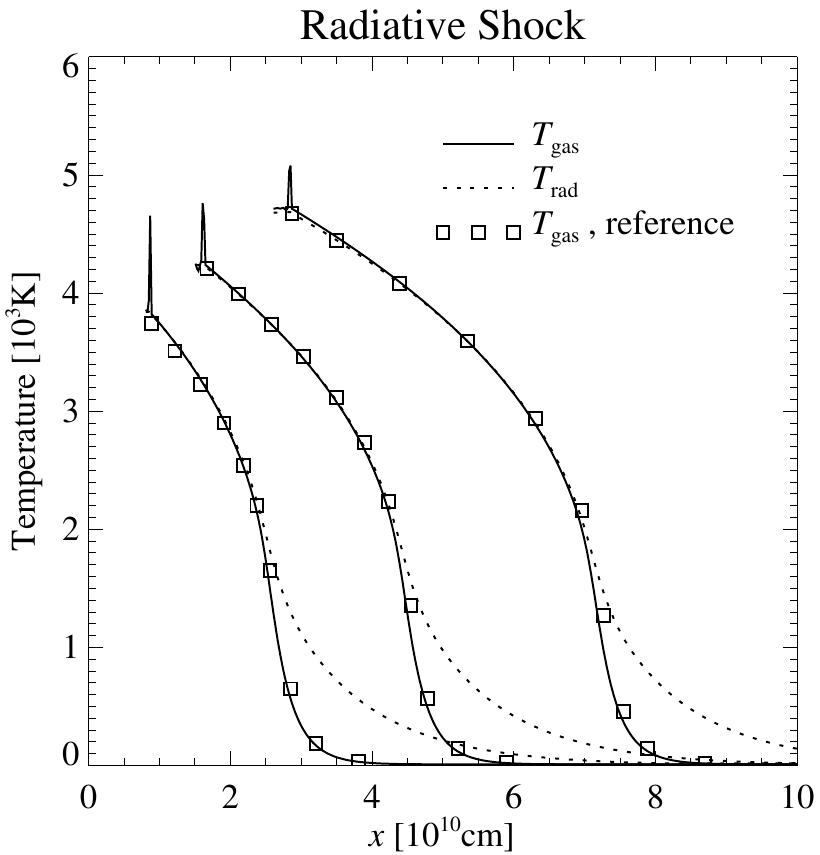}
  \caption{Supercritical radiative shock. We show the gas and radiation temperatures at three
    different times $t\in\{4\times 10^3,7.5\times 10^3, 1.3\times 10^4\}$\,s. Each curve is plotted in
    the frame in which the shock crosses $x=0$ at $t=0$ with a velocity
    $v_{\mathrm{s}}=2\times 10^6$\,cm\,s$^{-1}$. The square symbols denote points from the reference
    calculation by \citet{Vaytet2011}, which have been read off their Fig.~9.}
  \label{fig:radshock}
\end{figure}

\subsubsection{Supercritical radiative shock}\label{ssec:supercr-radi-shock}

Successively increasing the degrees of freedom taken into account, we now turn to a classical RHD
problem to test the accurate coupling between transport and hydrodynamics, the radiative shock
tube. Having been the subject of numerous investigations, both analytically
\citep[e.g][]{Zeldovich1966,Mihalas1984} and numerically \citep[e.g.][]{Ensman1994,Sincell1999},
radiative shock tubes repeatedly served as test problems for the development of new RHD codes as,
e.g., in \citet{Turner2001,Hayes2003,Gonz'alez2007,Vaytet2011,Jiang2012,Skinner2013}.

Since the detailed physical description of radiative shocks is out of the scope of our presentation,
we only briefly summarize their essential properties here. In contrast to purely hydrodynamic shocks
radiative shocks allow for energy transfer between the gas and radiation, effectively introducing
cooling of the post-shock and heating of the pre-shock material. Depending on the shock velocity,
heating of upstream material in front of the shock -- this region is called radiative precursor --
can become so efficient that the pre-shock temperature adapts to the post-shock temperature, in
which case the shock is called a supercritical radiative shock. In this case, both the up- and
downstream material is in radiative equilibrium close to the shock and separated by a sharp
non-equilibrium temperature spike (``Zeldovich spike''), which is roughly as wide as the local
mean-free path of radiation.

We initialize our model of a supercritical radiative shock using the same setup as
\citet{Vaytet2011}, who also employed a spectral AEF scheme fairly similar to ours. Moreover, to
facilitate the comparability with \citet{Vaytet2011} we apply the $M_1$ instead of the Minerbo
closure for this test. The only difference to the setup of \citet{Vaytet2011} is that we ignore here
the 3rd-moment terms in the evolution equation for the 1st moment (i.e. the (IV)-terms in
Eq.~\eqref{eq:tmtf1}). However, these terms vanish in the energy-integrated form of the RHD
equations and the fluid quantities, such as the temperature, should not be significantly influenced
by this measure, particularly in view of the relatively low velocities at hand. A Cartesian box of
length $10^{11}$\,cm is discretized by a uniform grid of 500 cells and initially homogeneously
filled with gas of density $\rho=7.78\times 10^{-10}$\,g\,cm$^{-3}$, temperature $T=10$\,K and grey
absorption opacity $\kappa_{\mathrm{a}}=3.1\times 10^{-10}$\,cm$^{-1}$. Initially, radiation is
everywhere in equilibrium with the gas, such that
$\bar{E}=\bar{E}^{\mathrm{eq}}=a_{\mathrm{rad}}T^4$, where
$a_{\mathrm{rad}}\approx 7.57\times 10^{-15}$erg\,cm$^{-3}$\,K$^{-4}$ is the radiation constant. The
gas pressure is computed as
$P_{\mathrm{g}} = (\gamma_{\mathrm{gas}}-1)e_{\mathrm{i}} = \rho k_{\mathrm{B}} T / m_{\mathrm{B}}$
with $\gamma_{\mathrm{gas}}=1.4$. We take the frame of the shock moving with
$v_{\mathrm{s}}=2\times 10^6$\,cm\,s$^{-1}$ relative to its preceding medium as our simulated
inertial frame. To this end, we let all matter in the computational domain initially move with
velocity $v=-v_{\mathrm{s}}$. The shock is induced by using a reflective boundary at $x=0$ and it is
maintained by feeding new material with the original properties at $x=10^{11}$\,cm into the
computational domain. We discretize the frequency space with $N_\eps=8$ evenly spaced bins between
$\eps\in[0,8\times 10^{14}]$\,Hz.

The results are shown in Fig.~\ref{fig:radshock} in form of the distributions of the gas temperature
$T_{\mathrm{gas}}$ and the radiation temperature, the latter being defined as
$T_{\mathrm{rad}}\equiv(\bar{E}/a_{\mathrm{rad}})^{1/4}$. Using an essentially similar physical
evolution model as \citet{Vaytet2011}, the results we obtain with our quasi-explicit numerical
method are in good agreement with the outcome of their implicit radiation solver which shows that
the coupling between the radiative and hydrodynamics systems is numerically robust and produces
accurate results.

\subsection{Two-dimensional idealized test problems}\label{sec:two-dimens-ideal}

\begin{figure}
  \includegraphics[width=84mm]{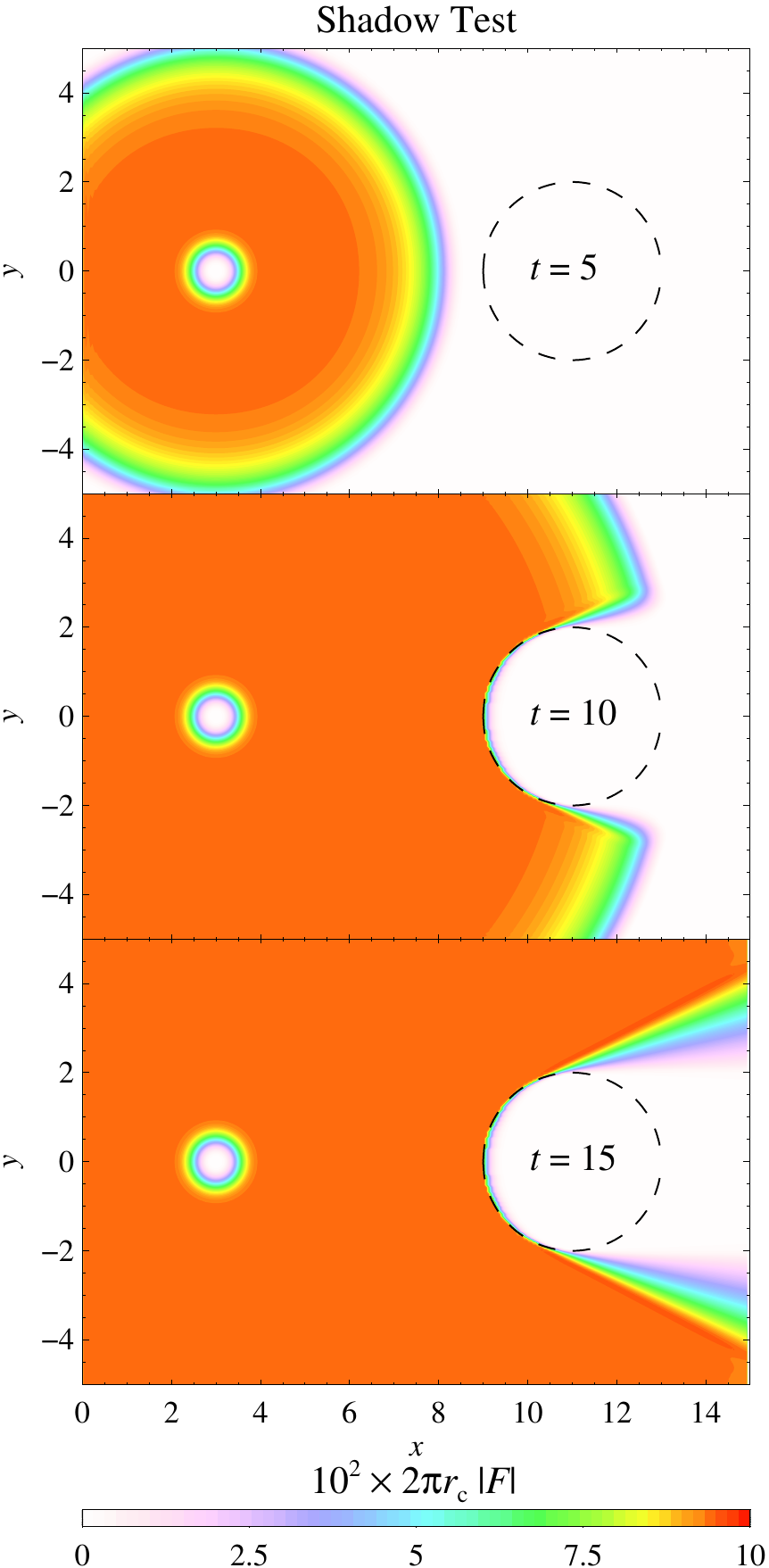}
  \caption{Shadow casting problem. The 2D luminosity at the three indicated times $t$ is plotted. The
    dashed line indicates the boundary of the absorbing gas cloud.}\label{fig:shadow}
\end{figure}

\begin{figure*}
  \includegraphics[width=0.99\textwidth]{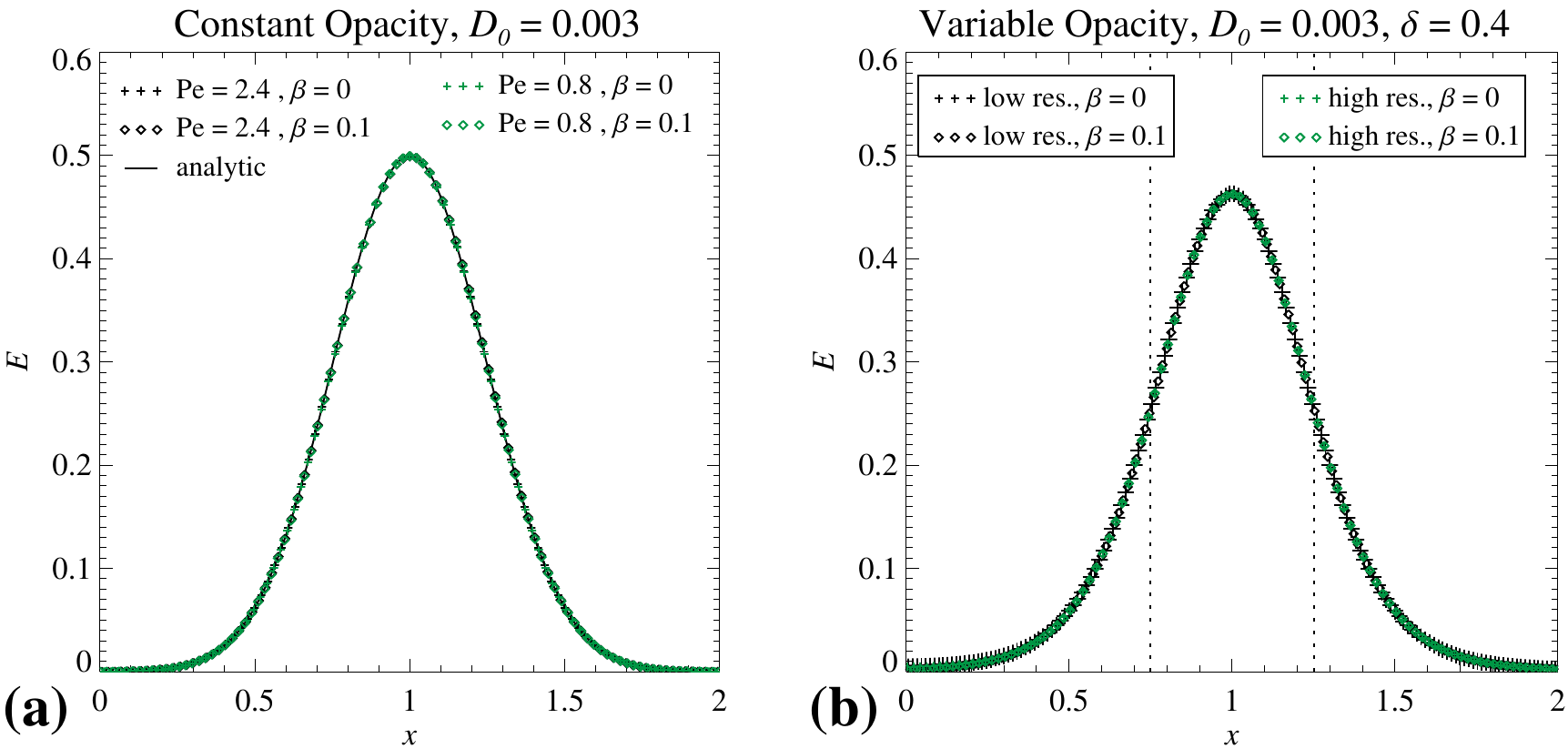}
  \caption{Static and dynamic diffusion. We show profiles of the energy densities along the slice
    $y=1$ after the simulation time $t-t_0=5$ of the Gaussian pulse centered at $\vecx_0=(1,1)$. In
    panel~(a) the results for the cases with spatially constant opacity are plotted. The models
    differ in spatial resolution, and therefore in the stiffness parameters $\mathcal{P}$, and in
    the uniform velocity field parallel to the $x$-direction with magnitude $\beta$. The results for
    the cases with $\beta>0$ were transformed to the $\beta=0$ frame and shifted to their initial
    position at $t=0$. For a better readability, only every second zone value is plotted for the
    high-resolution models. The results shown in panel~(b) are produced by the same initial Gaussian
    pulse but employing an opacity that declines with distance to the center of the pulse. The
    vertical lines denote the locations where the stiffness parameter of the low-resolution models
    is $\mathcal{P}=1$. In the high-resolution models, the stiffness parameter is $\mathcal{P}<1$
    everywhere. \label{fig:2ddiff2plot}}
\end{figure*}

\subsubsection{Shadow casting problem}\label{sec:shad-cast-probl}

We begin our presentation of 2D test problems with a rather qualitative test that puts one of the
generic advantages of a two-moment scheme into focus, namely the ability of an opaque object to
generate a shadow when being illuminated by radiation. In a two-moment scheme, the flux-density
vector is an evolved quantity, and the pressure tensor is generally non-isotropic. In contrast, in
an FLD scheme (see Sec.~\ref{sec:flux-limit-diff}), the flux direction is determined by the gradient
of the scalar energy density, which corresponds to an isotropic radiation pressure. This leads to
the unphysical effect that in the free-streaming regime, $f\rightarrow 1$, anisotropic features of
the radiation field, such as shadows, cannot be maintained and are quickly washed out.

As it has likewise been done before for various other radiative transfer/transport codes \citep[see
e.g.][and references therein]{Audit2002,Hayes2003,Iliev2006,Skinner2013} we set up a purely
absorbing gas cloud which is exposed to near free-streaming radiation to test the ability of
radiation to cast a shadow behind the gas cloud. Specifically, in a Cartesian domain with
$\vecx\equiv(x,y)\in[0,15]\times[-5,5]$ and resolved by $N_x\times N_y=300\times200$ cells, we
define one region in which the radiation field is generated, the circular region $\mathcal{S}$
centered around $\vecx_{\mathcal{S}}=(3,0)$ with radius $r_{\mathcal{S}}=3/2$, and we define another
circular region $\mathcal{A}$ centered around $\vecx_{\mathcal{A}}=(11,0)$ with radius
$r_{\mathcal{A}}=2$ to be the purely absorbing cloud. The absorption opacity $\kappa_{\mathrm{a}}$
and equilibrium energy density $E^{\mathrm{eq}}$ are defined as follows:
\begin{subequations}
\begin{eqnarray}
  \kappa_{\mathrm{a}}(\vecx) &=&
  \begin{cases}
    10 \exp\{ - (4|\vecx-\vecx_{\mathcal{S}}|/r_{\mathcal{S}})^2 \} & , \, \vecx\in\mathcal{S} \\
    10                            & , \, \vecx\in\mathcal{A} \\
    0                             & , \, \mathrm{else}       \quad ,
  \end{cases}
  \\
  E^{\mathrm{eq}}(\vecx) &=&
  \begin{cases}
    10^{-1}                        & , \, \vecx\in\mathcal{S} \\
    0                             & , \, \mathrm{else}       \quad .
  \end{cases}
\end{eqnarray}
\end{subequations}
The model is initialized with vanishing flux densities and a homogeneous distribution of negligibly
small energy densities.

From the numerical point of view, the present objective is to test the correct implementation of the
multidimensional hyperbolic part of the radiation moment equations, particularly of the angular
dependence of the signal speeds in the Riemann solver, cf. Eqs.~\eqref{eq:sigspeeds}. The signal
speeds determine the numerical fluxes between grid cells, cf. Eq.~\eqref{eq:HLLhyp}, and close to
free-streaming conditions both the signal speeds as well as the intercell fluxes should be strongly
suppressed orthogonal to the direction of the radiation flux.

The three snapshots in Fig.~\ref{fig:shadow} show for three consecutive times contours of the
isotropic luminosity $L$ emitted by the source, given in this two-dimensional geometry by
$L=2\pi r_{\mathrm{c}}|\vecF|$ with $r_{\mathrm{c}}\equiv|\vecx-\vecx_{\mathcal{S}}|$. One can see
that a clearly obscured region behind the gas cloud emerges. The luminosity behind the gas cloud is
not an ideal step-function in vertical direction but it changes rather continuously within a fan of
opening angle $\approx 20^{\circ}-30^{\circ}$. The reasons for this are, first, that radiation is
not emitted from a point-like but a spatially extended source, causing the flux-factor to be
$|\vecF|/E \sim 0.98 < 1$ at $r_{\mathrm{c}}=8$, and second, that the gas cloud is not perfectly
absorbing but has a finite value of $\kappa_{\mathrm{a}}$, which allows a small fraction of
radiation to pass through the gas cloud near its edges. Altogether our code performs well in this
test, the development and propagation of the multidimensional radiation field and its particular
feature to cast a shadow are consistently captured.

\begin{figure*}
  \includegraphics[width=0.99\textwidth]{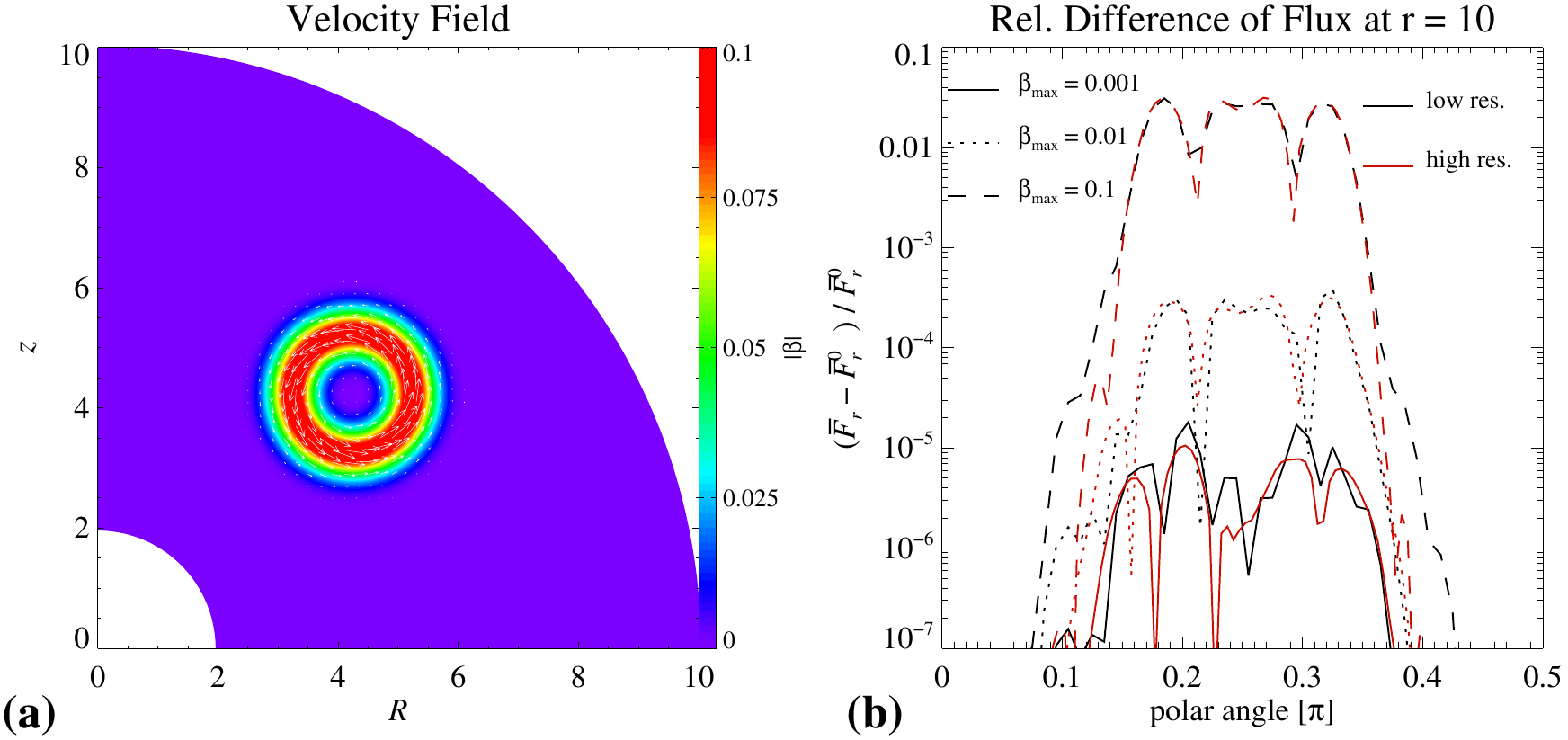}
  \caption{Radiation traversing variable velocity fields. In panel~(a) we show the absolute value
    (color coded) of the velocity and its polar component (arrows) for the high-velocity case with
    $\beta_{\mathrm{max}}=0.1$. In panel~(b) we plot the relative difference of the energy-integrated
    radial flux density, $\bar{F}_r$, with respect to the case of vanishing velocity, $\bar{F}_r^0$,
    measured at radius $r=10$ as function of polar angle.}\label{fig:2dvelplot}
\end{figure*}

\subsubsection{Static and dynamic diffusion}\label{sec:stat-dynam-diff}

A standard test for radiation codes allowing for the treatment of optically thick regions
\citep[e.g.][]{Gonz'alez2007,Swesty2009} is the scenario of an initially concentrated bulge of
radiation diffusing into its environment. Being conceptually based on the diffusion limit, an FLD
scheme usually performs well in this test, as long as the medium is sufficiently opaque. On the
other hand, in a two-moment AEF scheme the diffusion equation only results in the parabolic limit of
the otherwise hyperbolic equations. Therefore, it has to be checked that the numerical method chosen
to solve the AEF scheme consistently describes the parabolic diffusion limit and that the transition
to the hyperbolic regime proceeds in a numerically stable and accurate fashion
(cf. Secs.~\ref{ssec:hyperbolic} and~\ref{ssec:physsource}). Concerning the last point, we
explicitly want to ensure that no spurious, resolution-dependent features result from modifying the
numerical fluxes whenever the local stiffness parameter exceeds unity,
cf. Eq.~\eqref{eq:pecflux}. Finally, in this setup we also want to test the ability of the algorithm
to accurately describe dynamic diffusion, i.e. diffusion out of a moving medium.

We perform a set of calculations in a Cartesian box given by
$\vecx \equiv(x,y) \in[0,3]\times[0,2]$. All configurations are initialized at a fiducial time
$t_0=5$ with the following Gaussian pulse of radiation energy density and corresponding diffusive
flux density centered around $\vecx_0=(1,1)$:
\begin{equation}
  \label{eq:gaussdiffe}
  E(\vecx,t_0) = E_0 \exp\left\{ -\frac{|\vecx-\vecx_0|^2}{4D_0} \right\} \, , \quad
  \vecF(\vecx,t_0) = - D_0 \mathbf{\nabla} E \, ,
\end{equation}
where $E_0=1$ and $D_0=3\times10^{-3}$.

In the first configuration we define a spatially constant diffusion coefficient,
$D\equiv(3\kappa_{\mathrm{s}})^{-1} = D_0$, which allows us to compare the numerical results with an
analytic solution, given by:
\begin{equation}
  \label{eq:gaussdifff}
  E(\vecx,t) = E_0 \frac{t_0}{t_0+t}\exp\left\{ -\frac{|\vecx-\vecx_0|^2}{4D_0(t_0+t)} \right\} \, .
\end{equation}
For this configuration we switch between two resolutions $\{N_x,N_y\}=\{450,300\}$ and $\{150,100\}$
corresponding to which the stiffness parameters, $\mathcal{P}= \kappa_{\mathrm{s}}\Delta x$, are lower
and greater than 1, respectively. Additionally, for both resolutions we vary between a vanishing and
a non-vanishing spatially constant velocity in $x$-direction, i.e. $\beta=0$ and $0.1$,
respectively. For comparison with the $\beta=0$ models, we plot the radiation energy densities of
the $\beta>0$ models in the coordinate frame of the $\beta=0$ models.

In panel~(a) of Fig.~\ref{fig:2ddiff2plot}, we see that after a simulation time of $t-t_0=5$, when
the maximum value of $E$ has reached about half of its initial value $E_0=1$, all models agree well
with the analytic solution and we cannot identify unphysical numerical features in any of the
different cases.

We consider a second configuration with the same initial conditions, cf. Eqs.~\eqref{eq:gaussdiffe},
to test the correct transition from a stiff ($\mathcal{P}>1$) to a non-stiff ($\mathcal{P}<1$)
regime. To this end, instead of a constant opacity we now use a variable opacity that declines with
distance from the center as
\begin{equation}
  \label{eq:gausskappa}
  \kappa_{\mathrm{s}}(\vecx) = \frac{1}{3D_0} \exp\left\{ -\frac{|\vecx-\vecx_0|^2}{\delta^2} \right\}
\end{equation}
on a length scale of $\delta=0.4$. We display the results at the simulation time $t-t_0=5$ in
panel~(b) of Fig.~\ref{fig:2ddiff2plot}. We again compare four cases with two resolutions,
$\{N_x,N_y\}=\{450,300\}$ and $\{225,150\}$, and two homogeneous velocities in $x$-direction
$\beta=0$ and $0.1$. For the models with $\beta>0$, the opacity profile is advected with the fluid,
i.e. the opacity $\kappa_{\mathrm{s}}^\beta(\vecx,t) \equiv \kappa_{\mathrm{s}}(\vecx-\beta t)$ is
employed. In the high-resolution cases, which serve as references for the low-resolution cases, the
stiffness parameter is $\mathcal{P}<1$ everywhere, while in the low-resolution cases the stiffness
parameter crosses $\mathcal{P}=1$ at the locations indicated by the dotted lines. We observe no
numerical artifacts near the transition where $\mathcal{P}=1$ in the low-resolution cases, which
indicates that the modification of the numerical fluxes, Eq.~\eqref{eq:pecflux}, works accurately.

\subsubsection{Radiation traversing variable velocity fields}\label{sec:cons-comov-frame}

Since the quantities $E, \vecF$ evolved in our two-moment formulation are defined in the comoving
frame, they are subject to variations whenever radiation crosses regions of variable velocity even
without any interactions present. The net impact on the radiation properties after passing such
regions and returning back into the original frame would vanish in an exact calculation. In
practice, however, we encounter two limitations that spoil this feature to be fulfilled precisely:
First, our underlying scheme for the radiation moments neglects all contributions of order
$\mathcal{O}(v^2/c^2)$ in both evolution equations, which results in a loss of the property that a
transformation from one frame to another is exactly reversible -- instead such a transformation
generates errors of the disregarded order $\mathcal{O}(v^2/c^2)$. The second reason is that we do
not solve the evolution equations exactly but only numerically, i.e. all our solutions are beset
with truncation errors depending on the spatial and temporal resolution and on the numerical
algorithm.

In order to obtain a qualitative impression of how strongly both aforementioned effects can impact
the solution, we set up an arbitarily shaped velocity field and let it be traversed by a spherically
expanding radiation field. The radiation field is induced at the boundary at radius $r=2$ with an
energy density of $E(\eps)=\eps^3/(\eul^\eps-1)$, a radial flux density of $F_r(\eps)=E(\eps)/2$,
and vanishing non-radial flux components, $F_\theta=F_\phi=0$. The energy grid consists of
$N_\eps=10$ bins logarithmically distributed between $\eps\in[0,30]$ with an enlargement factor
$\Delta\eps_{\xi+1}/\Delta \eps_{\xi}=1.3$.  The velocity field in the polar plane,
$\vecv_{\mathrm{pol}}$, represents an eddy with radius $d_1=1$ circulating around its center at
$\vecx_0=(5,5)/\sqrt{2}$, while the toroidal velocity field $\vecv_{\mathrm{tor}}$ has the same
absolute magnitude as the poloidal field but points into the $\phi$-direction. Explicitly, we set
\begin{equation}
  \vecv_{\mathrm{pol}} = \frac{\beta(\vecx)}{\sqrt{2}} \mathbf{e}_{\mathrm{pol}} \quad , \quad
  \vecv_{\mathrm{tor}} = \frac{\beta(\vecx)}{\sqrt{2}} \mathbf{e}_{\phi} \, ,
\end{equation}
with
\begin{equation}
  \beta(\vecx) = \beta_{\mathrm{max}}
    \exp\left\{ - (|\vecx-\vecx_0|-d_1)^2/d_2^2 \right\} \, ,
\end{equation}
where $d_2=0.4$, $\mathbf{e}_{\phi}$ is the unit vector in $\phi$-direction, and
$\mathbf{e}_{\mathrm{pol}}$ is the unit vector perpendicular to both $\vecx-\vecx_0$ and
$\mathbf{e}_{\phi}$ and signed such to point in clockwise direction in the $R-z$ plane (see
panel~(a) of Fig.~\ref{fig:2dvelplot}). We vary the maximum value of the velocity between
$\beta_{\mathrm{max}}\in\{10^{-3},10^{-2},10^{-1}\}$ and we use the two spatial resolutions
$N_{r}=N_\theta\in\{50,100\}$ between $r\in[2,10]$ and $\theta\in[0,\pi/2]$. At $r=10$ we employ an
outflow BC.

In panel~(b) of Fig.~\ref{fig:2dvelplot} we compare the energy-integrated fluxes $\bar{F}_r$
obtained for each velocity field and resolution with the corresponding value $\bar{F}^0_r$ resulting
for a vanishing velocity field. If the two shortcomings mentioned in the beginning of this section
were absent, both fluxes would be exactly equal, i.e. $(\bar{F}_r-\bar{F}_r^0)/\bar{F}^0_r=0$.
Instead, in our numerical calculation we receive relative errors of up to
$\left.(\bar{F}_r-\bar{F}^0_r) / \bar{F}^0_r\right|_{\mathrm{max}} \sim
\{4\times10^{-2},4\times10^{-4}, 1\times10^{-5}\}$
for $\beta_{\mathrm{max}}=\{10^{-3},10^{-2},10^{-1}\}$. The outcome that the lines lie much closer
together for different resolutions than for different $\beta_{\mathrm{max}}$ shows that the low
resolution already sufficiently resolves the solution corresponding to the underlying moment
equations, at least for $\beta_{\mathrm{max}}>10^{-3}$. It can further be observed that the
leading-order error term representing missing components compared to the fully relativistic
formulation is not $\mathcal{O}(v/c)$, as can be inferred from the tendency of relative differences
to roughly decrease by two orders of magnitude for a one-order reduction of
$\beta_{\mathrm{max}}$. Hence, we deduce that in our implementation no significant contributions of
order $\mathcal{O}(v/c)$ are missing or are erroneously present since in any other case we would
have found an error scaling linearly with $\beta_{\mathrm{max}}$.

\subsection{One-dimensional problems including microphysics}\label{sec:test-probl-incl}

While the previous test problems are based on rather idealized setups, the two remaining
one-dimensional test problems specifically focus on neutrino transport in CCSNe. In particular, the
tests should address the question how the AEF scheme performs compared to the present standard
methods, the FLD and the Boltzmann-type solvers. To this end, in the first test in
Sec.~\ref{sec:neutr-radi-field} we keep the hydrodynamic background -- consisting of a typical
proto-NS configuration -- fixed and we only compare the stationary radiation field resulting from
the three different aforementioned types of methods. In the second test in
Sec.~\ref{sec:fully-dynam-evol} we then compare the results of a fully dynamic CCSN simulation with
two similar calculations performed with well-known Boltzmann-type neutrino-hydrodynamics codes.

\begin{figure*}
  \includegraphics[width=0.49\textwidth]{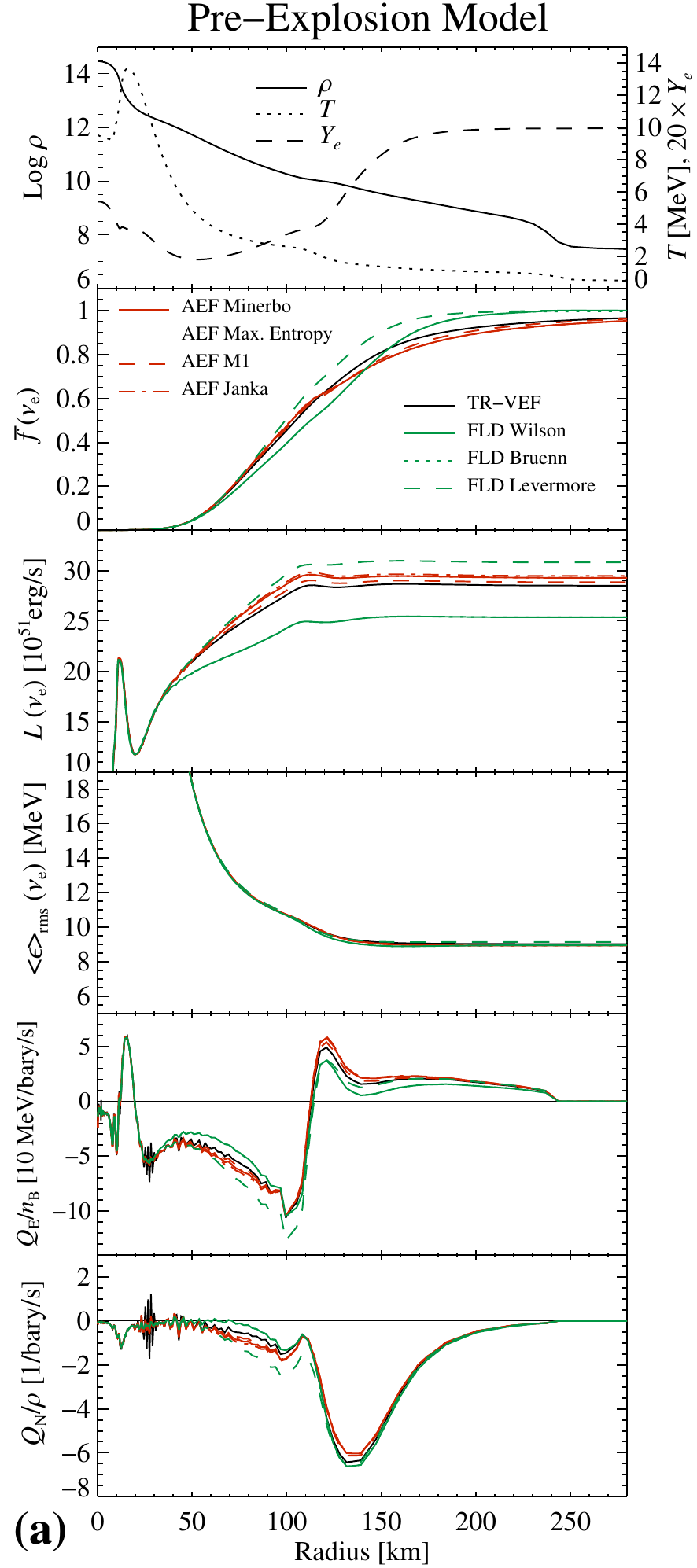}
  \includegraphics[width=0.49\textwidth]{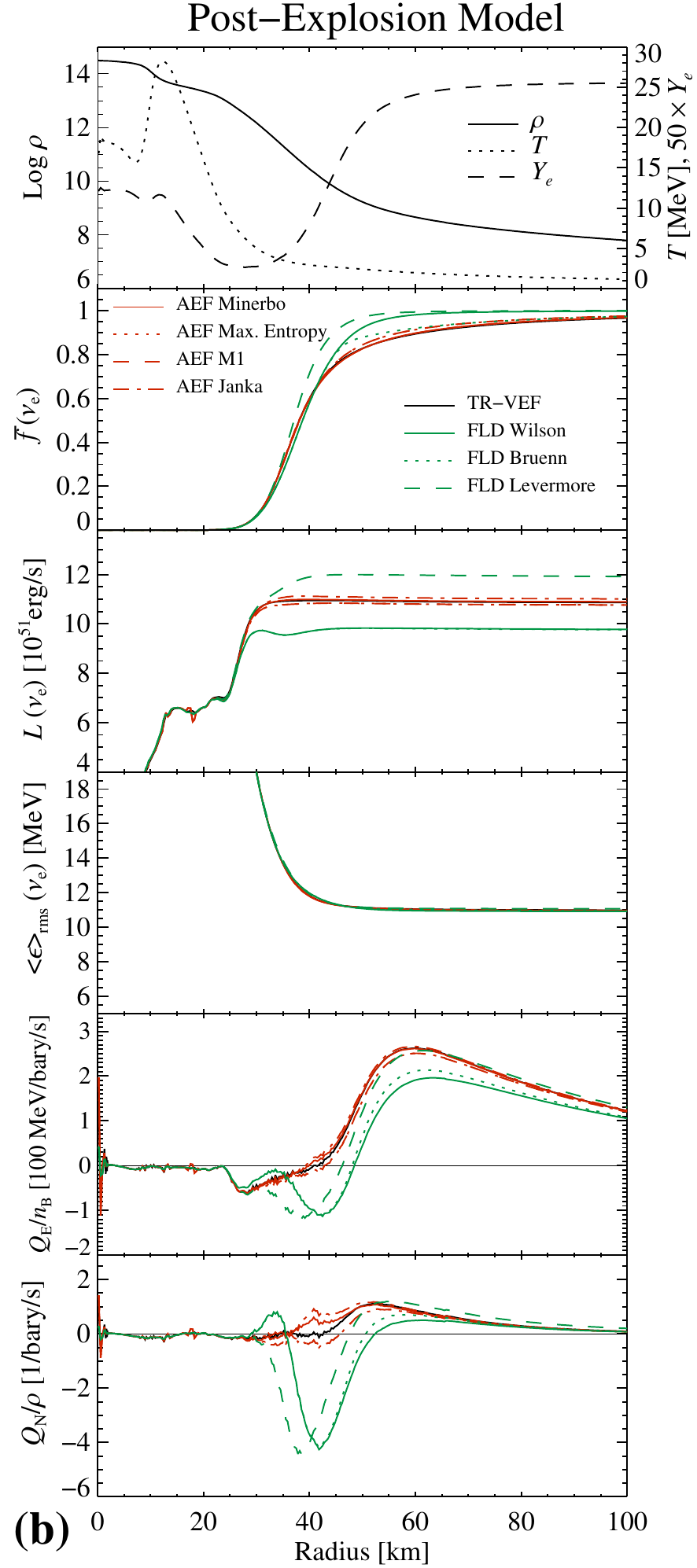}
  \caption{Neutrino radiation field of a static proto-neutron star. We compare for two models (see top
    title) between different transport schemes (see line labels in the second panel from the top). In
    the panels are displayed from top to bottom the properties (density $\rho$, temperature $T$ and
    electron fraction $Y_e$) of the hydrodynamic background, the mean flux-factor
    $\bar{f}\equiv \bar{F}/(\cl \bar{E})$, luminosity $L\equiv 4\pi r^2 \bar{F}$ and rms-energy
    $\langle\epsilon\rangle_{\mathrm{rms}} \equiv \sqrt{\int \eps E(\eps) \mathrm{d}\eps / \int
      N(\eps)\mathrm{d}\eps}$
    of electron neutrinos, and the source terms $Q_{\mathrm{E}}/n_{\mathrm{B}}, Q_{\mathrm{N}}/\rho$
    for the gas energy density (cf. Eq.~\eqref{eq:einsource}) and electron-number density
    (cf. Eq.~\eqref{eq:yesource}), respectively. Note that in cases when the dotted green line is
    invisible it is superimposed by the solid green line.}\label{fig:pnscomp1}
\end{figure*}

\subsubsection{Neutrino radiation field of a static proto-neutron star}\label{sec:neutr-radi-field}

To compare the different neutrino transport schemes AEF, FLD and Boltzmann with each other in the
CCSN context in a manner that is independent of the hydrodynamics part of the numerical method, it
is instructive to evolve only the radiation field in a proto-NS background that is held constant
during the evolution. As background configurations, we take two profiles of hydrodynamic data
obtained from two different simulations performed with the VERTEX code. The hydrodynamic data for
our first model (called ``pre-explosion model'' hereafter) is represented by a snapshot taken at
time $150\,$ms after bounce in the accretion phase of model ``N13'', which was investigated in the
course of \citet{Liebendorfer2005}\footnote{See Sec.~\ref{sec:fully-dynam-evol} for more details
  about this model, which also served as reference model in the fully dynamic CCSN simulation
  presented in that section.}. The snapshot for our second model (called ``post-explosion model''
hereafter) is taken from the model ``Sr'' in \citet{Hudepohl2010}\footnote{The data was provided via
  \\\textsf{ http://www.mpa-garching.mpg.de/ccsnarchive/ .}} in the neutrino-driven wind phase at
time $300\,$ms after core bounce. In both models we adopt the spatial and energy grids and the
profiles of the density $\rho$, temperature $T$ and electron fraction $Y_e$ from the reference
calculations, while the remaining quantities needed for the opacities have been obtained by applying
the EOS of \cite{Lattimer1991}. Note that since the data in the pre-explosion model is partially
rather noisy we smoothed the original profiles of $\hat{\rho}^0,\hat{T}^0,\hat{Y}_e^0$ as given by
the reference calculation by making the replacement
$\hat{u}_\hi = ( \hat{u}^0_{\hi-1} + 2\hat{u}^0_{\hi} + \hat{u}^0_{\hi+1}) / 4$ for
$\hat{u}\in\{ \hat{\rho}, \hat{T}, \hat{Y}_e \}$ at each radial grid point $\hi$ in the
pre-explosion model. In both models, we ignore frame-dependent effects by setting the velocity
$\vecv = 0$ everywhere. Regarding the neutrino interactions, we only take into account the nucleonic
$\beta$-processes ($n+e^+\leftrightharpoons p+\bar\nu_e$ and $p+e^-\leftrightharpoons n+\nu_e$) and
isoenergetic scattering of (anti-)neutrinos off free nucleons as described in
\citet{Bruenn1985}. Only eletron-type neutrinos are evolved.

For comparison of the AEF scheme with the two remaining schemes, we additionally implemented an FLD
scheme and a Boltzmann solver. The FLD scheme was obtained by modifying our existing time-dependent
AEF scheme while for the Boltzmann-type solution we implemented a separate algorithm in which the
time-independent two-moment system is solved with the closure being provided by a tangent-ray
scheme. The details about both methods can be found in Appendix~\ref{sec:append-a:-numer-1}.

For each of the two background models eight calculations using different methods or closures were
performed: The reference solution is represented by the calculation conducted with the
Boltzmann-type tangent-ray variable-Eddington-factor scheme (TR-VEF), since this method should yield
the most accurate solution. In four additional simulations AEF schemes together with the Eddington
factors in Eqs.~\eqref{eq:closures} were employed, while in the remaining three simulations FLD
schemes together with the flux-limiters in Eqs.~\eqref{eq:fluxlimiters} were applied. For computing
the flux-limiter $\Lambda_{\mathrm{Bruenn}}$, cf. Eq.~\eqref{eq:bruennlimiter}, the
(energy-dependent) neutrinosphere radii $r_\nu$, defined by
\begin{equation}
  \label{eq:optdepth}
  \tau_\nu(r_\nu,\eps)\equiv \int_{r_\nu}^\infty \kappa_{\mathrm{tra}}(r',\eps) \mathrm{d}r' = 1 \, ,
\end{equation}
were used.

In Fig.~\ref{fig:pnscomp1}, we compare for the different calculation methods the radial profiles of
the mean flux-factor, luminosity, and rms-energy of electron neutrinos, as well as the hydrodynamic
source terms corresponding to heating/cooling and (de-)leptonization resulting from the interaction
with both electron neutrinos. We observe the following properties: (1) Concerning the luminosities
and flux-factors, the accuracy of the AEF schemes is throughout slightly higher than that of the FLD
schemes. While the FLD luminosities and flux factors exhibit errors up to $\sim 10\,\%$, the
corresponding AEF quantities are throughout accurate to less than $\sim 3\,\%$. (2) The rms-energies
are accurate to $\la 1\,\%$ in all schemes. (3) The accuracy to which the hydrodynamic source terms
are reproduced is comparable for AEF and FLD in the pre-explosion model, with the heating rates in
the gain region being somewhat underestimated for FLD and overestimated for AEF. In contrast, in the
post-explosion model the AEF schemes perform clearly better than the FLD schemes. The main
difference between both background models is that the semi-transparent region is more extended in
the pre-explosion model. (4) The overall results appear to be more sensitive to different
flux-limiters when using FLD than to different Eddington factors when using AEF.

The results of this comparison indicate that AEF methods can perform at least equal to, if not
better than, FLD schemes in a 1D proto-NS environment. However, since we have only investigated two
stationary snapshots we cannot be sure about the universality of the observed levels of accuracy
with respect to time- and model variations. For this reason we avoid at this point to speculate
about the hydrodynamic impact of the observed differences in the heating- and deleptonization-rates
for the various methods. Nevertheless, the comparison test discussed in the next section
demonstrates that AEF schemes can in fact compete with Boltzmann solvers even in the fully dynamic
case.

\begin{figure*}
  \centering
  \includegraphics[width=0.99\textwidth]{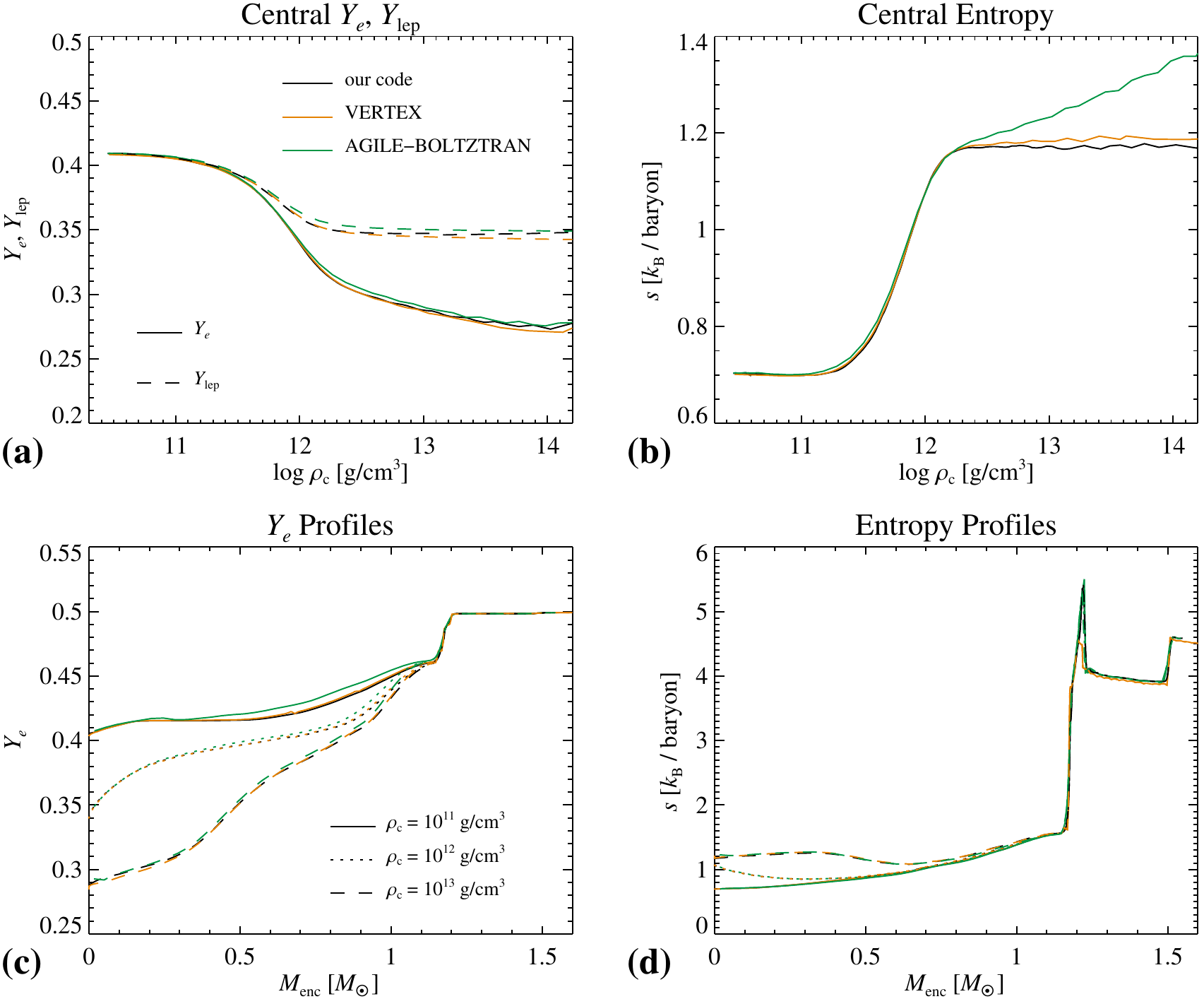}
  \caption{
    Collapse dynamics in the core-collapse test. All black, orange and green lines display results
    obtained with our code, VERTEX and AGILE-BOLTZTRAN, respectively.  The panels~(a) and (b) show
    the evolution of the central electron (solid lines) and lepton fraction (dashed lines) and the
    central entropy, respectively, as functions of the central density during collapse. The increase
    of the central entropy for central densities $\rho_{\mathrm{c}}\ga 10^{12}\,$g\,cm$^{-3}$ in the
    AGILE-BOLTZTRAN run is a numerical artifact which has no impact on the subsequent physical
    evolution (see \citealp{Liebendorfer2005} for more details). The panels~(c) and (d) display
    profiles of the electron fraction and the entropy as functions of the enclosed mass at times
    when the core reaches the central densities $\rho_{\mathrm{c}}$ shown in the legend in
    panel~(c).}
  \label{Fig:SN-collapse}
\end{figure*}

\begin{figure*}
  \centering
  \includegraphics[width=0.99\textwidth]{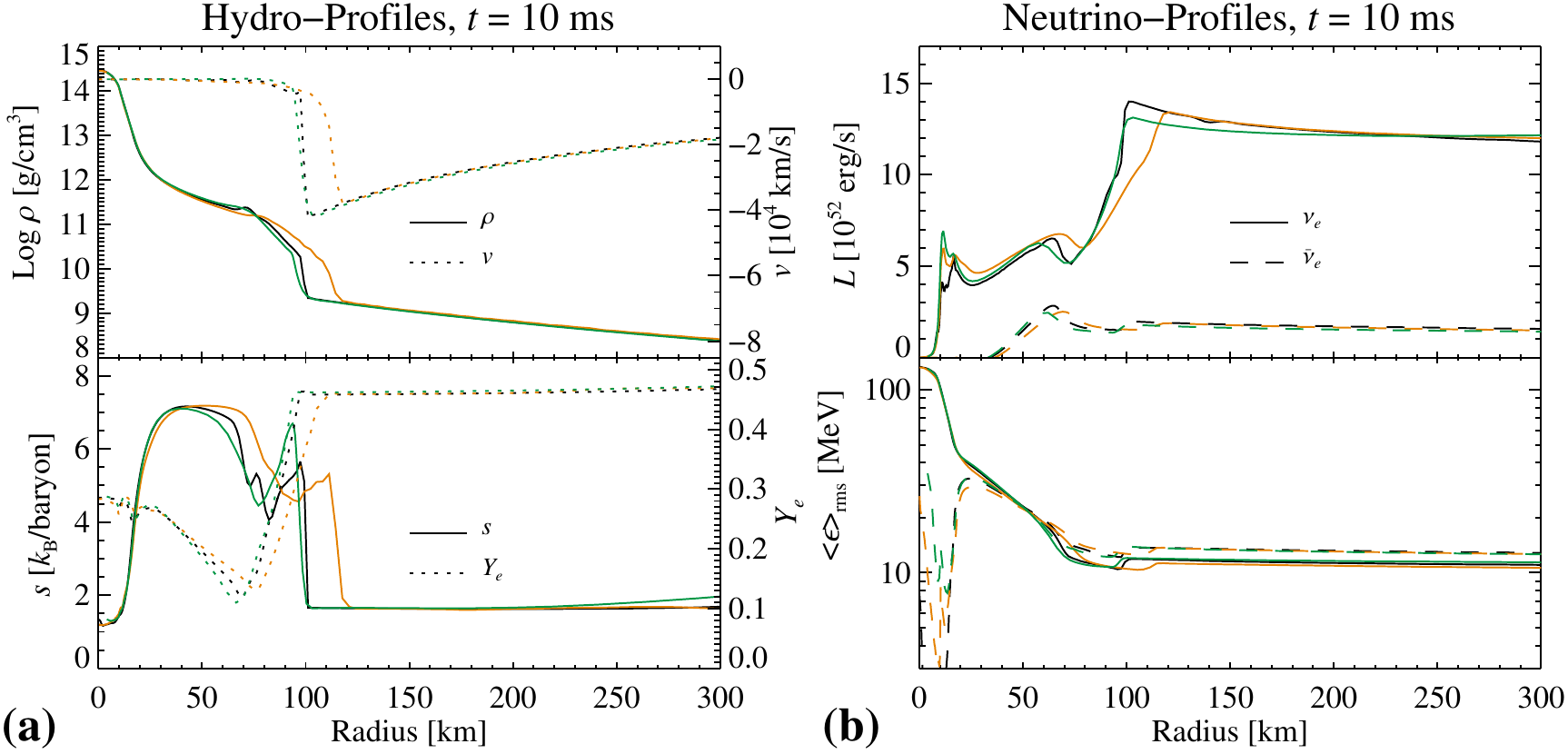}
  \includegraphics[width=0.99\textwidth]{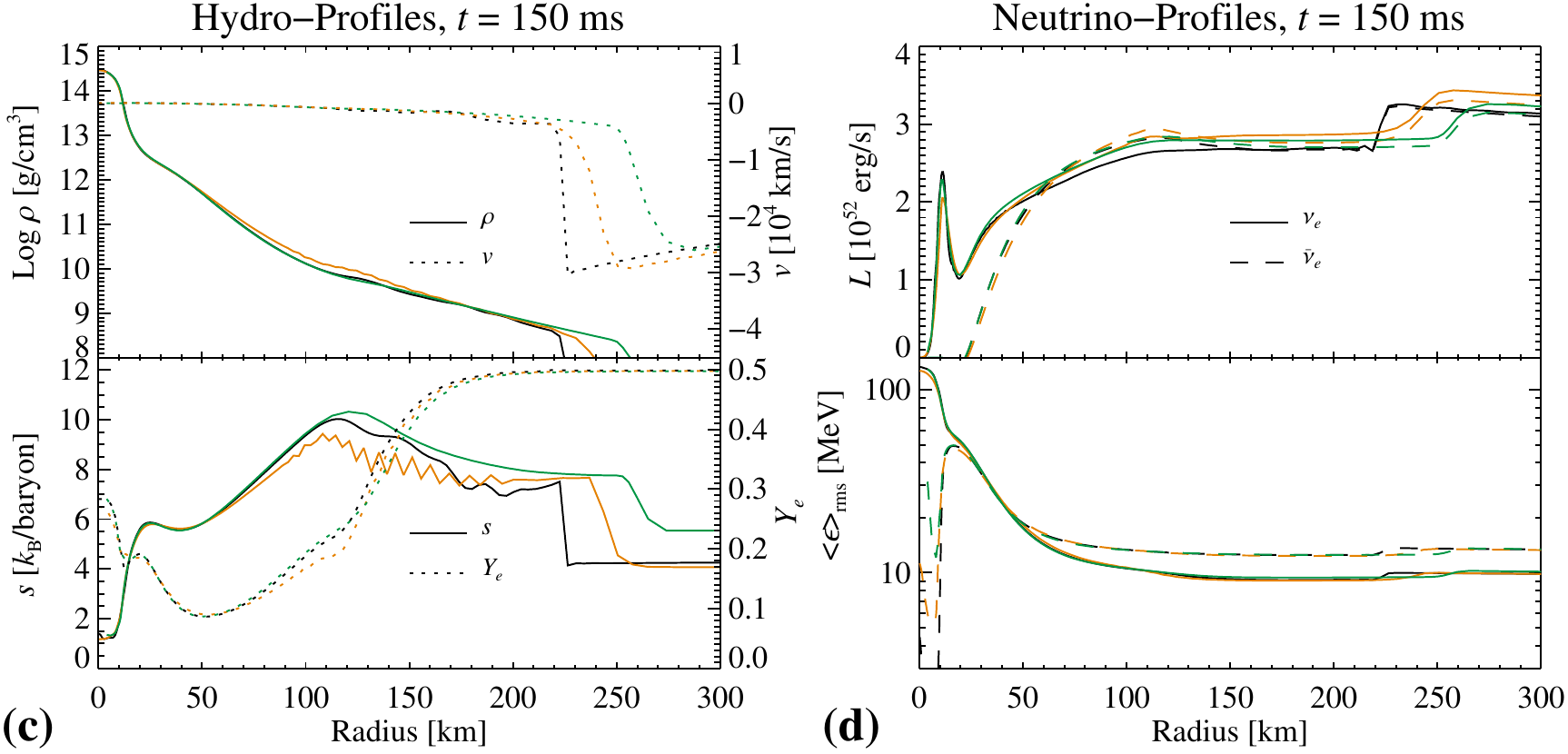}
  \caption{
    Radial profiles in the post-bounce phase of the core-collapse test. All black, orange and green
    lines display results obtained with our code, VERTEX and AGILE-BOLTZTRAN, respectively. In
    panel~(a) we show the density and velocity (top), and the entropy per baryon and electron
    fraction (bottom), while in panel~(b) we plot the luminosities (top) and rms-energies (bottom)
    at a post-bounce time of $t=10\,$ms. In panels~(c),(d) the same respective quantities as in
    panels~(a),(b) are plotted, but at a post-bounce time of $t=150\,$ms.}
  \label{Fig:postbounce_prof}
\end{figure*}

\begin{figure*}
  \centering
  \includegraphics[width=0.99\textwidth]{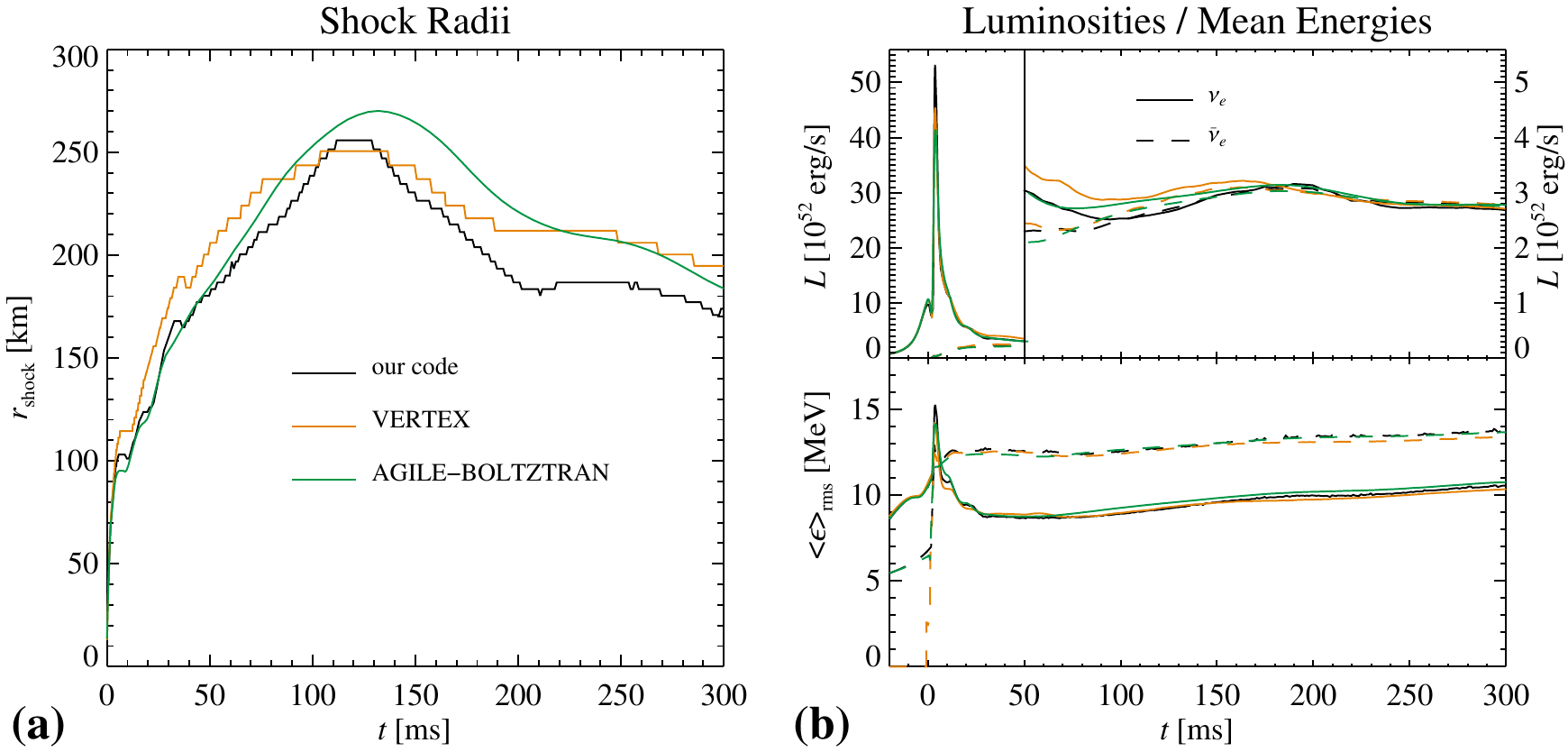}
  \caption{
    Post-bounce evolution in the core-collapse test. All black, orange and green lines display
    results obtained with our code, VERTEX and AGILE-BOLTZTRAN, respectively.  In panel~(a) we show
    the shock radii as functions of the post-bounce time. The top and bottom of panel~(b) depict the
    luminosities and mean energies, respectively, of electron neutrinos (solid lines) and electron
    antineutrinos (dashed lines). Note that for the luminosities the scaling on the left $y$-axis
    applies for $t<50\,$ms, while the scaling on the right $y$-axis applies for $t>50\,$ms. }
  \label{Fig:postbounce_time}
\end{figure*}

\subsubsection{Fully dynamic collapse and post-bounce evolution of a $13\,M_\odot$
  progenitor}\label{sec:fully-dynam-evol}

Our final and most comprehensive test comprises the self-consistent collapse and subsequent
post-bounce evolution of the core of a star with approximate main-sequence mass of $13\,\Msol$
\citep{Nomoto1988}. This model was already investigated in the comparison in
\citet{Liebendorfer2005} (labeled ``N13'' there) between the neutrino-hydrodynamics codes
AGILE-BOLTZTRAN \citep{Liebendorfer2004} and VERTEX-PROMETHEUS \citep{Rampp2002}. The comparison
revealed a good agreement between the two Boltzmann-type codes, which allows for a straightforward
assessment of the accuracy of our method by direct comparison with the results of
\citet{Liebendorfer2005}\footnote{The results discussed in \citet{Liebendorfer2005} and used here as
  reference data are publically available under
  \textsf{http://iopscience.iop.org/0004-637X/620/2/840/fulltext/datafiles.tar.gz}}. Note in passing
that \citet{OConnor2014} recently conducted a similar comparison but using the AEF-type code GR1D
and inspecting model ``G15'' from \citet{Liebendorfer2005}. The main finding of \citet{OConnor2014},
namely the very good agreement of the AEF scheme with the Boltzmann codes, is in consensus with
ours. Finally, using additional, slightly modified simulation setups we will check the robustness of
our results with regard to the numerical scheme and the choice of the closure prescription.

\paragraph{Model setup}

We start the spherically symmetric evolution at the onset of collapse when the core has a central
density of $\rho_{\mathrm{c}} = 3.16 \times 10^{10} \, \gccm$ and we follow the system up to a
post-bounce time of $t = 300 \, \ms$. This test involves all types of neutrino interactions listed
in Sec.~\ref{ssec:sourcesandhydro}. Thus, except for our omission of the processes
$\nu_e\bar\nu_e\leftrightharpoons e^-e^+$ \footnote{However, these reactions are subdominant in both
  the collapse and post-bounce evolution by at least an order of magnitude compared to the dominant
  reactions, see e.g. \citealp{Buras2006}.}, we employ the same neutrino microphysics as in
\citet{Liebendorfer2005}. Above a threshold density of
$\rho_{\mathrm{LS}} = 6 \times 10^{7} \, \gccm$ the stellar gas is modeled by the nuclear EOS of
\cite{Lattimer1991}, with a compressibility modulus of $K = 180$\,MeV. Below this threshold, we use
a low-density EOS that includes photons, arbitrarily relativistic and arbitrarily degenerate
electrons and positrons, and a non-relativistic Boltzmann gas of baryons. For
$\rho>\rho_{\mathrm{LS}}$ the nuclear composition is in nuclear statistical equilibrium, which only
depends on $\rho, T$ and $Y_e$. For $\rho<\rho_{\mathrm{LS}}$ we assume the baryonic composition to
be given by pure $^{28}\mathrm{Si}$ for temperatures $T<0.44\,$MeV, or by pure $^{56}\mathrm{Ni}$
for temperatures $T>0.44\,$MeV.

We employ a Eulerian radial grid with $N_r = 384$ zones distributed logarithmically between the
origin and an outer radius of $\approx 7,000 \, \mathrm{km}$ with a minimum grid width of
$200 \, \mathrm{m}$. The neutrino energy space is discretized into 21 energy bins that are roughly
logarithmically distributed between 0\,MeV and 400\,MeV. For the initialization of the model we take
the profiles of velocity, density, electron fraction and entropy from the corresponding
AGILE-BOLTZTRAN run. For the time integration of the source terms we ignore the implicit time
dependence of the hydrodynamic quantities. Specifically, we use the integration modes~\emph{c)}
and~\emph{b)} described in Sec.~\ref{ssec:physsource} for densities lower and higher than
$10^{11}\,$g\,cm$^{-3}$, respectively. With the CFL factor set to 0.7, our whole simulation required
the calculation of about $700,000$ time steps.

\paragraph{Collapse dynamics} 

The core collapses in a time of $t_{\mathrm{coll}} \simeq 95\, \mathrm{ms}$ to reach a maximum
density of $\rho_{\mathrm{max}} = 3.6 \times 10^{14} \, \gccm$. We show the evolution of the central
values of the electron fraction $Y_{e}$, lepton fraction
$Y_{\mathrm{lep}}\equiv (n_{e^-} - n_{e^+} + \bar{N}_{\nu_e} - \bar{N}_{\bar\nu_e}) /
n_{\mathrm{B}}$
and entropy per baryon $s$ as functions of the central density $\rho_{\mathrm{c}}$ during collapse
in the top panels and the profiles of $Y_e$ and $s$ as functions of the enclosed mass coordinate
$M_{\mathrm{enc}}(r)\equiv 4\pi \int_0^r\rho\tilde{r}^2\;\mathrm{d}\tilde{r}$ in the bottom panels
of Fig.~\ref{Fig:SN-collapse}, respectively. Starting at central densities of about
$\rho_{\mathrm{c}} \sim 10^{11} \, \gccm$, inelastic electron scattering reduces the mean energy of
the neutrinos escaping from the core, leading to an accelerated deleptonization and to an increase
of the central entropy of the gas. Neutrino trapping sets in at a central density of
$\rho_{\mathrm{c}} \approx 2 \times 10^{12} \, \gccm$, above which the central values of
$Y_{\mathrm{lep}}$ and $s$ remain roughly constant until core bounce. In all variables, we find a
nearly perfect agreement of our results with both reference solutions.

\paragraph{Post-bounce evolution} 

Once the core reaches nuclear densities, a shock wave is formed at a mass coordinate of
$M_{\mathrm{sh}} \approx 0.67 \, \Msol$, consistent with the reference results. We show the radial
profiles of various quantities at two different post-bounce times in Fig.~\ref{Fig:postbounce_prof}
and the time dependence of the shock radius as well as the neutrino luminosities and rms-energies in
Fig.~\ref{Fig:postbounce_time}. All quantities are defined as in \cite{Liebendorfer2005}. About
$\approx 5$\,ms after bounce, the prompt shock reaches a radius of
$r_{\mathrm{shock}}\sim 100 \, \mathrm{km}$, where it stalls for almost 10\,ms. At this point the
slightly stronger prompt shock in VERTEX leads to somewhat larger shock radii in the VERTEX runs
compared to our simulation and to the AGILE-BOLTZTRAN run. Subsequently, neutrino heating increases
the entropy of the matter in the gain region behind the shock and causes the shock to propagate
further outward. However, the energy deposition by neutrinos is not sufficient to drive an explosion
and we observe a continuous decrease of the shock radius after it has reached its maximum value of
$r_{\mathrm{shock}} \approx 250\, \mathrm{km}$ at $t \approx 120 \, \mathrm{ms}$. The shock
trajectory $r_{\mathrm{shock}}(t)$ (cf. panel~(a) of Fig.~\ref{Fig:postbounce_time}) of our
simulation is in good but not perfect agreement with the reference results which, however, are also
not exactly consistent with each other. The quantitative differences could to some degree stem from
slight differences of the thermodynamic treatment of the low-density regime
$\rho<\rho_{\mathrm{LS}}$ to which \citet{Liebendorfer2005} already attributed the discrepancies of
$r_{\mathrm{shock}}(t)$ between the two reference models: Differences in the treatment of nuclear
burning in the low-density regime lead to different entropies of the material that falls into the
shock and thereby to different post-shock entropies. Besides further potentially significant
differences in the hydrodynamic treatment of the three codes, the small underestimation of
$r_{\mathrm{shock}}(t)$ in our AEF run at late times could also be the result of the approximate
closure. However, even though the situation remains unclear, we do not consider this behavior to be
very significant, given the fact that even both Boltzmann codes exhibit differences in
$r_{\mathrm{shock}}(t)$ of roughly the same size.

\begin{figure*}
  \centering
  \includegraphics[width=0.99\textwidth]{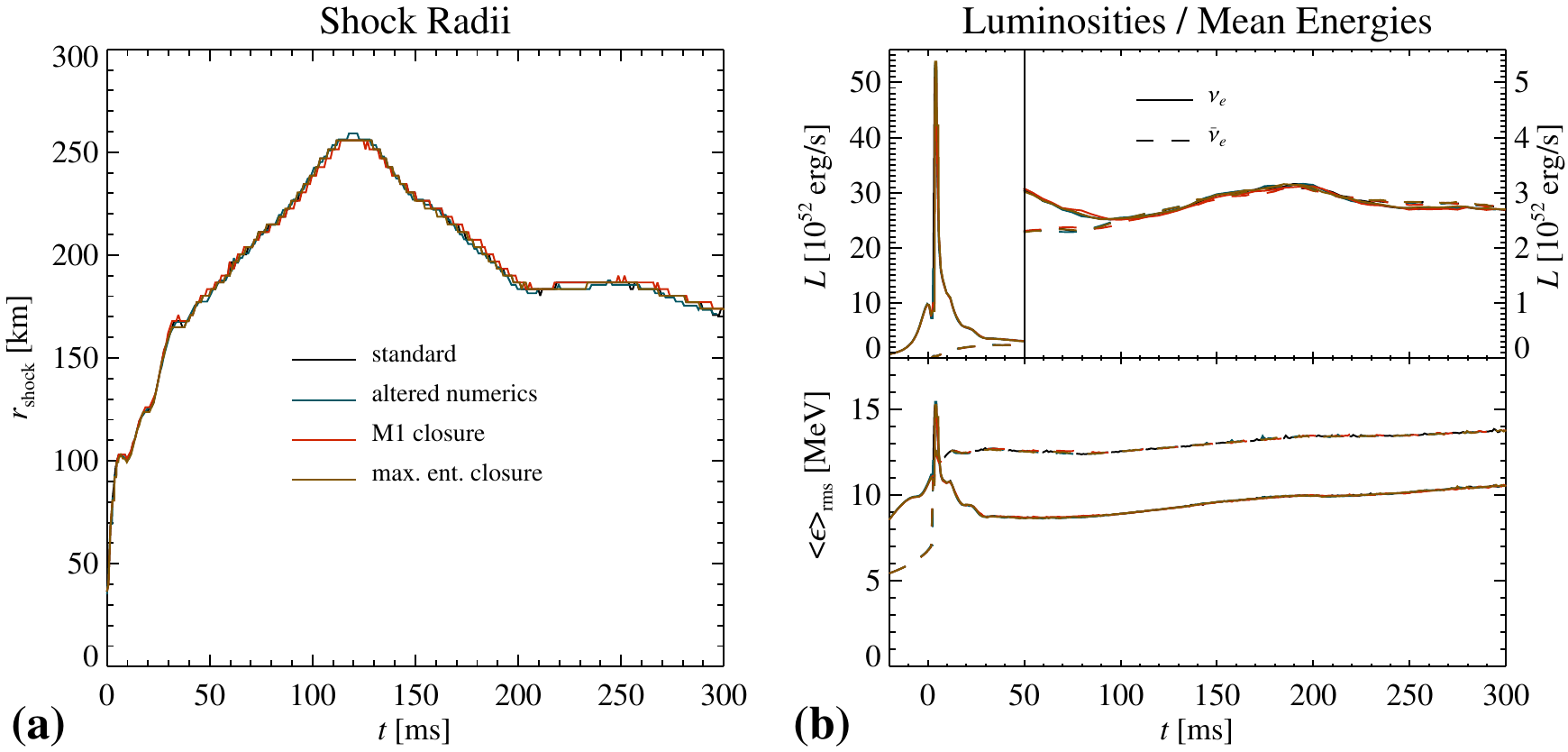}
  \caption{
    Same as Fig.~\ref{Fig:postbounce_time} but for the runs ``standard'', ``altered numerics'',
    ``$M_1$ closure'', and ``maximum entropy closure'' described in text. The maxima of the neutrino
    bursts in these models are $L_{\mathrm{max}}\approx \{5.3,5.3,5.1,5.4\}\times 10^{53}\,\ergs$,
    respectively.}
  \label{Fig:postbounce_comp}
\end{figure*}

\paragraph{Neutrino emission} 

At bounce, the core emits a short, intense burst of electron neutrinos. The neutrino burst in our
simulation exhibits a higher maximum luminosity of
$L_{\mathrm{max}} \approx 5.3 \times 10^{53} \, \ergs$ than in the two reference solutions
($L_{\mathrm{max}} \approx 4.14 \times 10^{53} \, \ergs$ and
$L_{\mathrm{max}} \approx 4.55 \times 10^{53} \, \ergs$ for AGILE-BOLTZTRAN and VERTEX,
respectively). However, the shorter duration of the burst in our model compensates for this, and the
integrated energy emitted during the burst,
$E_{\mathrm{burst}} \equiv \int_{0}^{0.02 \, \mathrm{s}} L_{\nu_e}(t) \, \mathrm{d} t$, is almost
equal for all three codes: $E_{\mathrm{burst}} = 2.80$, $2.80$, and $2.85 \times 10^{51} \erg$ for
our model, AGILE-BOLTZTRAN, and VERTEX, respectively. The reason for our neutrino burst being
sharper may be found in the numerical treatment: Our explicit time integration (at least for the
hyperbolic terms which describe the propagation of radiation) combined with a high-order spatial
discretization is certainly less dissipative than both fully-implicit reference schemes, which
facilitates the accurate evolution of narrow radiation peaks. Note that this explanation is
supported by the fact that \citet{OConnor2014} obtain a similarly enhanced neutrino burst using
their explicit AEF scheme; see their Fig.~6.

After the end of the neutrino burst, the luminosity of electron neutrinos drops quickly while the
luminosity of electron antineutrinos increases, and subsequently neutrinos of both flavors are
emitted at about equal luminosities.  After the mean energies of both flavors peak at the neutrino
burst they remain approximately constant with a slow trend towards higher values. Due to their
larger interaction rates with matter, electron neutrinos decouple at lower densities and
temperatures than electron antineutrinos, leading to a mean energy that is $\sim 3$\,MeV below that
of the antineutrinos.

In summary, most of our results lie well within the tolerance region spanned by the results of both
reference calculations, such that no obvious deficiencies of the AEF method can be identified.

\paragraph{Variations of the calculation method}

In this last study, we address various issues to check the robustness of the results presented above
and therefore the reliability of the whole algorithm described in this paper. To this end, we
perform additional test runs that are described below. In what follows, we denote the simulation
presented above as the ``standard'' run. In Fig.~\ref{Fig:postbounce_comp} we show similar
quantities as in Fig.~\ref{Fig:postbounce_time} but we compare the ``standard'' run with the test
runs described below.

We set up a simulation identical to the ``standard'' run described above but with altered numerical
specifications, called ``altered numerics''. The number of spatial grid points is increased to 576
(with an unchanged innermost grid width of 200\,m) and the time step is reduced according to
CFL=0.3. Moreover, the integration mode for the source terms~\emph{a)} is used instead of~\emph{b)}
for densities $\rho>10^{11}\,\gccm$ (cf. Sec.~\ref{ssec:physsource}). As a final difference to the
``standard'' run, we additionally take into account all the velocity-/acceleration- dependent terms
that have been dropped when deriving the moment equations, Eqs.~\eqref{eq:tmt1}, from the transfer
equation, Eq.~\eqref{eq:transfer1}, in Sec.~\ref{sec:equations}. For the numerical implementation of
these terms we adopt methods in close analogy to the existing ones,
cf. Sec.~\ref{sec:numerics}. This model yields very similar results compared to the standard case,
as is exemplarily shown for selected functions of time in Fig.~\ref{Fig:postbounce_comp}. This
similarity between the two models proves the robustness of our integration method regarding several
aspects, at least for physical conditions similar to the investigated case of a one-dimensional
CCSN: Besides convergence regarding the spatial resolution the test evinces that the mixed
explicit--implicit time-integration scheme does not produce time-step dependent numerical
artefacts. Furthermore, the test justifies the time-explicit treatment of the hydrodynamic
quantities in the source-term integration. Finally, the test verifies that the velocity-dependent
terms dropped in Eqs.~\eqref{eq:tmt1} are truly insignificant.

Two additional calculations with different closure prescriptions are conducted: In one run, ``$M_1$
closure'', the closure in Eq. \eqref{eq:m1closure} and in the other run, ``maximum entropy
closure'', the closure in Eq.~\eqref{eq:maxentclosure} is used to express the Eddington factor
$\chi$. For simplicity, we keep using the Minerbo prescription for the 3rd-moment factors $q$,
cf. Eq.~\eqref{eq:defqsmall}. Since the 3rd-moment terms are of minor relevance compared to the
terms including the 2nd moments (recall that the 3rd-moment terms vanish in the energy-integrated
1st-moment equation) this is a justified approximation to test the dominant impact of using a
different closure prescription. All three simulations that make use of different closures give
almost identical results, as is shown in Fig.~\ref{Fig:postbounce_comp} for selected
quantities. This important result is consistent with the findings in Sec.~\ref{sec:neutr-radi-field}
and it suggests that the AEF algorithm is sufficiently self-consistent when applied to CCSNe, in the
sense that the outcome of the calculation is rather insensitive to the precise shape of the closure.


\section{Summary}\label{sec:conc}

We presented the neutrino-transport code ALCAR that was developed to perform multidimensional
simulations of CCSNe and (different stages of) NS-mergers. The energy-dependent neutrino-transport
scheme is based on the multi-group evolution (with full energy-bin coupling) of the first two
angular moments of the specific intensity defined in the frame comoving with the fluid, and it takes
into account the dominant $\mathcal{O}(v/c)$ terms describing fluid-advection, aberration and
Doppler shift. The resulting system of equations for the neutrino energy density and the three
components of the flux density (cf. Eqs.~\eqref{eq:tmt1}) is closed by an approximate prescription
for the Eddington tensor, which assumes that the specific intensity is axisymmetric around the
direction of the flux-density vector and that the remaining single parameter $\chi$ is given by an
algebraic function of the evolved radiation moments (cf. Eqs.~\eqref{eq:closures}). Thereby, the
resulting AEF method circumvents the computationally demanding task to solve for the detailed
angular dependence of the specific intensity as it is done in Boltzmann-solvers. In contrast to the
standard FLD method, the AEF method consistently evolves the 1st moments and, hence, it is
potentially more accurate (see, e.g., the test in Sec.~\ref{sec:neutr-radi-field}), it allows
radiation anisotropies to be described (cf. Sec.~\ref{sec:shad-cast-probl}), and it ensures the
conservation of the total (radiation plus fluid) momentum and energy up to $\mathcal{O}(v/\cl)$
\citep{Baron1989, Cernohorsky1990}. Finally, a computationally relevant difference between the AEF
and FLD methods is that the time step in the case of a time-explicit advection scheme is required to
be considerably smaller in the FLD than in the AEF formulation, in practice forcing realizations of
the former method to employ fully implicit time-integration algorithms
(cf. Sec.~\ref{ssec:closures}).

Our numerical scheme essentially follows the ideas presented in \citet{Pons2000} and
\citet{Audit2002}, which have recently been implemented also in a number of photon-transport
\citep[e.g.][]{Hayes2003, Aubert2008, Scadowski2013, Skinner2013, McKinney2014} and
neutrino-transport \citep[][]{Shibata2012, Takahashi2013a, OConnor2014} codes. The basic strategy is
to utilize the hyperbolic nature of the moment equations to employ a Godunov-type scheme, in which
the advection fluxes between grid cells are given as solutions of Riemann problems. Thanks to a
neutrino-number conservative scheme developed by \citet{Muller2010} to handle the Doppler shift
terms we avoid the simultaneous evolution of the number-related moments together with the
energy-based moments. A distinctive feature of the presented scheme compared to existing FLD and
most Boltzmann-type neutrino-transport solvers is that all except the source terms are integrated
explicitly in time. Although in this case the time step resulting from the Courant condition is
comparable to the light-crossing times of single grid cells, it will usually only be marginally (or
at least not several orders of magnitude) lower than the already small fluid time steps in the hot
and dense physical systems this code is supposed to be applied to. To capture the transition to the
stiff parabolic limit of the moment equations in a numerically consistent and stable fashion, the
source terms are handled time-implicitly and the upwind-type Riemann solver switches to a
central-type solver in optically thick regions. Since the implicitly handled source terms are only
functions of the local neutrino-gas properties, a computationally convenient feature of this
explicit--implicit integration scheme is that the method can be parallelized with high efficiency.

We conducted a series of tests to assess the quality of the AEF method and to check the correct
implementation of the velocity-dependent terms, the source terms and the coupling to the
hydrodynamics part of the code. By means of one- and two-dimensional test problems it was shown that
the AEF method allows for a stable and self-consistent evolution of the radiation field in the full
range between isotropic diffusion and free-streaming, including the accurate description of
frame-dependent effects such as Doppler shift and diffusion in static and moving media. Although
this was done here in two dimensions, the code readily generalizes to three dimensions.

Two additional tests specifically focused on (one-dimensional) neutrino transport in CCSNe. In the
first test the hydrodynamic background, consisting of a proto-NS configuration, was held fixed to
compare the neutrino fields resulting from an AEF scheme with different Eddington factors with the
outcomes of an FLD scheme with different flux-limiters and of a more accurate Boltzmann scheme. The
essential findings were that the AEF solvers reproduced the results of the Boltzmann solver slightly
more accurately than the FLD scheme and that using different closure prescriptions has less impact
on the solution in an AEF scheme than in an FLD scheme. In the last test we performed a fully
dynamic core-collapse simulation up to $300\,$ms post bounce and we found very good agreement with
the corresponding results obtained with the Boltzmann-type RHD codes VERTEX-PROMETHEUS and
AGILE-BOLTZTRAN. For this scenario we conducted additional test runs which checked the robustness
with respect to our integration algorithm and that revealed the convenient outcome that the physical
results only marginally depend on the actual choice of the closure relation.

Although in this paper we investigated many cases in which the AEF method yields results comparable
to the Boltzmann equation, one should keep in mind that the computational advantages of the AEF
method compared to a Boltzmann solver do not come for free. That is, the closure relation between
angular moments cannot be fulfilled to the same degree for arbitrarily complex radiation fields. A
related, particular shortcoming of the AEF method is that even in the optically thin limit radiation
fronts interfere with each other, which is an immediate result of the closure being a generally
non-linear function of the evolved moments (for consequences of this features in the case of a
post-merger BH-torus system, see the appendix of \citealp{Just2015a}). Nevertheless, we consider
these deficiencies to be not overly restrictive for our purposes since the present code is primarily
designed to describe systems in which a single extended source dominantly determines the evolution
of the radiation field.

We have already started to operate the described code to examine the combined neutrino- and
magnetic-field effects in two-dimensional CCSNe \citep{Obergaulinger2014} and to study the impact of
neutrino transport on outflows from post-NS merger BH-accretion tori \citep{Just2015a}. Since here
we only discussed test problems which are simplified in one way or another we refer the reader to
these mentioned papers for more specific discussions of results obtained with the AEF scheme in
multidimensional applications. In the future we plan to improve the presented code by supplementing
the coevolution of $\mu$- and $\tau$-neutrinos, refining the set of neutrino-interaction channels,
and adding relativistic corrections.


\section*{Acknowledgments}
We are grateful to Bernhard M\"{u}ller for helpful discussions about various neutrino-transport
issues, to Lorenz H\"{u}depohl for providing his \textsc{Vertex-Prometheus} data used for the test
in Sec.~\ref{sec:neutr-radi-field}, to Bruno Peres for useful comments, and to the anonymous referee
for helping to improve the clarity of the manuscript. We acknowledge support by the
Max-Planck/Princeton Center for Plasma Physics (MPPC), by the Deutsche Forschungsgemeinschaft
through the Transregional Collaborative Research Center SFB/TR 7 ``Gravitational Wave Astronomy'',
by the Computational Center for Particle and Astrophysics (C2PAP) as part of the Cluster of
Excellence EXC 153 ``Origin and Structure of the Universe'', by the European Research Council
through ERC-AdG No.~341157-COCO2CASA at Garching and grant CAMAP-259276 at Valencia, and by the
Spanish Ministerio de Ciencia e Innovaci{\'o}n through grant AYA2013-40979-P Astrof{\'i}sica
Relativista Computacional at Valencia. The computations were performed at the Rechenzentrum
Garching, the Leibniz-Rechenzentrum, and at the cluster Lluisvives of the University of Valencia.



\appendix

\section{Derivation of the 3rd-Moment Factor for the Minerbo Closure}\label{sec:deriv-3rd-moment}

Here we outline the derivation of Eq.~\eqref{eq:defqsmall}, which expresses the 3rd-moment factor
$q$ as a function of the flux factor $f$ assuming that the radiation field obeys the Minerbo
closure. The Minerbo closure can be found by maximizing the entropy functional for a particle
configuration with a low number density (\citealp{Minerbo1978}; see also
\citealp{Cernohorsky1994}). This leads to the axisymmetric distribution function
\begin{equation}\label{eq:appq_F}
  \mathcal{F(\mu)} = \frac{1}{4\pi}\eul^{a\mu-\eta}\, ,
\end{equation}
in which $\mu$ is the cosine of the angle between the direction of particle momentum and the
symmetry axis, and the parameters $a, \eta$ are two Lagrange multipliers. The latter can be
eliminated using the definition of the 0th and 1st moments (cf. Eq.~\eqref{eq:momdef1}),
\begin{subequations}\label{eq:appq_ef}
\begin{align}
  & E = \left(\frac{\eps}{h \cl}\right)^3 \int\mathrm{d}\Omega \mathcal{F}
  = \left(\frac{\eps}{h \cl}\right)^3 \frac{\eul^{-\eta}}{a} \sinh{a} \, ,\\
  & \frac{1}{c}F = \left(\frac{\eps}{h \cl}\right)^3 \int\mathrm{d}\Omega \mu \mathcal{F} =
  \left(\frac{\eps}{h \cl}\right)^3 \frac{\eul^{-\eta}}{a} \left(\cosh{a} -
    \frac{\sinh{a}}{a}\right) \, ,
\end{align}
\end{subequations}
which results in
\begin{equation}\label{eq:appq_f}
 f\equiv \frac{F}{c E}=\coth{a} - \frac{1}{a}=L(a)
\end{equation}
and, hence, in $a=L^{-1}(f)$, where $L$ and $L^{-1}$ are the Langevin function and its inverse,
respectively. Since $L$ cannot be inverted analytically, we employ for all practical purposes the
following approximation of $L^{-1}$ \citet{Cernohorsky1994}:
\begin{equation}\label{eq:appq_linv}
 a=L^{-1}(f) \approx \frac{15f}{5-3f^2 + f^3 - 3f^4} \, ,
\end{equation}
of which the error is at most a few per cent. Now, the 2nd-moment factor $\chi$,
Eq.~\eqref{eq:minclos}, and 3rd-moment factor $q$, Eq.~\eqref{eq:defqsmall}, are obtained from their
definition in Eqs.~\eqref{eq:defchi} and~\eqref{eq:defqfac}, respectively, using
Eqs.~\eqref{eq:appq_F}--\eqref{eq:appq_linv} and $\vecn\cdot \vecn_\vecF=\mu$.

\section{Multidimensional Characteristic Wave Speeds}\label{sec:mult-char-wave}

\begin{figure*}
  \includegraphics[width=0.49\textwidth]{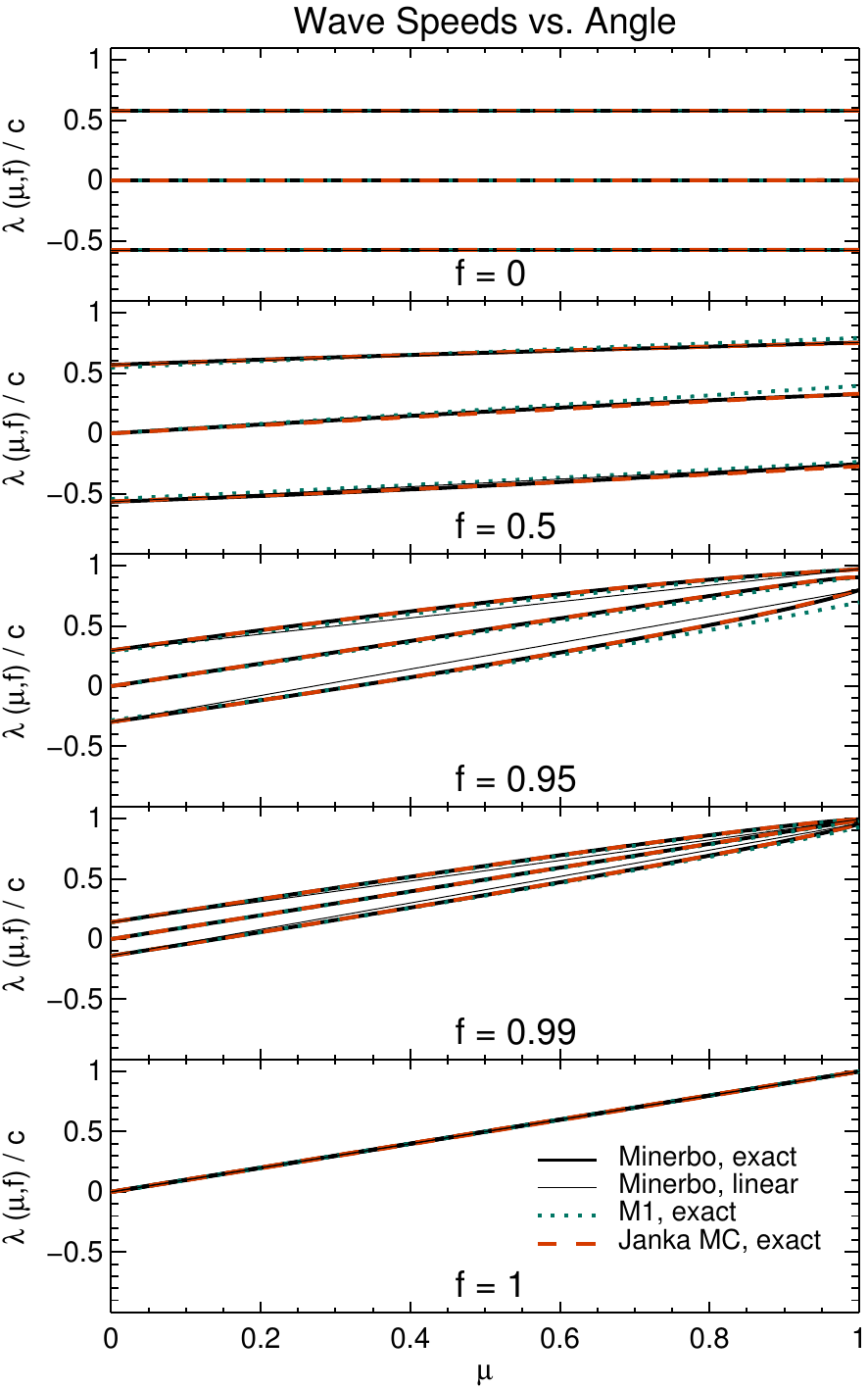}
  \includegraphics[width=0.49\textwidth]{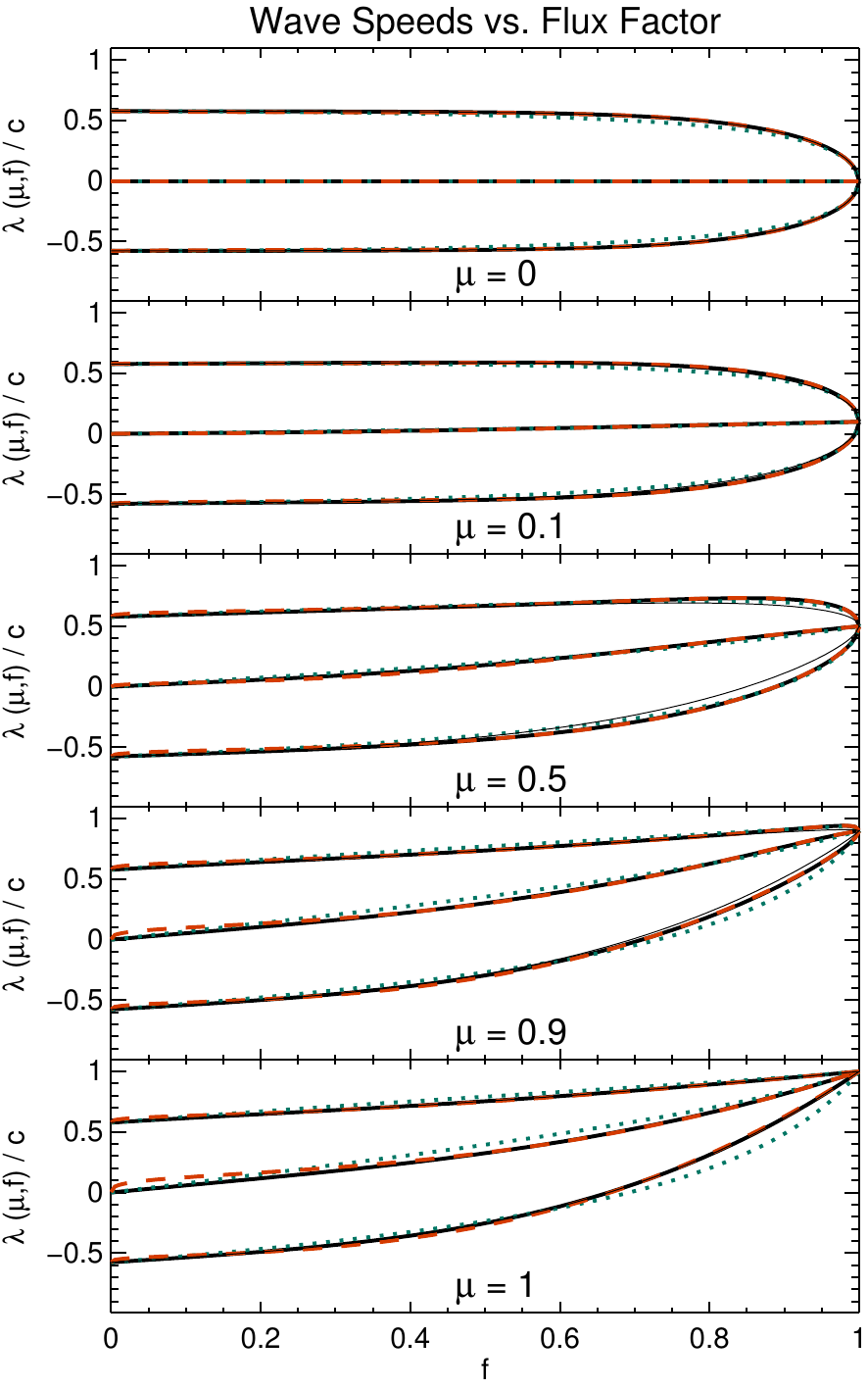}
  \caption{Comparison of characteristic speeds $\lambda_{\pm,0}(\mu,f)$ (with
    $\lambda_- < \lambda_0 < \lambda_+$) as functions of the direction cosine $\mu$ (left) and flux
    factor $f$ (right) for the Minerbo, $M_1$ and Monte-Carlo closure
    (cf. $\chi_{\mathrm{Minerbo}},\chi_{M_1}$ and $\chi_{\mathrm{Janka}}$ in
    Eqs.~\eqref{eq:closures}, respectively). The thick lines denote the exact wave speeds while the
    thin black lines depict for the Minerbo closure the linear expansion in $\mu$,
    cf. Eq.~\eqref{eq:sigspeedsalpha}.}
  \label{fig:wavespeeds}
\end{figure*}

In this section we provide supplementary information about the speeds $\lambda_{\pm,0}$
(cf. Sec.~\ref{ssec:hyperbolic}) of the characteristic waves associated with the hyperbolic part of
the multidimensional two-moment system closed by an algebraic Eddington tensor. Without loss of
generality, we consider characteristic waves propagating along the $x$-direction and assume that the
1st-moment vector $\vecF = (F_x,F_y) = (\mu,\sqrt{1-\mu^2})|\vecF|$ lies in the $x-y$ plane and
forms an angle $\alpha_\vecF$, where $\mu\equiv \cos\alpha_\vecF$, with the $x$-axis. The wave
speeds are then obtained as eigenvalues of the Jacobi matrix whose components are given in terms of
a general closure $\chi(e,f)$ as follows:
\begin{subequations}\label{eq:jacobi}
\begin{align}
  & \left( \frac{\partial F_x}{\partial E},\frac{\partial F_x}{\partial F_x}, \frac{\partial
      F_x}{\partial F_y} \right) =  \big( 0, 1, 0 \big) \, , \displaybreak[1]\\
  & \cl^2\frac{\partial P_{xx}}{\partial E} =\frac{\cl^2}{2} \left[1 - \mu^2 + ( 3\mu^2 - 1)\left(
      e\frac{\partial\chi}{\partial e} - f\frac{\partial\chi}{\partial f} + \chi \right)\right] \, , \displaybreak[1]\\
  & \cl^2\frac{\partial P_{xx}}{\partial F_x} = \frac{\cl\mu}{2f} \left[f\frac{\partial\chi}{\partial
      f} \left(3 \mu^2 - 1\right) - 2 (\mu^2-1) (3 \chi - 1)\right] \, , \displaybreak[1] \\
  & \cl^2\frac{\partial P_{xx}}{\partial F_y} = \frac{\cl\sqrt{1-\mu^2}}{2f}\left[
    f\frac{\partial\chi}{\partial f} (3\mu^2-1) - 2 \mu^2 (3\chi-1)\right] \, , \displaybreak[1] \\
  & \cl^2\frac{\partial P_{xy}}{\partial E} = \frac{\cl^2\mu\sqrt{1-\mu^2}}{2}\left[ 3 \left( 
      e\frac{\partial\chi}{\partial e} - f\frac{\partial\chi}{\partial f} + \chi\right) - 1\right] \, , \displaybreak[1] \\
  & \cl^2\frac{\partial P_{xy}}{\partial F_x} = \frac{\cl\sqrt{1-\mu^2}}{2f}\left[ 3\chi - 1 + 
    \mu^2 \left(2 + 3f\frac{\partial\chi}{\partial f} - 6 \chi\right) \right] \, , \displaybreak[1] \\
  & \cl^2\frac{\partial P_{xy}}{\partial F_y} = \frac{\cl\mu}{2f}\left[ 3 f\frac{\partial\chi}{\partial f} 
    (1-\mu^2) + (2 \mu^2 - 1) (3\chi - 1 )\right] \, .
\end{align}
\end{subequations}
We plot in Fig.~\ref{fig:wavespeeds} the resulting wave speeds as functions of $\mu$ and $f$ for
three different closures together with the linear expansion in $\mu$, Eq.~\eqref{eq:sigspeedsalpha},
for the Minerbo closure. It can be seen that all considered closures lead to rather similar wave
speeds. Moreover, the $\mu$-dependence of the wave speeds is very close to linear, which suggests
that using the linearized versions instead of the exact wave speeds as signal speeds for the Riemann
solver, cf. Eq.~\eqref{eq:HLLhyp}, is a justified approximation.

\section{Numerical Methods Used for the Comparison in
  Sec.~4.3.1}\label{sec:append-a:-numer-1}

\subsection{Flux-limited diffusion solver}\label{sec:flux-limit-diff-1}

To construct the spherically symmetric FLD solver used to find the corresponding solutions in
Sec.~\ref{sec:neutr-radi-field} we start from the presented AEF scheme and essentially drop the
evolution equation for the radial flux density $F_r=F$ and all velocity-dependent terms. What
remains to be determined is a suitable numerical representation of the flux $F_{\mathrm{FLD}}$
(cf. Eq.~\eqref{eq:fldflux}) at each cell interface. To this end, we first compute the cell-centered
version of the flux, for which we need a cell-centered representation of $\partial_r E$. The latter
is discretized as
\begin{equation}\label{eq:app_fld1}
  \partial_r E \longrightarrow
  (\hat{E}_{\hi+1} - \hat{E}_{\hi-1})/(2\Delta r_{\hi})\, ,
\end{equation}
where $\hi$ denotes the radial grid index and $\Delta r_{\hi}\equiv
r_{\hi+\half}-r_{\hi-\half}$. Out of the resulting cell-centered values $\hat{F}_{\mathrm{FLD},\hi}$
of the flux density, we compute the cell-interface values as
\begin{equation}\label{eq:app_fld2}
  \hat{F}_{\mathrm{FLD},\hi+\half}  \equiv \lambda_{\mathrm{FLD},\hi+\half} \hat{F}_{\mathrm{FLD},\hi} 
  + (1-\lambda_{\mathrm{FLD},\hi+\half}) \hat{F}_{\mathrm{FLD},\hi+1} \, ,
\end{equation}
where the interpolation parameter $\lambda_{\mathrm{FLD}}\in[0,1]$ is introduced to ensure that the
numerical method is based on central-type fluxes in the parabolic regime, where
$\lambda_{\mathrm{FLD}}\rightarrow 1/2$ should hold, and on upwind-type fluxes in the hyperbolic
regime, where $\lambda_{\mathrm{FLD}}\rightarrow 1$ for $f\equiv F_{\mathrm{FLD}}/(\cl E)\rightarrow
1$ and $\lambda_{\mathrm{FLD}}\rightarrow 0$ for $f \rightarrow -1$ should hold. We compute
$\lambda_{\mathrm{FLD}}$ as a function of the signed, average flux-factor $f_{\mathrm{\hi+\half}}\equiv
(\hat{F}_{\mathrm{FLD},\hi}/\hat{E}_{\hi} + \hat{F}_{\mathrm{FLD},\hi+1}/\hat{E}_{\hi+1}) /
(2\cl)$ as
\begin{equation}\label{eq:app_fld3}
\lambda_{\mathrm{FLD},\hi+\half}\equiv \max \{ \: \min \{ \: f_{\mathrm{\hi+\half}}+1/2, 1 \},\: 0 \} \, .
\end{equation}
The time step $\Delta t$ for the explicit integration is chosen to fulfill $\Delta t <
\min_\hi\{\tau_{\mathrm{FLD},\hi}\}$, cf. Eq.~\eqref{eq:FLDtimestep} with $\alpha\approx 0.5$.

\subsection{Tangent-ray Boltzmann-solver}\label{sec:append-a.2:-boltzm}

For calculating the reference solution in Sec.~\ref{sec:neutr-radi-field} we employ a
time-independent tangent-ray variable-Eddington-factor (TR-VEF) scheme closely analog to what is
described in Chap.~83 of \citet{Mihalas1984} and in \citet{Rampp2002}. For the discretization of the
two-moment system (cf. Eqs.~\eqref{eq:tmt1} with $\partial_t=0$, $\vecv=0$ and the source terms
expressed as in Eq.~\eqref{eq:sourceterms}) we use a finite-difference scheme in which we interprete
the energy densities $\hat{E}_{\hi}$ to be located at cell centers and the flux densities
$\hat{F}_{\hi+\half}$ to be located at cell interfaces. Using first-order differences for the radial
derivatives, this leads to the following linear system of equations:
\begin{subequations}\label{eq:app_tr1}
\begin{align}
  & \frac{\hat{F}_{\hi+\half} - \hat{F}_{\hi-\half}}{\Delta r_\hi} + \frac{2\hat{F}_{\hi}}{r_{\hi}}
  + \cl\hat{\kappa}_{\mathrm{a},\hi}\hat{E}_{\hi} = \cl\hat{\kappa}_{\mathrm{a},\hi}\hat{E}^\mathrm{eq}_{\hi} \, ,\\
  & \frac{\hat{\chi}_{\hi+1}\hat{E}_{\hi+1} - \hat{\chi}_{\hi}\hat{E}_{\hi}}{\Delta r_{\hi+\half}} +
  \frac{\hat{E}_{\hi+\half}}{r_{\hi+\half}} \left(3\hat{\chi}_{\hi+\half}-1\right) +
  \cl\hat{\kappa}_{\mathrm{tra},\hi+\half}\hat{F}_{\hi+\half} = 0 
\end{align}
\end{subequations}
for $\hi=1,\ldots,N_r$, where $\Delta r_{\hi+\half} \equiv r_{\hi+1} - r_{\hi}$ and
$\hat{E}_{\hi+\half}, \hat{F}_{\hi}$ are linear interpolations of the nearest neighbors on the
corresponding staggered grid. The Eddington factors $\chi$ required to solve Eqs.~\eqref{eq:app_tr1}
have to be obtained from the Boltzmann equation. For the considered type of interactions, the latter
reads \citep[e.g.][]{Cernohorsky1989}
\begin{multline}\label{eq:app_tr2}
  \mu\partial_r \mathcal{I}(r,\mu) + \frac{1-\mu^2}{r}\partial_\mu\mathcal{I}(r,\mu) = \\
  \kappa_{\mathrm{a}} \mathcal{I}^{\mathrm{eq}} + 
  \frac{1}{4\pi}\left( \kappa_{\mathrm{s}}^0\,\cl\, E + \kappa_{\mathrm{s}}^1\,\mu \,F \right) 
  - (\kappa_{\mathrm{a}}+\kappa_{\mathrm{s}}^0)\mathcal{I}(r,\mu) \, ,
\end{multline}
where $\mu$ is the cosine of the angle between the radial direction and the direction of neutrinos
in momentum space, $\mathcal{I}^{\mathrm{eq}}$ is the specific intensity corresponding to neutrinos
being in thermal equilibrium, and the scattering opacity
$\kappa_{\mathrm{s}}=\kappa_{\mathrm{s}}^0-\kappa_{\mathrm{s}}^1/3$ is decomposed into an isotropic
(superscript ``0'') and an anisotropic (superscript ``1'') contribution. For solving
Eq.~\eqref{eq:app_tr2} we first make a change of variables
$(r,\mu)\rightarrow(s\equiv \mu\,r, p\equiv \sqrt{1-\mu^2}r)$ and we use
\begin{subequations}\label{eq:app_tr3}
\begin{align}
  j(s,p) &\equiv (\mathcal{I}(\mu)+\mathcal{I}(-\mu))/2 \, ,\\
  h(s,p) &\equiv (\mathcal{I}(\mu)-\mathcal{I}(-\mu))/2
\end{align}
\end{subequations}
to rewrite Eq.~\eqref{eq:app_tr2} into the following system of equations
\begin{subequations}\label{eq:app_tr4}
\begin{align}
  \partial_sh(s,p) + (\kappa_{\mathrm{a}}+\kappa_{\mathrm{s}}^0)j(s,p) &= \kappa_{\mathrm{a}}\mathcal{I}^\mathrm{eq} 
  + \kappa_{\mathrm{s}}^0\frac{\cl}{4\pi}\,E \, ,\\
  \partial_sj(s,p) + (\kappa_{\mathrm{a}}+\kappa_{\mathrm{s}}^0)h(s,p) &= \kappa_{\mathrm{s}}^1\frac{\cl}{4\pi}\, \mu \,F .
\end{align}
\end{subequations}
Equations~\eqref{eq:app_tr4} are independently solved along each tangent-ray characterized by a
constant impact parameter $p_\hi=r_\hi$ $(\hi=1,\ldots,N_r)$. We employ a finite-difference
prescription in which we interprete $\hat{j}_{\hi}$ to be located at cell centers and
$\hat{h}_{\hi+\half}$ to be located at cell interfaces of the corresponding tangent-ray grid. The
linear system of difference equations to be solved then follows in an analog fashion as for the
two-moment system, cf. Eq.~\eqref{eq:app_tr1}. For the numerical angular integration of the specific
intensity we use the quadrature weights of \citet{Yorke1980}.

In practice, we obtain the initial values of $\chi$ by solving the Boltzmann equation with $E=F=0$.
Subsequently, the two-moment system and the Boltzmann equation are iteratively solved until
convergence is achieved.

\label{lastpage}

\end{document}